\documentclass[10pt,aps,a4paper,prd,floatfix,onecolumn,notitlepage,nofootinbib]{revtex4-1}

\usepackage[left=2cm,right=2cm,top=2cm,bottom=2cm]{geometry}

\usepackage[utf8]{inputenc}
\usepackage[british,UKenglish,USenglish,american]{babel}
\usepackage{amsmath}

\usepackage{graphicx}
\usepackage{subcaption}

\usepackage{slashed}

\usepackage{mathrsfs}
\usepackage{amssymb}
\usepackage{bbm}

\usepackage{nicefrac}

\usepackage{microtype}

\usepackage{physics}

\usepackage{xcolor}

\usepackage[normalem]{ulem}

\newcommand\numberthis{\addtocounter{equation}{1}\tag{\theequation}}

\DeclareMathOperator{\diag}{diag}

\newcommand{\p}{\prime}

\renewcommand{\bar}{\overline}

\usepackage{bm}\renewcommand{\vec}{\bm}

\newcommand{\e}{\mathrm{e}}

\newcommand{\fermi}{\mathrm{F}}

\newcommand{\bose}{\mathrm{B}}

\newcommand{\DD}[1]{ \mathcal{D}#1 \,} 

\newcommand{\sumeta}{\sum_{\eta=\pm 1}}
\newcommand{\sumi}{\sum_{i=0,1}}

\begin{document}

\title{
Functional renormalization group study of the critical region of the quark-meson model with vector interactions
}

\author{Renan C\^amara Pereira$^1$, Rainer Stiele$^{2,3,4}$ and Pedro Costa$^1$  }
\affiliation{$^1$ CFisUC, Department of Physics, University of Coimbra, P-3004 - 516  Coimbra, Portugal}
\affiliation{$^2$ INFN -- Sezione di Torino, Via Pietro Giuria 1, I-10125 Torino, Italy}
\affiliation{$^3$ Univ~Lyon, ENS de Lyon, D\'epartement de Physique, F-69342, Lyon, France }
\affiliation{$^4$ Univ~Lyon, Univ Claude Bernard Lyon 1, CNRS/IN2P3, IP2I Lyon, F-69622, Villeurbanne, France}
\email{renan.pereira@student.uc.pt}

%
%
%
%

\date{\today}

\begin{abstract}
The critical region of the two flavour Quark-Meson model with vector interactions is explored using the Functional Renormalization Group technique, a non-perturbative method that takes into account quantum and thermal fluctuations. Special attention is given to the low temperature and high density region of the phase diagram, which is very important for the evolution of proto-neutron stars and to construct the equation of state of compact stars. 
As in previous studies, without repulsive vector interaction, an unphysical region of negative entropy density is found near the first order chiral phase transition. We explore the connection between this unphysical region and the chiral critical region, especially the first order line and spinodal lines, using also different values for vector interactions. 
The effect of finite vector interactions in the critical region is explored. We find that the unphysical negative entropy density region appears because the $s=0$ isentropic line, near the critical region, is displaced from its $T=0$ location. For certain values of vector interactions this region is pushed to lower temperatures and high chemical potentials in such way that the negative entropy density region on the phase diagram of the model can even disapear.
In the case of finite vector interactions, the location of the critical end point has a non-trivial behaviour in the $T-\mu_B$ plane, which differs from that in mean-field calculations.
\end{abstract}

\maketitle

\section{Introduction}
\label{introduction}

The phase diagram of Quantum Chromodynamics (QCD) is a widely studied topic by both experimental and theoretical physics and much has been learned about its properties since its first conjecture by N. Cabibbo and G. Parisi \cite{CABIBBO197567}. However, the phase structure at low temperatures and high baryonic density remains a mystery, e.g., the existence of a first order phase transition and the critical end point (CEP).

Heavy ion collision (HIC) experiments conducted by the STAR Collaboration in the Relativistic Heavy Ion Collider (RHIC) at Brookhaven National Laboratory \cite{Adamczyk:2014fia,Adamczyk:2017wsl,Adamczyk:2017iwn}  and by NA61/SHINE Collaboration in the Super Proton Synchrotron (SPS) at CERN \cite{Aduszkiewicz:2015jna,Grebieszkow:2017gqx}, are currently, not only studying the properties of the quark-gluon plasma (QGP), but also trying to map the phase boundary of QCD. In the future, other facilities like the Nuclotron based Ion Collider fAcility (NICA) at Joint Institute for Nuclear Research \cite{NICAWP}, Facility for Antiproton and Ion Research (FAIR) at GSI Helmholtzzentrum für Schwerionenforschung \cite{Ablyazimov:2017guv} and J-PARC Heavy Ion Project at Japan Proton Accelerator Research Complex (J-PARC) \cite{Sako:J-PARC}, will also join the collective effort to better understand the properties of nuclear and quark matter under extreme conditions of temperature, density and in the presence of magnetic fields.

The low temperature and high density region of the phase diagram, where the CEP might exist, is not only interesting for nuclear and particle physics studies, but also extremely important for astrophysical applications, namely to study the evolution and properties of neutron stars. Since the equation of state for nuclear matter derived from first principles is still unknown, the core composition of these objects is still an open question and several options have been proposed. Some model calculations propose different neutron star core compositions such as hyperon matter, pion or kaon condensates and quark matter \cite{Weber:2007ch,Vidana:2018lqp}.

From the theoretical side, lattice QCD (LQCD), a first principles method, is not able to shed light on these questions since it is not yet possible to do LQCD calculations at finite chemical potential due to the famous sign problem which renders the importance sampling needed in Monte Carlo simulations ineffective \cite{Schmidt:2017bjt}. Recently, different approaches have been tried to circumvent this problem like reweighing, Taylor series expansions, imaginary chemical potential and Complex Langevin dynamics \cite{Schmidt:2017bjt,Seiler:2017wvd,Iwami:2015eba}. Due to these shortcomings, to seek for qualitative and increasingly quantitative understanding of QCD matter, other theoretical tools have been applied to study the phase diagram, such as Dyson-Schwinger equations and effective model calculations. Some of these calculations predict a first-order phase transition and a CEP in the low temperature and high density region of the phase diagram \cite{Fischer:2013eca,Fischer:2014ata,Herbst:2010rf,Gupta:2011ez,Costa:2010zw,Stiele:2016cfs,Costa:2019bua}.

Model calculations, using the Nambu$-$Jona-Lasinio (NJL) model or the Quark-Meson (QM) model, can be improved by going beyond the common mean field approximation (MF) \cite{Nikolov:1996jj,Nemoto:1999qf,Oertel:2000jp,Baacke:2002pi,Andersen:2008qk,Muller:2010am,YAMAZAKI201319,Zacchi:2017ahv,CamaraPereira:2020ipu}. Usually, when dealing with these chiral effective models, the quark contribution to the path integral can be reduced to a quadratic interaction and be exactly integrated out. The remaining path integrals, usually related to meson degrees of freedom, are not quadratic and some approximation has to be performed in order to obtain an effective action. In the MF approximation the remaining path integrals are calculated by the saddle-point approximation which effectively means that the only field configuration taken into account is the classical one, all quantum fluctuations to the remaining fields are left aside.

One way to go beyond the MF approximation is using an application of the renormalization group to continuous field theories, the Functional Renormalization Group (FRG), a powerful non-perturbative method which allows to incorporate quantum and thermal fluctuations in a field theory. The renormalization group is an important tool in theoretical physics since it allows the study of physical phenomena in different scales of distance and/or energy with enormous range of applications such as: studying the strong interaction, the electroweak phase transition, effective models of nuclear physics, condensed matter physics systems and quantum gravity \cite{Avila:2015iza,Saueressig:2015xua,Safari:2017irw}. Some of its most important applications in the history of physics are the  elimination of ultraviolet divergences in renormalizable quantum field theories and its application to explain the universality properties of continuous phase transitions. The FRG has been extensively used to study the QCD phase diagram using chiral effective models beyond the MF, like the NJL model \cite{Fukushima:2012xw,Braun:2011pp,Aoki:2014ola,Aoki:2015hsa} and the QM model \cite{Schaefer:2004en,Herbst:2013ail,Fu:2015naa,Herbst:2013ufa,Tripolt:2013jra,Jung:2016yxl,Andersen:2013swa}. For detailed reviews on the FRG method see \cite{Gies:2006wv,Pawlowski:2005xe,Delamotte:2007pf}.

The application of the FRG method to the 2-flavour QM model leads to the presence of an unphysical negative entropy density region in the low temperature and high density region of the phase diagram, near the critical region where a first order chiral phase transition and CEP are predicted by the model. This behaviour was first discussed in detail by R. Tripolt et al. in \cite{Tripolt:2017zgc}, although previous FRG studies have reported decreasing pressures with increasing temperatures \cite{Herbst:2013ail,Fu:2015naa}. The authors have put forward some explanations for this unphysical region: the truncation used to derive the QM flow equation is not enough to define a thermodinamically consistent model beyond the mean field approximation or the specific choice of regulator, used to account for fluctuations in the model, is not appropriate. They also discuss the possibility that the source for such behaviour is physical like a pairing transition to a color superconducting phase or to the existence of inhomogeneous phases. For more details see \cite{Tripolt:2017zgc}.

In this work, we will consider the 2-flavour QM model with vector interactions to explore the connection between these vector degrees of freedom with the critical region predicted by the model and the unphysical negative entropy density region. Vector interactions are very important to describe in-medium properties and are widely used to describe neutron stars \cite{Zacchi:2015lwa,Pereira:2016dfg,Otto:2019zjy}, study the curvature of the critical line \cite{Bratovic:2012qs} and vector meson masses. The general effect of these interactions on the phase diagram, in MF calculations, is to drive the first order phase transition and CEP towards lower temperatures and higher chemical potentials. For high enough vector couplings, the critical region disappears leaving a smooth crossover for the chiral transition for all values of temperature and chemical potential. The $\omega_0$ and $\rho_0^3$ vector mesons will be considered. The $\omega_0$ vector is known to stiffen the equation of state of quark matter while the $\rho_0^3$ can be very important in isospin asymmetric systems, acting as an isospin restoring interaction. Hence the inclusion of these degrees of freedom can be essential to describe certain physical systems at high densities like neutron stars.

This paper is organized as follows. In Section II the 2-flavour QM model, including vector interaction and the FRG formalism are presented. The vector degrees of freedom are frozen and the flow equations for the effective potential and entropy density are laid out. In Section III the results are presented and the effect of the vector interactions on the critical region and on the unphysical negative entropy density are discussed. Finally, in Section IV conclusions are drawn and further work is proposed.

\section{Model and Formalism}
\label{model_and_formalism}

The 2-flavour Quark-Meson model is invariant under chiral symmetry i.e., $SU\qty(2)_L \times SU\qty(2)_R$. It can be built by considering a quark field $\psi$, interacting with dynamical meson fields via symmetry conserving terms at the Lagrangian level. Considering the scalar and pseudoscalar fields, $\sigma$, $\vec{\pi}$ and the isoscalar-vector and isovector-vector fields, $\omega_\mu$ and $\vec{\rho}_\mu$, the following symmetry conserving Lagrangian density, in Minkowski spacetime, can be written:
\begin{align*}
\mathcal{L} 
& =  
\bar{\psi} 
\qty[ i\slashed{\partial} - 
g_S \qty( \sigma + i \vec{\tau} \cdot \vec{\pi} \gamma_5 ) -
g_\omega \slashed{\omega} - g_\rho \vec{\tau} \cdot \vec{\slashed{\rho}}
+ \hat{\mu} \gamma_0
]\psi 
\\
&
+ \frac{ 1 }{ 2 } \qty( \partial_\mu \sigma )^2 
+ \frac{ 1 }{ 2 } \qty( \partial_\mu \vec{\pi} )^2 
- \frac{ 1 }{ 4 } F_{\mu \nu} F^{\mu \nu}
- \frac{ 1 }{ 4 } \vec{R}_{\mu \nu} \vec{R}^{\mu \nu}
\\
&
- U \qty( \sigma, \vec{\pi}, \omega_\mu, \vec{\rho}_\mu ) .
\numberthis
\label{eq_QMvec_lagrangian}
\end{align*}
Here, the quark field $\psi$ is a $N_c$-component vector in flavour space, where each component is a Dirac spinor and $\vec{\tau}$ are the three Pauli matrices. To study the system at finite density, a diagonal quark chemical potential matrix, $\hat{\mu}=\diag \qty(\mu_u,\mu_d)$, was also included. The field strength tensors $F_{\mu \nu}$ and $\vec{R}_{\mu \nu}$ are used to define the kinetic terms for the $\omega_\mu$ and $\vec{\rho}_\mu$ fields, respectively, and are given by: 
\begin{align}
F_{\mu \nu} & = \partial_\mu \omega_\nu - \partial_\nu \omega_\mu ,
\\
\vec{R}_{\mu \nu} & = \partial_\mu \vec{\rho}_\nu - \partial_\nu \vec{\rho}_\mu 
- g_\rho \vec{\rho}_\mu \times \vec{\rho}_\nu .
\end{align}
The potential $U \qty( \sigma, \vec{\pi}, \omega_\mu, \vec{\rho}_\mu )$, must be invariant under chiral symmetry except for an explicit chiral symmetry breaking term, that tilts the potential to give a finite mass to the Goldstone mode, the pion. At the mean field level, for the the $\sigma$ and $\vec{\pi}$ fields, this potential can include arbitrary powers of the chiral invariant $\phi^2 = \sigma^2 + \vec{\pi}^2$. Regarding the vector field contributions to the potential, several types of terms can be included, as long as the symmetries are respected. Due to the nature of the FRG calculation, one has only to specify the potential at the ultraviolet scale. 

The effect of current quark masses is to explicitly break chiral symmetry at the Lagrangian level, giving rise to a (slightly) massive Goldstone mode. This can be accomplished in the QM model by adding to the potential a non-vanishing expectation value for the $\sigma$ field, 
\begin{align}
U \qty( \sigma, \vec{\pi}, \omega_\mu, \vec{\rho}_\mu ) \to U \qty( \sigma, \vec{\pi}, \omega_\mu, \vec{\rho}_\mu  ) - c\sigma .
\end{align}
This field will behave as an order parameter for the chiral transition.

Finite temperature can be included using the Matsubara formalism in which a Wick rotation to Euclidean space-time is applied to the action. 
To simplify the notation, we introduce the fields, $\phi^i = \qty{ \sigma, \vec{\pi} }$ and $V^i_\mu = \qty{ \omega_\mu, \vec{\rho}_\mu }$. Using the Euclidean action $\mathcal{S}_E =
\int_0^{\nicefrac{1}{T}} \dd{\tau} \int \dd[3]{x} 
\mathcal{L}$, the generating functional of the fully connected Green’s functions, for a given temperature ($T$) and chemical potential ($\mu$), is defined as:
\begin{align}
\mathcal{W} \qty[ T, \mu; J^i, j^i_\mu ]
& = 
\ln
\int
\DD{\psi}
\DD{\bar{\psi}}
\DD{\phi^i}
\DD{V^i_\mu}
\e^{
- \mathcal{S}_E \qty[T, \mu; \psi,\bar{\psi},\phi^i,V^i_\mu] 
+ \int_0^{\nicefrac{1}{T}} \dd{\tau} \int \dd[3]{x} \qty( J^i \phi^i + j^i_\mu V^i_\mu  )
},
\end{align}
where we have included sources for the scalar fields $\qty(J^i)$ and for the vector fields $\qty(j^i_\mu)$, omitting the sources for the fermion fields which can be integrated out. First we are only interested in dealing with the path integral over the vector fields hence, we write:
\begin{align}
\mathcal{W} \qty[ T , \mu ; J^i, j^i_\mu ]
& = 
\ln
\int
\DD{V^i_\mu}
\e^{
- \mathcal{S}_V \qty[ T , \mu ; J^i, V^i_\mu ] 
+ \int_0^{\nicefrac{1}{T}} \dd{\tau} \int \dd[3]{x} 
j^i_\mu V^i_\mu 
} .
\end{align}

We have defined the effective action for vector degrees of freedom as:
\begin{align}
\mathcal{S}_V \qty[ T , \mu ; J^i, V^i_\mu] 
& =
-
\ln
\int
\DD{\psi}
\DD{\bar{\psi}}
\DD{\phi^i}
\e^{
- \mathcal{S}_E \qty[  T , \mu ; \psi,\bar{\psi},\phi^i,V^i_\mu ] 
+ \int_0^{\nicefrac{1}{T}} \dd{\tau} \int \dd[3]{x} J^i \phi^i  
}
\label{eq:vector_action}
\end{align}

Writing explicitly only the functional dependence on $j^i_\mu$, the effective action can be computed by Legendre transforming $\mathcal{W} \qty[j^i_\mu]$ as follows:
\begin{align}
\Gamma [ \tilde{V}^i_\mu ] = 
- \mathcal{W} [ j^i_\mu ] +
\int_0^{\nicefrac{1}{T}} \dd{\tau} \int \dd[3]{x}  j^i_\mu \tilde{V}^i_\mu  ,
\end{align}
where $\tilde{V}^i_\mu$ is  the expectation value of the vector fields $V^i_\mu$, in the presence of an external source $j^i_\mu$ and it is defined as:
\begin{align}
\frac{\delta \mathcal{W} \qty[ j^i_\mu ]}{ \delta j^i_\mu (x) }  = \tilde{V}^i_\mu (x) .
\end{align} 

The effective action can be written as \cite{das1993field}:
\begin{align}
\e^{ - \Gamma [ \tilde{V}^i_\mu ]  } = 
\int \DD{V^i_\mu}
\exp \qty{ 
-\mathcal{S}_V \qty[ V^i_\mu + \tilde{V}^i_\mu  ] 
+ \int_0^{\nicefrac{1}{T}} \dd{\tau} \int \dd[3]{x}
\fdv{ \Gamma [ \tilde{V}^i_\mu ] }{ \tilde{V}^i_\mu }
{V}^i_\mu 
}.
\label{master_eq_vector}
\end{align}

In a chiral effective model, the most important dynamics comes from chiral symmetry breaking. This means that the dynamics of the more massive fields, will play a secondary role. Hence, following previous approaches \cite{Drews:2014spa}, the vector fields will be used to model unknown degrees of freedom at short distances. This allows the use of the saddle-point approximation to solve the path integral in Eq.~(\ref{master_eq_vector}): the classical trajectories will be the most important for such fields, effectively freezing these heavier modes.

Using the saddle-point approximation, the main contribution to the integral will come from the minimum of the action $\mathcal{S}_V [ V^i_\mu  ] $ . Taylor expanding the action $\mathcal{S}_V [ V^i_\mu + \tilde{V}^i_\mu  ] $ around $\tilde{V}^i_\mu$ one can get the effective action in the mean field approximation,
\begin{align}
\e^{
-\Gamma [ \tilde{V}^i_\mu ] } 
\approx 
\e^{ -\mathcal{S}_V \qty[ \tilde{V}^i_\mu  ] }.
\label{MFA_exp_actions}
\end{align}
Where the mean field configuration is calculated from:
\begin{align}
\left .
\pdv{ \mathcal{S}_V ( {V}^i_\mu ) }{ {V}^i_\mu }
\right|_{ {V}^i_\mu = \tilde{V}^i_\mu }  = 
\left .
\pdv{ \Gamma ( {V}^i_\mu ) }{ {V}^i_\mu }
\right|_{ {V}^i_\mu = \tilde{V}^i_\mu } = 0
\label{eq:vec_stationary_cond}
\end{align}

Due to rotational invariance, the spatial components of the mean fields $\tilde{V}^i_j$, vanish \cite{Drews:2014spa}. Since we are not interested in studying the condensation of mean fields that change the properties of the vacuum, the non-diagonal elements $\tilde{{\rho}}^1=\tilde{{\rho}}^2=0$ will also be zero. Therefore, only the fields $\tilde{\omega}_0$ and $\tilde{\rho}_0^3$ can have non-zero values. These fields can be absorbed in the definition of the effective quark chemical potential matrix, $\tilde{\mu}$, as:
\begin{align}
\tilde{\mu} = 
\hat{\mu} - 
g_\omega \tilde{\omega}_0 - 
g_\rho\tilde{\rho}_0^3 \tau^3 . 
\end{align}
As expected, the mean field $\tilde{\rho}_0^3$, introduces an isospin asymmetry \cite{Drews:2014spa}.

Using Eqs.~(\ref{eq:vector_action}) and (\ref{MFA_exp_actions}), we can write the effective action as:
\begin{align}
{
\Gamma [ T, \mu ; J^i ] 
}
=
-
\ln
\int
\DD{\psi}
\DD{\bar{\psi}}
\DD{\phi^i}
\e^{
- \mathcal{S}_E \qty[ T,\tilde{\mu} ; \psi,\bar{\psi},\phi^i ] 
+ \int_0^{\nicefrac{1}{T}} \dd{\tau} \int \dd[3]{x}  J^i \phi^i 
} .
\label{Z_QM_mean_vectors}
\end{align}
The same approximation could be performed in the remaining meson path integrals and the quarks can be integrated out exactly, yielding the quark-meson model in the mean field approximation. However, in the present work, we will go beyond mean field by taking into account quantum fluctuations of the $\sigma$ and $\vec{\pi}$ fields using the FRG method. Modifying Eq.~(\ref{Z_QM_mean_vectors}) with a regulator term, the effective average action can be defined through a modified Legendre transformation \cite{Gies:2006wv}.

\subsection{The FRG method}

In the formalism of the FRG, the central object is the average effective action, $\Gamma_k$. This object depends explicitly on a momentum scale $k$ and has well defined limits: at the momentum scale $k=\Lambda$, we have the classical action to be quantized, $\mathcal{S}$, at the momentum scale $k=0$, all quantum fluctuations have been included and we obtain the full quantum effective action, $\Gamma$ i.e., 
\begin{align*}
\Gamma_{ k \to \Lambda} & = \mathcal{S},
\\
\Gamma_{ k \to 0} & = \Gamma.
\end{align*}
The average effective action interpolates these regimes in the space of all possible actions. The behaviour of this quantity during the renormalization group flow is governed by the so-called Wetterich equation \cite{Wetterich:1992yh}. For boson fields this equation is given by:
\begin{align}
\partial_t \Gamma_k \qty[ \phi ] = \frac{1}{2} \tr \qty{ \partial_t R_k^\mathrm{B} \qty( \Gamma_k^{(2)} \qty[ \phi ] + R_k^\mathrm{B} )^{-1} }  ,
\label{wetterichboson}
\end{align}
while, for fermions, it can be written as:
\begin{align}
\partial_t \Gamma_k \qty[ \psi ] = -  \tr \qty{ \partial_t R_k^\mathrm{F} \qty( \Gamma_k^{(1,1)} \qty[ \psi ] + R_k^\mathrm{F} )^{-1} }.
\label{wetterichfermion}
\end{align}
Here, $t=\ln \frac{k}{\Lambda}$ is the adimensional renormalization time with respect to some cutoff momentum $\Lambda$. The derivatives of the average effective action, $\Gamma_k^{(2)}$ and $\Gamma_k^{(1,1)}$, follow the usual notations for boson and fermion fields derivatives. The function $R_k$ is the so-called regulator function and it can be interpreted as a scale dependent mass term. As long as the interpolation between the ultraviolet and the infrared is correct, the regulator can take any functional form since it will only interfere in the arbitrary path taken, between these points in the theory space. Of course, since from the numerical point of view it is impossible to reach $k=0$ \cite{Gies:2006wv}, a finite infrared cut-off has to be applied meaning that different regulators might lead to different infrared effective actions.

These equations are exact functional differential equations for the effective average action which, in principle, can be solved given a set of initial conditions. Solving exactly the Wetterich is an impossible task due to the infinitely high coupled behaviour of the equation and some approximation scheme is needed. There are two widely used approximations schemes to solve this equation: the vertex expansion and the operator expansion. In the present work we will use the latter approach in the so-called local potential approximation (LPA), by building an effective average action based on the operator expansion with increasing mass dimension.

To use Wetterich's equation, a regulator function, which respects the interpolating limits of the effective average action, has to be chosen. We employ the so-called optimized or Litim regulator function \cite{Litim:2001up}, for bosons and fermions, respectively given by:
\begin{align}
R_k^\mathrm{B} \qty( \vec{p}^2 ) & = 
\qty( k^2 - \vec{p}^2 ) \theta \qty( k^2 - \vec{p}^2 ) ,
\\
R_k^\mathrm{F} \qty( \vec{p}^2 ) & =
- \vec{p} \cdot \vec{\gamma} \qty( \sqrt{ \frac{k^2}{\vec{p}^2} }  - 1)  \theta \qty( k^2 - \vec{p}^2 )  .
\end{align}

After solving the flow equation one can relate the effective action in the minimum, with the grand canonical potential, $\Gamma_{k=\mathrm{IR}}\qty(T,\mu )_{\text{min}} =\beta V \Omega\qty(T,\mu)$, to calculate several thermodynamic quantities of interest such as the pressure ($P$), particle ($\rho_i$), entropy ($s$) and energy densities ($\epsilon$),  using the following relations \cite{Kapusta:2006pm}:
\begin{align}
P \qty(T,\mu)  - P_0 & =  - \Omega \qty(T,\mu), 
\label{eq:def.pressao}
\\
\rho_i \qty(T,\mu) & =  -  \left( 
\frac{\partial \Omega \qty(T,\mu) }{\partial \mu_i} 
\right)_{T} , 
\label{eq:def.densidade}
\\
s \qty(T,\mu) & =  -  
\left(
\frac{\partial \Omega \qty(T,\mu)}{\partial T} \right)_ {\mu} ,
\label{eq:def.entropia}
\\
\epsilon \qty(T,\mu)  & = -P\qty(T,\mu) + Ts\qty(T,\mu) + \sum_i \mu_i \rho_i\qty(T,\mu). 
\label{eq:def.energia}
\end{align}
The constant $P_0$ is the vacuum pressure i.e., $P_0=P \qty(0,0)$.

\subsection{The flow equations}

In the lowest order of LPA, only the potential is scale dependent and using Eq.~(\ref{Z_QM_mean_vectors}), the imaginary-time average effective action can be written as:
\begin{align*}
\Gamma_k \qty[T, \mu] = 
\int_0^{\nicefrac{1}{T}} \dd{\tau} \int \dd[3]{x}
\bigg\{ &
\bar{\psi} 
\qty[ 
\slashed{\partial} +
g_S \qty( \sigma + i \vec{\tau} \cdot \vec{\pi} \gamma_5 ) 
- \tilde{\mu} \gamma_0
]\psi 
+ \frac{ 1 }{ 2 } \qty( \partial_\mu \sigma )^2 
+ \frac{ 1 }{ 2 } \qty( \partial_\mu \vec{\pi} )^2 
+ U_k \qty( \sigma, \vec{\pi}, \omega_0, \rho_0^3 )  
\bigg\} 
\numberthis
\label{effective_actionLPA}
\end{align*}
with the scale dependent  grand potential, $U_k$,  written in terms the chiral invariant $\phi^2 = \sigma^2 + \vec{\pi}^2$ and of the mean vector fields:
\begin{align}
U_k \qty( \sigma, \vec{\pi}, \omega_0, \rho_0^3 )   = 
U_k^\chi \qty( \phi^2 ) + 
U_k^V \qty( \sigma, \vec{\pi}, \omega_0, \rho_0^3 ) .
\end{align}
The contribution $U_k^\chi$ is a function of the chiral invariant only and the term $U_k^V$ represents the contribution from the vector degrees of freedom. While the functional dependence of the chiral part of the potential is calculated during the flow, a mean field approximation is performed in the vector channels, and a functional dependence for $U_k^V$ must be chosen. In this work we use:
\begin{align}
U_k^V \qty( \sigma, \vec{\pi}, \omega_0, \rho_0^3 )  =
- \frac{m_\omega^2}{2} {\omega}_{0}^2
- \frac{m_\rho^2}{2} {\rho}_{0}^2 .
\label{pot_vec_def}
\end{align}

Applying the stationary condition of Eq.~(\ref{eq:vec_stationary_cond}) to the effective average action in Eq.~(\ref{effective_actionLPA}) is equivalent to requiring that the potential $U_k \qty( \sigma, \vec{\pi}, \omega_0, \rho_0^3 )$ is minimal with respect to the vector fields, at each momentum scale $k$ \cite{Drews:2014spa,Zhang:2017icm}, i.e.,
\begin{align}
\left.
\pdv{ U_k ( \sigma, \vec{\pi}, \omega_{0}, \rho_{0}^3 ) }{ \omega_{0} } 
\right|_{\omega_{0} = \tilde{\omega}_{0,k} } 
=
\left.
\pdv{ U_k ( \sigma, \vec{\pi}, \omega_{0}, \rho_{0}^3 ) }{ \rho_{0}^3 } 
\right|_{\rho^3_{0} = \tilde{\rho}^3_{0,k} }
= 0 .
\label{eq:vec_stationary_cond_pot}
\end{align}
Hence, the vector fields acquire an implicit dependence on the RG scale $k$. This requirement ensures that the flow equation follows a path, in theory space, where the effective potential is always in the minimum with respect to the vector fields. 

Since no pion condensation will be considered, only the radial direction of the field, $\phi = \qty{ \sigma, \vec{0} } $, will contribute and we can switch variables to $\sigma$. 

Calculating the scale derivative of Eq.~(\ref{effective_actionLPA}) and using the stationary conditions for the vector fields given in Eq.(\ref{eq:vec_stationary_cond_pot}), yields:
\begin{align}
\partial_t U_k \qty( T, \mu ; \sigma, \tilde{\omega}_{0,k} , \tilde{\rho}^3_{0,k}  ) =
\partial_t U_k^\chi \qty( T, \mu ; \sigma,\tilde{\omega}_{0,k} , \tilde{\rho}^3_{0,k} ) .
\label{pot_flow_eq_after_vec_gap}
\end{align}
Hence, ensuring at each momentum shell that the vector fields stationary conditions hold, one can simply solve the flow equation for $U_k^\chi$ with effective quark chemical potentials modified by the vector fields. Putting everything together leads to the dimensionful LPA flow equation for the effective potential $U_k^\chi$:
\begin{align*}
\partial_t U_k^\chi \qty( T, \mu ; \sigma , \tilde{\omega}_{0,k} , \tilde{\rho}^3_{0,k} ) =
\frac{k^5}{12\pi^2} \Bigg\{ 
& 
\frac{1}{E_\sigma} \qty[ 1 + 2 n_\bose \qty( E_\sigma ) ]
+\frac{3}{E_\pi} \qty[ 1 + 2 n_\bose \qty( E_\pi ) ]
\\
&
-  
\frac{4 N_c }{E_\psi} \sumi
\qty(
1  
- \sumeta n_\fermi \qty( E_\psi - \eta \tilde{\mu}_{k,i} ) 
)
\Bigg\} . 
\numberthis
\label{eq:QM_pot_vec_flow_EQ}
\end{align*}
Here, the effective chemical potential, $\tilde{\mu}_{k,i}$, is defined as:
\begin{align}
\tilde{\mu}_{k,i} = \mu_i - v_{k,i}. 
\label{eff_mu_def}
\end{align}
With, $i=0$ for up quarks and $i=1$ for down quarks. The vector contribution, $v_{k,i}$, is defined as:
\begin{align}
v_{k,i} = 
g_\omega \tilde{\omega}_{0,k}
+ \qty(-1)^i 
g_\rho   \tilde{\rho}^3_{0,k} .
\label{def_vi}
\end{align}
The functions, $n_\bose \qty( E )$ and $n_\fermi \qty( E )$ are the Bose-Einstein and Fermi-Dirac distribution functions respectively and,
\begin{align}
E_\sigma^2 & =  k^2 + \partial_\sigma^2 U_k^\chi ,
\\
E_\pi^2 & = k^2 + \frac{ 1 }{ \sigma } \partial_\sigma U_k^\chi  ,
\\
E_\psi^2 & = k^2 + g_S^2 \sigma^2 .
\end{align}
After solving the above flow equation, one has access to  $U_{k=\text{IR}}^\chi $. The full potential in the infrared, $U_{k=\text{IR}}$, containing the contribution coming from vector fields, can easily be calculated with ${k=\text{IR}}$:
\begin{align}
U_{k=\text{IR}} \qty( T,\mu ; \sigma , \tilde{\omega}_{0} , \tilde{\rho}^3_{0} ) =
U_{k=\text{IR}}^\chi \qty( T, \mu ; \sigma , \tilde{\omega}_{0} , \tilde{\rho}^3_{0} )  
+
U_{k=\text{IR}}^V \qty( T, \mu ; \sigma , \tilde{\omega}_{0} , \tilde{\rho}^3_{0} ) .
\end{align}
The contribution coming from the vector fields can be calculated using Eq.~(\ref{pot_vec_def}) in the infrared.

Using Eq.~(\ref{eq:vec_stationary_cond_pot}) and the flow equation (\ref{eq:QM_pot_vec_flow_EQ}), applying the substitution $\partial_\mu \to - (\eta E/p) \partial_p $ and performing an integration by parts \cite{drewsthesis}, the following self-consistent equations for the vector fields can be written:
\begin{align}
g_\omega \tilde{\omega}_{0,k} \qty( T, \mu ; \sigma , \tilde{\omega}_{0,k} , \tilde{\rho}^3_{0,k} ) 
& = 
g_\omega \tilde{\omega}_{0,\Lambda} +
\frac{ 4 N_c }{ 12 \pi^2 } \qty( \frac{ g_\omega }{ m_\omega } )^2
\sumeta
\sumi
I_{k,\eta i} \qty( T, \mu ; \sigma , \tilde{\omega}_{0,k} , \tilde{\rho}^3_{0,k} ) ,
\label{omega_equation}
\\
g_\rho   \tilde{\rho}_{0,k}^3 \qty( T, \mu ; \sigma , \tilde{\omega}_{0,k} , \tilde{\rho}^3_{0,k} ) 
& = 
g_\rho \tilde{\rho}_{0,\Lambda}^3 + 
\frac{ 4 N_c }{ 12 \pi^2 } \qty( \frac{ g_\rho}{ m_\rho } )^2
\sumeta
\sumi
\qty(-1)^i
I_{k,\eta i} \qty( T, \mu ; \sigma , \tilde{\omega}_{0,k} , \tilde{\rho}^3_{0,k} )  . 
\label{rho_equation}
\end{align}
Where we have defined:
\begin{align}
I_{k,\eta i}\qty( T, \mu ; \sigma , \tilde{\omega}_{0,k} , \tilde{\rho}^3_{0,k} ) 
=
3\int_k^\Lambda \dd{p} 
\eta p^2 n_\fermi \qty( E_\psi - \eta \tilde{\mu}_{k,i} )
-
\qty[
\eta p^3 n_\fermi \qty( E_\psi - \eta \tilde{\mu}_{k,i} )
]_k^\Lambda .
\label{I_k,eta}
\end{align}

Since the equations depend only on the product $g_\omega \tilde{\omega}_{0,k}= \tilde{\omega}_k$ and $g_\rho \tilde{\rho}_{0,k}^3=\tilde{\rho}_k$, we take this combined quantity as variables. Likewise, the equations depend only on the combination $\frac{ g_\omega }{ m_\omega }=G_\omega$ and $\frac{ g_\rho }{ m_\rho }=G_\rho$, we take these ratios as parameters.

We are also interested in studying the entropy of the system including quantum fluctuations in order to understand what happens to the low temperature behaviour of this quantity. Hence, a flow equation for the entropy must be derived. Using Eqs.~(\ref{eq:def.entropia}) and (\ref{eq:QM_pot_vec_flow_EQ}) and considering that the temperature derivative commutes with the scale derivative, the following dimensionful flow equation for the chiral contribution to the average entropy density, $s_k^\chi$, can be derived:
\begin{align*}
\partial_t s_k^\chi \qty( T, \mu ; \sigma , \tilde{\omega}_{0,k} , \tilde{\rho}^3_{0,k} )  = 
- \frac{k^5}{12\pi^2} \Bigg\{ 
& 
2 n_\bose \qty( E_\sigma ) \qty[ 1 + n_\bose \qty( E_\sigma ) ]
\qty[
\frac{1}{T^2} + \frac{ \partial^2_\sigma s_k^\chi }{ 2 T E_\sigma^2 }
]
+
\partial^2_\sigma s_k^\chi 
\frac{ \qty[ 1 + 2 n_\bose \qty( E_\sigma ) ] }{ 2 E_\sigma^3 } 
\\
+ 
& 
6 n_\bose \qty( E_\pi ) \qty[ 1 + n_\bose \qty( E_\pi ) ]
\qty[
\frac{1}{T^2} + \frac{ \partial_\sigma s_k^\chi }{ 2 \sigma T E_\pi^2 }
]
+
3 \partial_\sigma s_k^\chi
\frac{ \qty[ 1 + 2 n_\bose \qty( E_\pi ) ] }{ 2 \sigma E_\pi^3 } 
\\
+ & 
\frac{ 4 N_c }{ E_\psi } \sum_{i=u,d} \sumeta 
\frac{ n_\fermi \qty( E_\psi - \eta \tilde{\mu}_{k,i} )  }{ T^2 }
\qty[ 
1 - n_\fermi \qty( E_\psi - \eta \tilde{\mu}_{k,i} )
]
\qty[ E_\psi - \eta \tilde{\mu}_{k,i} - \eta T \partial_T v_{k,i} ]
\Bigg\} . 
\numberthis
\label{eq:QM_entropy_vec_flow_EQ}
\end{align*}
If considering finite vector interactions, there is an extra contribution coming from the temperature dependence of the vector fields, at each momentum shell, $\partial_T v_{k,i}$. For the detailed calculation of this quantity, see Appendix \ref{vector_temperature_derivatives}.

The entropy density in the infrared, $s_{k=\text{IR}}$, containing the contribution coming from vector fields, can easily be calculated after solving the system of flow equations through:
\begin{align}
s_{k=\text{IR}} \qty( T, \mu ; \sigma , \tilde{\omega}_{0} , \tilde{\rho}^3_{0} ) =
s_{k=\text{IR}}^\chi \qty( T, \mu ; \sigma , \tilde{\omega}_{0} , \tilde{\rho}^3_{0} )  
- \pdv{ U_{k=\text{IR}}^V ( T, \mu ; \sigma , \tilde{\omega}_{0} , \tilde{\rho}^3_{0} ) }{ T }  .
\end{align}
The contribution coming from the vector fields can be calculated using the stationary conditions for the vector fields given by Eqs.~(\ref{omega_equation}) and (\ref{rho_equation}).

The system of coupled, partial differential equations, for the effective average action and average entropy, given in Eqs.~(\ref{eq:QM_pot_vec_flow_EQ}) and (\ref{eq:QM_entropy_vec_flow_EQ}) alongside the self-consistent equations for the vector fields, (\ref{omega_equation}) and (\ref{rho_equation}), must be solved numerically. One way to do so, is to use a Taylor expansion around the scale-dependent minimum of the effective potential $U_k$. This method however, is not well suited to study the low temperature and high density regime of the phase diagram, where for certain parametrizations, a first order chiral phase transition is expected and two minima co-exist. In the present work we use the grid method, a much more powerful technique that provides full access to the effective potential, in a given range of the $\sigma$ field. This allows the study of the phase diagram around a first order phase transition. In this numerical approach, the field variable $\sigma$ is discretized in an one-dimensional grid, and the first and second derivatives of the effective potential with respect to $\sigma$ are calculated using finite differences. For more information about the used numerical approach, see the Appendix (\ref{appendix_numerical_details}).

The value of the vector fields, $\tilde{\omega}_{0,k}$ and $\tilde{\rho}_{0,k}^3$, are calculated using Eqs.~(\ref{omega_equation}) and (\ref{rho_equation}) at each momentum shell $k$. In practice, by following this approach, the flow of the effective potential and the entropy density are automatically in the minimum with respect to to the vector fields.

In the MF calculation, the self-consistent equation for the $\tilde{\omega}_{0}$ field is directly related to the sum of the quark densities while the one for the $\tilde{\rho}_{0}^3$ field is related to the difference of the quark densities. This means that the $\tilde{\rho}_{0,k}^3$ field is zero for symmetric matter  ($\tilde{\rho}_{0,k}^3=0$), i.e.~if considering $\mu_u=\mu_d$. The FRG calculation leads to a similar scenario. Indeed, in \cite{Drews:2014spa} it was shown that neglecting the boson quantum fluctuations, the MF results can be recovered. However, the choice of non-zero ultraviolet value of the $\tilde{\rho}_{0}^3$ vector field, $\tilde{\rho}_{0,\Lambda}^3$, would lead to an explicit isospin breaking interaction and to a non-zero $\tilde{\rho}_{0}^3$ field, even for symmetric matter. Indeed, in \cite{Zhang:2017icm}, non-zero values for $\tilde{\omega}_{0,\Lambda}$ were considered and their effect on the phase diagram was studied. However there is no reason to consider a ultraviolet potential with explicit isospin breaking by the $\tilde{\rho}_{0}^3$ field. Hence, in the present work, we will consider $\tilde{\rho}_{0,\Lambda}^3=0$.

In order to investigate the effect of the $\tilde{\rho}_{0}^3$ field on the phase diagram and the unphysical negative entropy density region, an asymmetry between the quark flavours has to be considered. Following \cite{Buballa:2003qv}, we allow for different chemical potentials for each quark flavour,
\begin{align}
\mu_u & = \mu + \delta \mu ,
\\
\mu_d & = \mu - \delta \mu . 
\end{align}
In principle, upon considering a finite $\delta \mu$, pion condensation could happen. This means that the effective potential would dependent on two chiral invariants, $\phi^2 = \sigma^2 + \pi_3^2$ and $\xi^2 = \pi_1^2 + \pi_2^2$ \cite{Kamikado:2012bt}. In such a scenario, not only the flow equations would be much more complicated but a two dimensional grid would have to be considered since there are two distinct chiral invariants. Following previous works \cite{Drews:2014spa,Otto:2019zjy}, to simplify the calculations, we will neglect the possibility of pion condensation and work only with one chiral invariant. To make this approximation valid, a very small difference between quark chemical potentials of $\delta \mu=-30 \text{ MeV}$ will be considered \cite{Buballa:2003qv}. For such a value of $\delta \mu$, we will be describing matter with more down quarks then up quarks, a very important scenario to study neutron stars, for example.

\section{Results}
\label{results}

In this section we present the phase diagram of the 2-flavour Quark-Meson model, calculated by solving the flow equation (\ref{eq:QM_pot_vec_flow_EQ}), for different values of temperature and chemical potential. Different vector couplings are considered in order to study their effect on the phase diagram. We also present, for the same scenarios, the results of solving the flow equation for the entropy, given in Eq.~(\ref{eq:QM_entropy_vec_flow_EQ}). From this calculation we are able to study the behaviour of the entropy density near the critical region with and without vector interactions where an unphysical region, of negative entropy density, is expected from previous calculations \cite{Tripolt:2017zgc}. We also test the thermodynamical consistency by checking if the numerical temperature derivative of the effective potential agree with the result coming from solving the flow equation for the entropy density.

Regarding the numerical calculation, solving the system of coupled flow equations in a grid is very computationally demanding. In fact, the computational time is not only dictated by the grid size and the infrared cutoff, $k_{\mathrm{IR}}$ but, in the case of finite vector couplings, of consistently solving Eqs.~(\ref{omega_equation}) and (\ref{rho_equation}) for each grid point at every momentum shell (see the Appendix (\ref{appendix_numerical_details})). 

In order to make the numerical computations more efficient within the scope of the present work, we decided to use a higher infrared cutoff of $k_{\mathrm{IR}}=80$ MeV then the one used in \cite{Tripolt:2017zgc} of $k_{\mathrm{IR}}=40$ MeV. We verified, by solving the flow equations for different values of $k_{\mathrm{IR}}$, that this change does not influence the results qualitatively, allowing the study of the qualitative effect of different vector interactions in the phase diagram and in the unphysical negative density entropy region, using less computing resources. Using a finite value for the infrared cutoff $\qty(k_{\mathrm{IR}})$ physically means neglecting, in the numerical calculation, low momentum modes of the meson fields at the level of the path integral.

Different grid sizes were also studied and after some analysis we decided to use a $80$-point grid size in $\sigma \in \qty[2, 122]$ MeV. As for the infrared cut-off, using a thinner grid does not change the qualitative behaviour of the results and since the same grid is used for every scenario, considering a given grid size represents a systematic uncertainty.

The system of flow equations can then be solved from the UV scale, $k=\Lambda$, down to the infrared scale, $k=$IR, to yield $U_{k=\mathrm{IR}}$ and $s_{k=\mathrm{IR}}$, the effective potential and entropy density in the infrared. In order to solve this system of coupled partial differential equations, a set of initial conditions has to be provided. In the case of Eqs.~(\ref{eq:QM_pot_vec_flow_EQ}) and (\ref{eq:QM_entropy_vec_flow_EQ}), these correspond to the effective action and entropy density in the $k=\Lambda$ momentum shell. The effective potential in the UV, $U_{k=\Lambda}$, is chosen in such a way that it respects the symmetries of the system and to yield, in the infrared, the experimental values for pion mass and decay constant. In this work we use the usual potential:
\begin{align}
U_{k=\Lambda} \qty( T, \mu ; \sigma ) & =
\frac{1}{2} m_\Lambda^2 \sigma^2 + \frac{1}{4} \lambda_\Lambda \sigma^4, 
\label{UV_potential}
\end{align}
with the parameters given in Table \ref{table_FRG_parameter_set}.

\begin{table}[ht!]
\begin{tabular}{c|c|c|c|c}
$\Lambda$ {[}MeV{]} & $m_\Lambda / \Lambda$ & $\lambda_\Lambda$ & $c / \Lambda^3$ & $g_S$ \\ \hline
1000                & 0.969                 & 0.001             & 0.00175         & 4.2  
\end{tabular}
\caption{Used parameter set \cite{Tripolt:2017zgc}. It yields in the vacuum, for $k_{\mathrm{IR}}=80\text{ MeV}$, the following observables: $f_\pi=92.4$ MeV, $m_\pi=137.6$ MeV, $m_\sigma=606.7$ MeV and $m_q=388.2$ MeV.}
\label{table_FRG_parameter_set}
\end{table}

The vector fields in the UV, $\tilde{\omega}_{0,\Lambda}$ and $\tilde{\rho}_{0,\Lambda}^3$, in this work, were chosen to be zero,
\begin{align}
g_\rho \tilde{\rho}_{0,\Lambda}^3 \qty( T, \mu ; \sigma ) & = 0, 
\\
g_\omega \tilde{\omega}_{0,\Lambda} \qty( T, \mu ; \sigma ) & = 0. 
\end{align}
In \cite{Zhang:2017icm} also the effect of a $\sigma$ dependence for the $\tilde{\omega}_{0,\Lambda}$ vector field was studied, which ended up not changing the phase structure significantly. We will consider the vector coupling constants, $G_\omega=g_\omega/m_\omega$ and $G_\rho=g_\rho/m_\rho$ as free parameters and study the influence of different values on the structure of the phase diagram. In \cite{Drews:2014spa}, these parameters were considered as bounded by $G_\omega=G_\rho=0.001-0.01$ MeV$^{-1}$. These bounds were obtained using vacuum properties, by considering the vector fields as massive, $m_\omega, m_\rho \sim 1$ GeV and $g_\omega=g_\omega=1-10$ \cite{Drews:2014spa}. However these parameters might be density dependent and in-medium modifications could change their magnitudes.

Due to the fact that there is no temperature dependence in the UV potential, the UV entropy density, $s_{k=\Lambda}$, is simply given by:
\begin{align}
s_{k=\Lambda} \qty( T, \mu ; \sigma ) & = 0  .
\label{UV_entropy}
\end{align}
Since the UV scale is fixed at a finite value, there is no reason why the effective potential in the UV, should be temperature and chemical potential independent \cite{Herbst:2013ufa}. Indeed, in \cite{Strodthoff:2013cua}, only the purely thermal flow equation was solved, effectively generating a temperature and chemical potential UV potential.

\begin{figure}[ht!]
\centering
\includegraphics[width=0.45\linewidth]{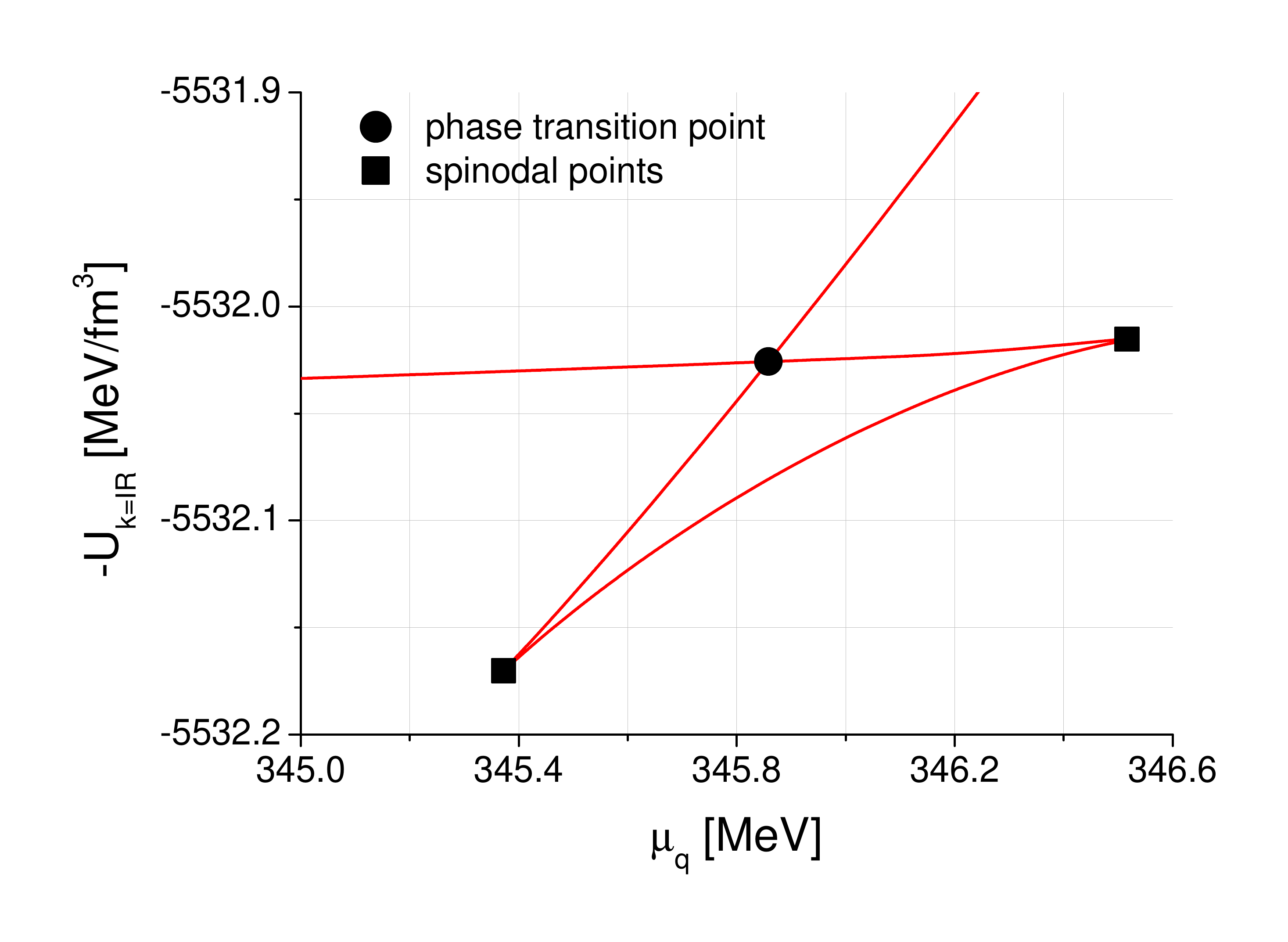}  
\caption{Extrema of the infrared effective potential as a function of the quark chemical potential for $T=20$ MeV. The phase transition point is represented by the dot while the squares are the spinodal points.}
\label{fig:maxwell_construction}
\end{figure}

In the presence of a first order chiral phase transition, the effective potential has two minima. The phase transition in this case will be defined through the Maxwell construction: when the effective potential has several minima, the one with lowest energy represents the stable phase. In Fig.~\ref{fig:maxwell_construction}, we present such a construction at $T=20$ MeV for the QM model using the FRG method. The dot is the chiral transition chemical potential while the squares are the chemical potentials of spinodal points. 

\begin{figure}[ht!]
\includegraphics[width=0.45\linewidth]{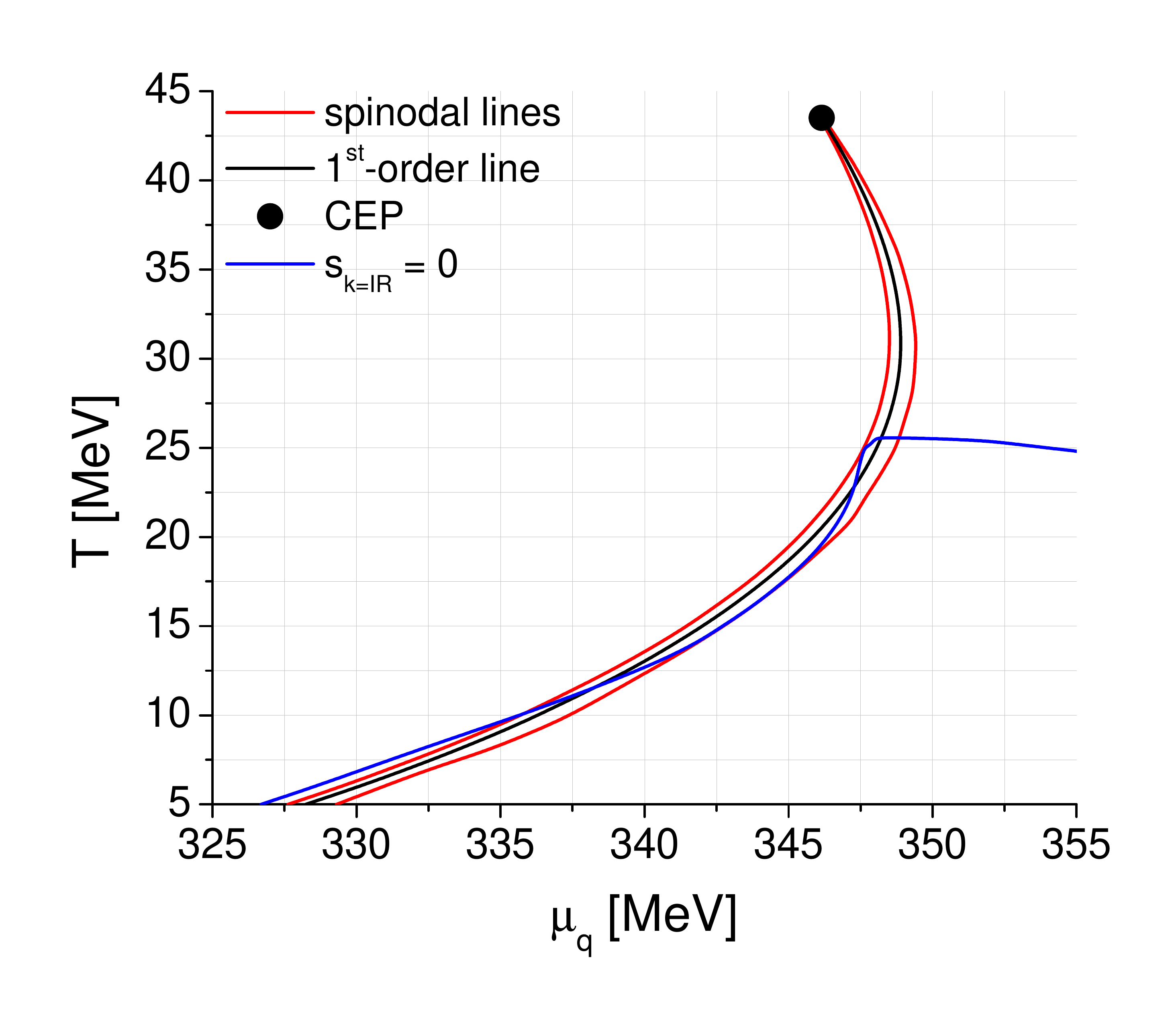}
\caption{First order phase transition of the QM model without vector interactions. The red lines are the spinodals, the black line is the first order phase transition line and the CEP is the black dot. The blue line corresponds to the $s_{k=\text{IR}}=0$ line and below this line entropy density is negative.}
\label{phase_diagram_QM_model}
\end{figure}

We start our study by considering the QM model without vector interactions. In Fig.~\ref{phase_diagram_QM_model} we present the first order phase transition of the model, the $s_{k=\text{IR}}=0$ line and the CEP. One can see that the region in-between spinodal lines is very narrow, differently from MF calculations. Indeed, for $T=20$ MeV one can analyse Fig.~\ref{fig:maxwell_construction} and observe that the overall size of the region in-between spinodal points is less then $1.5$ MeV.  We also present the line $s_{k=\text{IR}}=0$ which, trivially, separates the region of positive and negative entropy densities. This result is very similar to the one presented in \cite{Tripolt:2017zgc}. However, in \cite{Tripolt:2017zgc}, a region of negative entropy density is only discussed 
on the right side of the first order phase transition. Here, we find such an unphysical region on both sides of the 
phase transition line.
This apparent difference can be of numerical origin. While we solved the flow equation for the entropy density, it can also be calculated as the derivative $\left.-\partial U_{k=\text{IR}}/\partial T\right|_\mu$, after solving the flow equation for the effective potential.

More, we see that the $s_{k=\text{IR}}=0$ line behaves like an isentropic line that crosses the first order phase transition: in \cite{Costa:2016vbb,Ferreira:2017wtx} when an isentropic line crosses the first order phase transition it enters the critical region, touches each spinodal line once and exits the critical region.
The branches entering from outside the spinodal region until touching the phase-transition line correspond to the entropy in the stable minimum of the potential. The two parts of the line between the phase transition line and touching the spinodal lines correspond to the entropy in the respective local minimum. Finally, the branch between touching both spinodal lines follows the solution in the maximum of the potential.\\
Thermodynamically, the zero-entropy density line must be located at the zero-temperature axis.
This leads us to observe 
that the $s_{k=\text{IR}}=0$ isentropic line is displaced from its $T=0$ location in this model within the FRG approach.

A possible origin for the displacement of this isentropic line and consenquently the existence of the negative entropy density region is that finite chemical potential effects are not correctly accounted in the model beyond mean field. Upon considering an UV potential which is independent of the temperature and chemical potential, one is considering that the initial conditions to solve the differential equations are the same for every point in the phase diagram. Such case may not be true and considering temperature and chemical potential dependences in the UV potential are known to modify the thermodynamics and the phase structure \cite{Strodthoff:2013cua}. Hence, building a temperature and/or chemical potential UV effective potential could provide more insights on the origin of the negative entropy density region. Such a study is beyond the scope of the present work and is left as future work.

\begin{figure}[t!]
\begin{subfigure}[b]{0.35\textwidth}
\includegraphics[width=\textwidth]{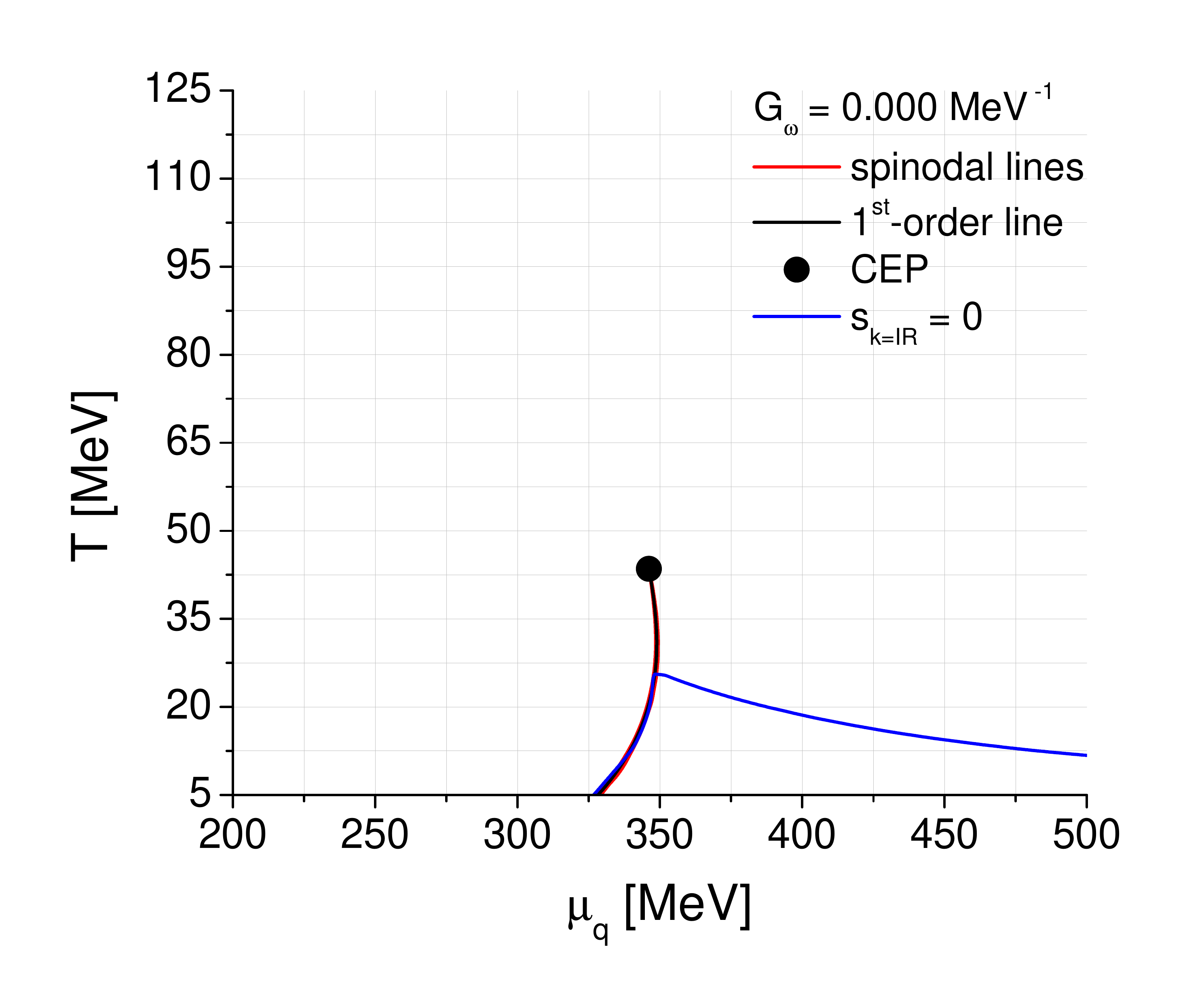}
\caption{}
\label{PD_gw_0000}
\end{subfigure}
\hspace{-0.05\textwidth}
\begin{subfigure}[b]{0.35\textwidth}
\includegraphics[width=\textwidth]{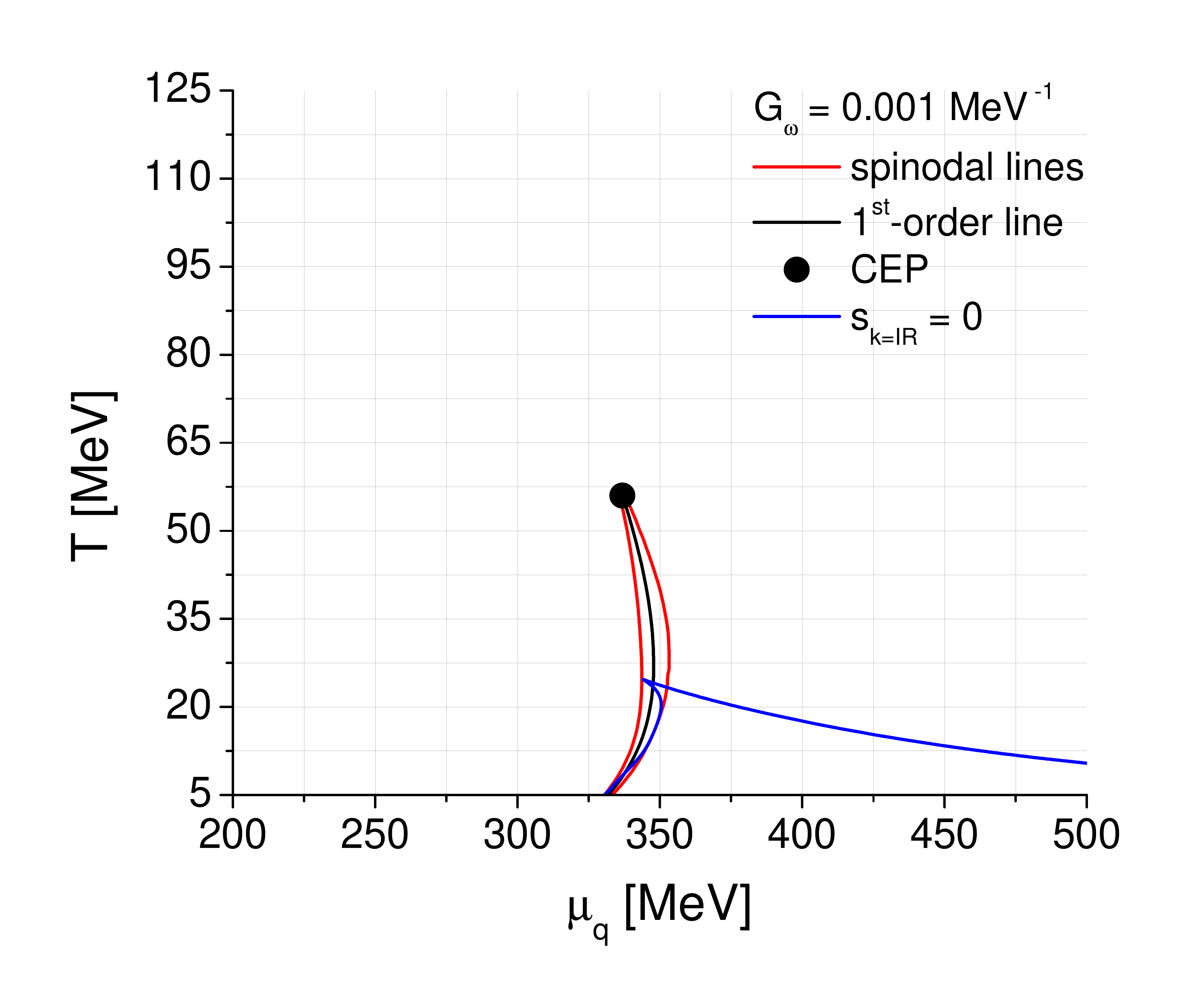}
\caption{}
\label{PD_gw_0001}
\end{subfigure}
\hspace{-0.05\textwidth}
\begin{subfigure}[b]{0.35\textwidth}
\includegraphics[width=\textwidth]{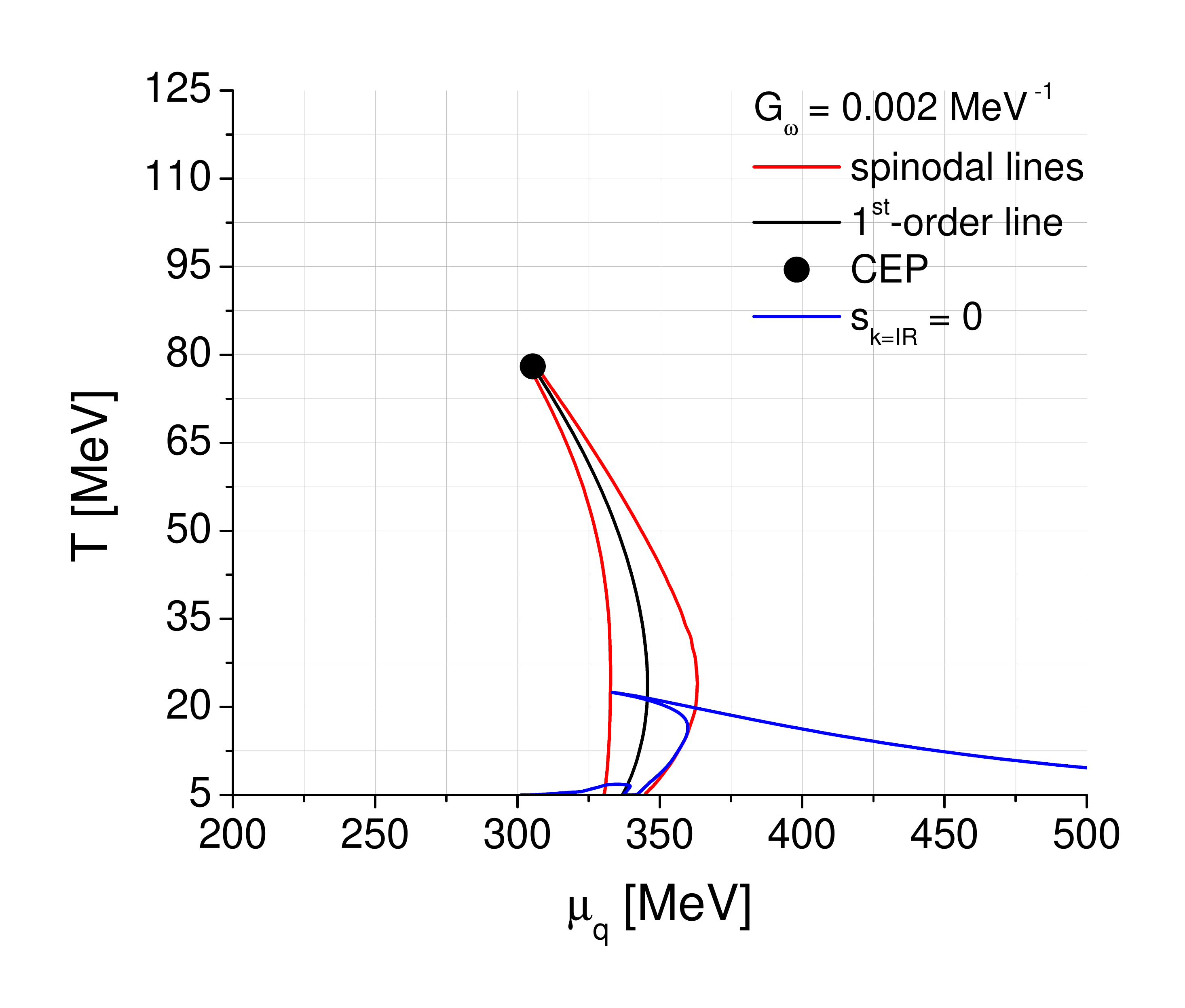}
\caption{}
\label{PD_gw_0002}
\end{subfigure}
\\ 
\begin{subfigure}[b]{0.35\textwidth}
\includegraphics[width=\textwidth]{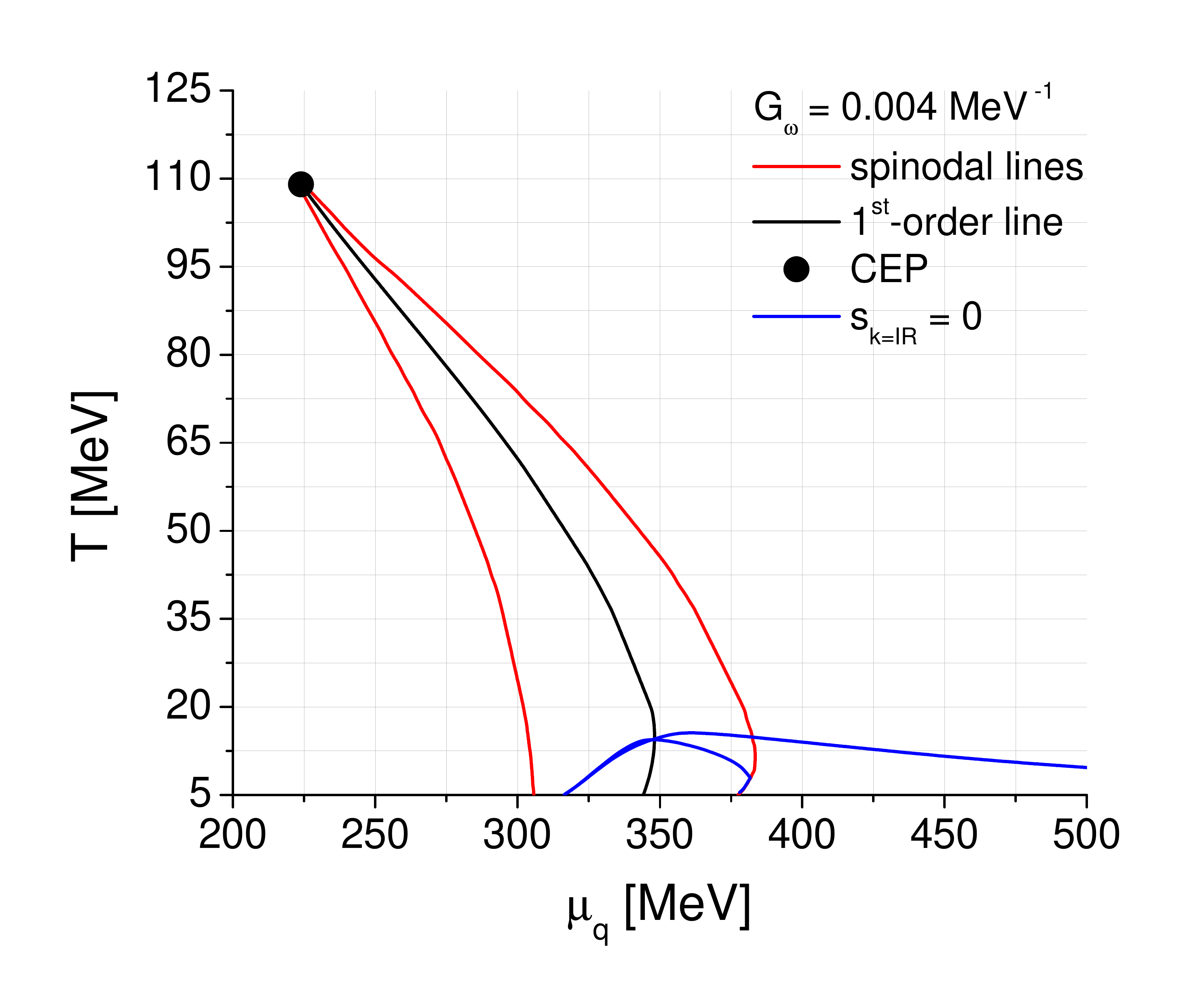}
\caption{}
\label{PD_gw_0004}
\end{subfigure}
\hspace{-0.05\textwidth}
\begin{subfigure}[b]{0.35\textwidth}
\includegraphics[width=\textwidth]{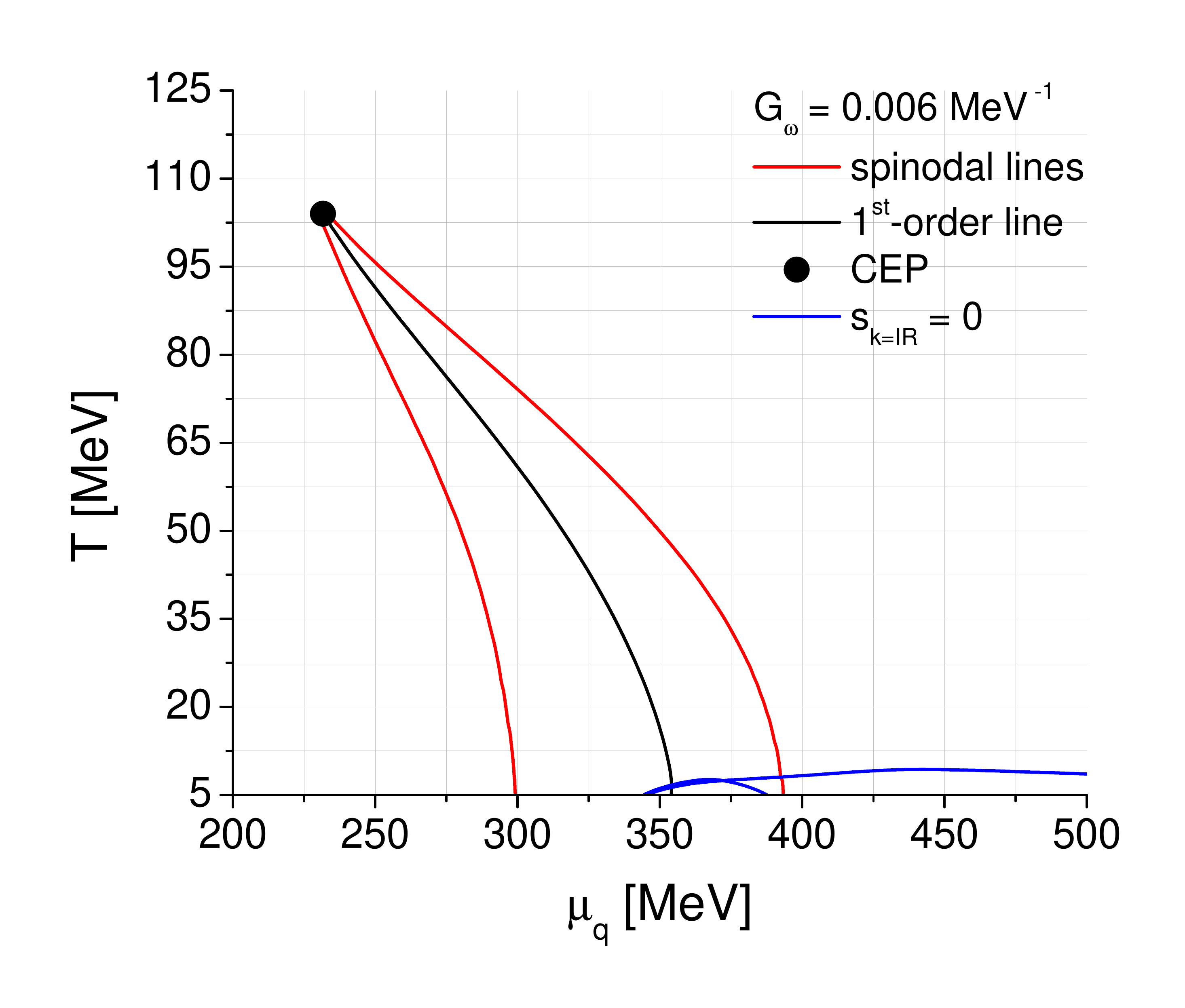}
\caption{}
\label{PD_gw_0006}
\end{subfigure}
\hspace{-0.05\textwidth}
\begin{subfigure}[b]{0.35\textwidth}
\includegraphics[width=\textwidth]{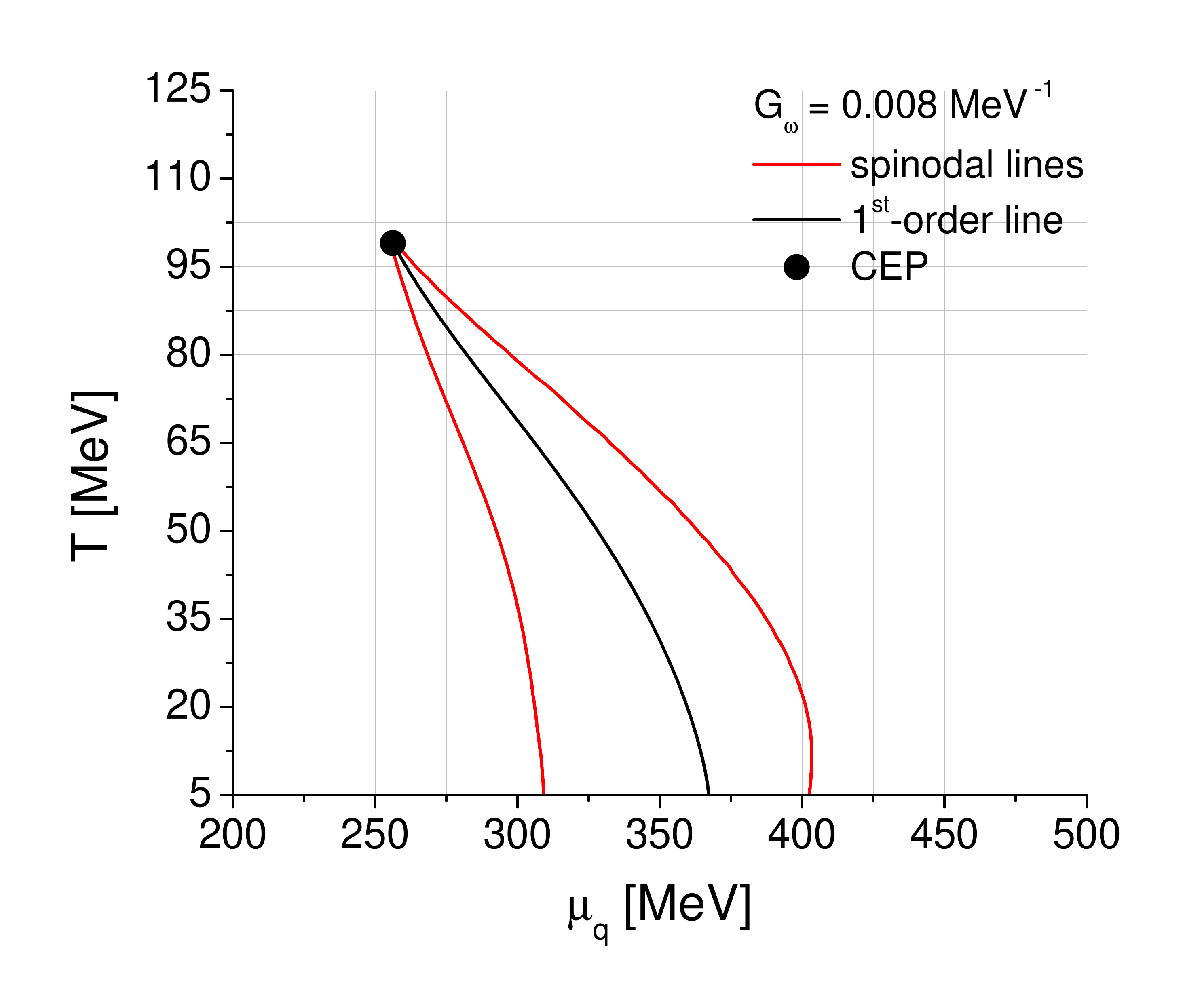}
\caption{}
\label{PD_gw_0008}
\end{subfigure}
\\
\begin{subfigure}[b]{0.35\textwidth}
\includegraphics[width=\textwidth]{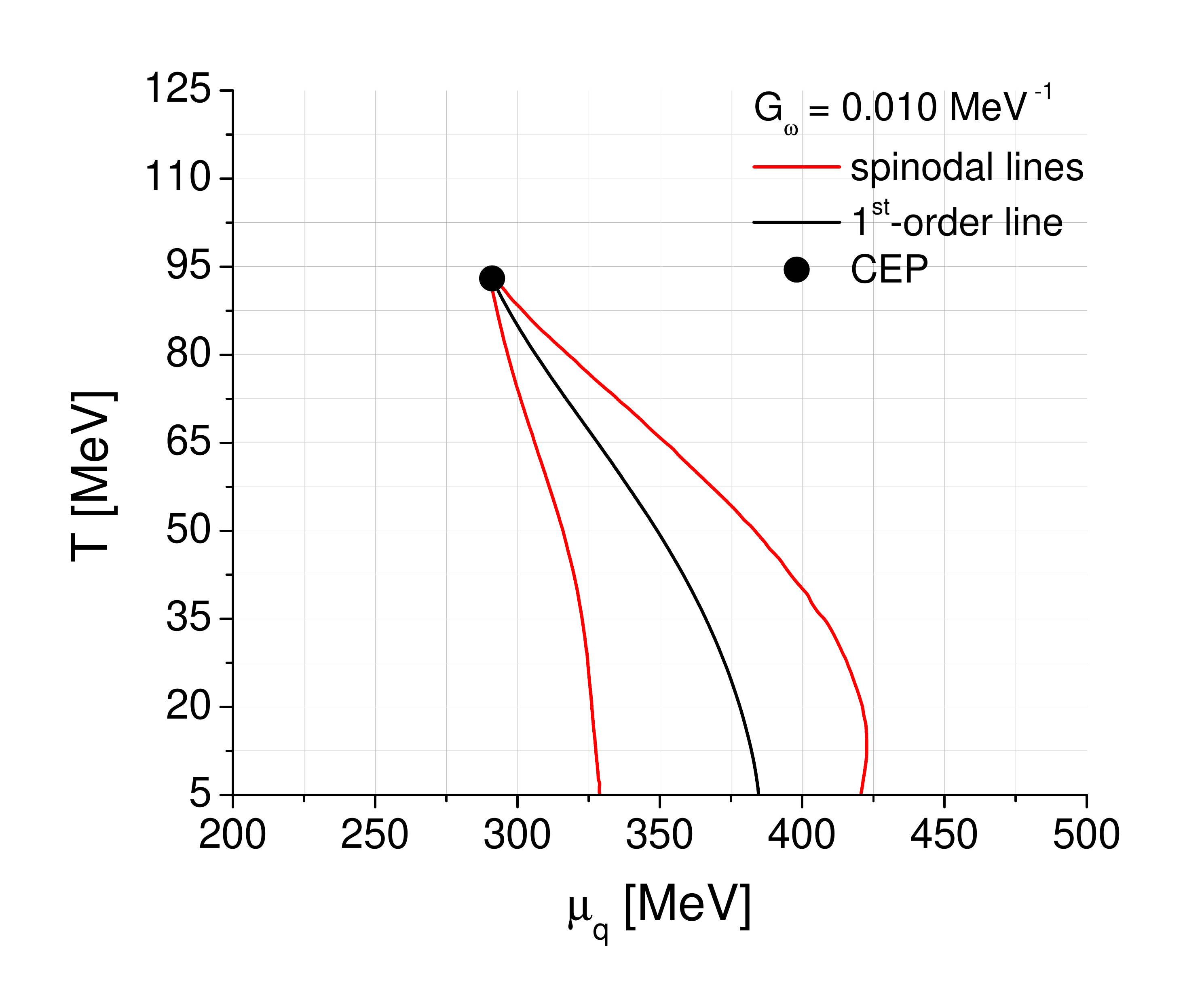}
\caption{}
\label{PD_gw_0010}
\end{subfigure}
\hspace{-0.05\textwidth}
\begin{subfigure}[b]{0.35\textwidth}
\includegraphics[width=\textwidth]{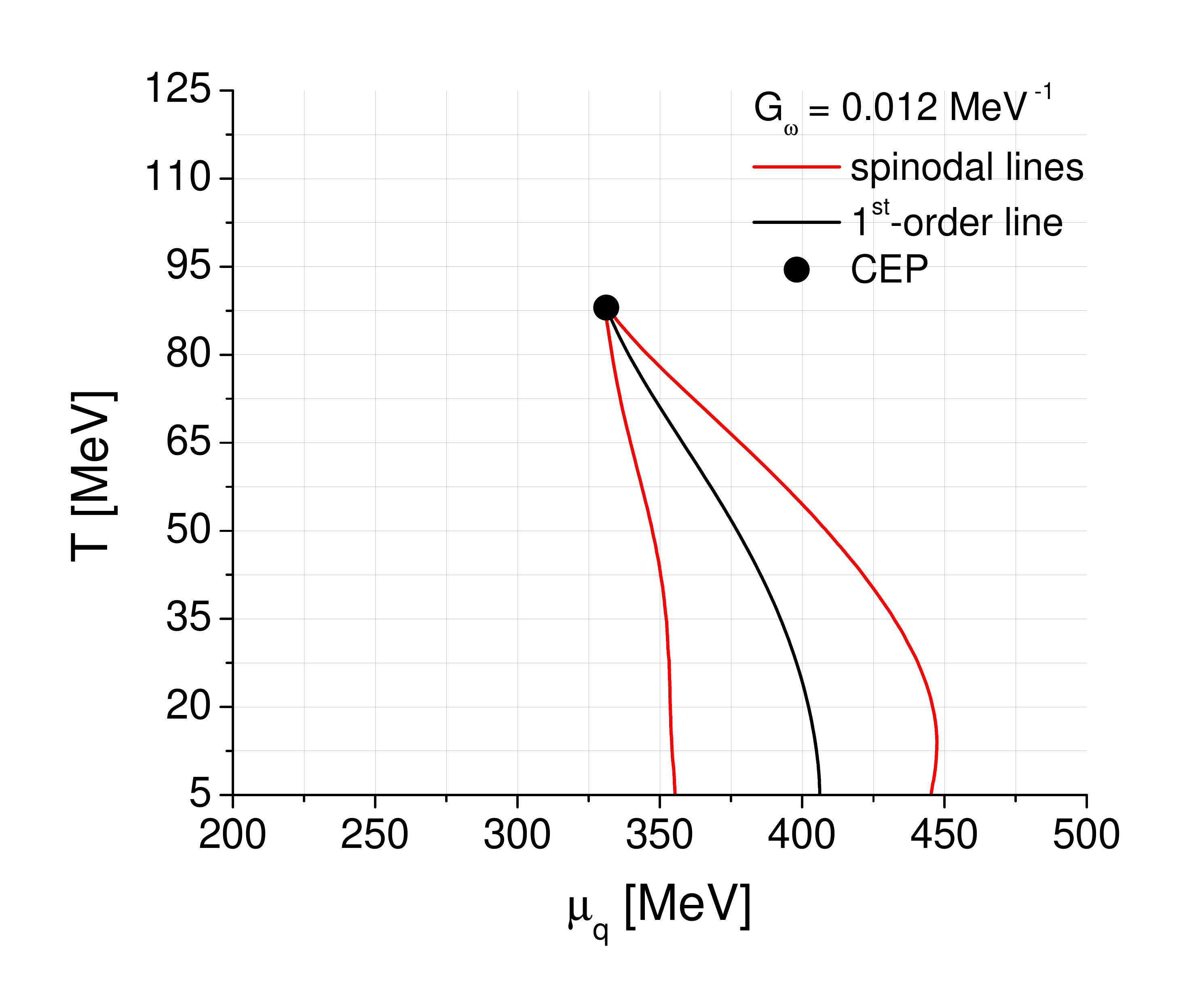}
\caption{}
\label{PD_gw_0012}
\end{subfigure}
\hspace{-0.05\textwidth}
\begin{subfigure}[b]{0.35\textwidth}
\includegraphics[width=\textwidth]{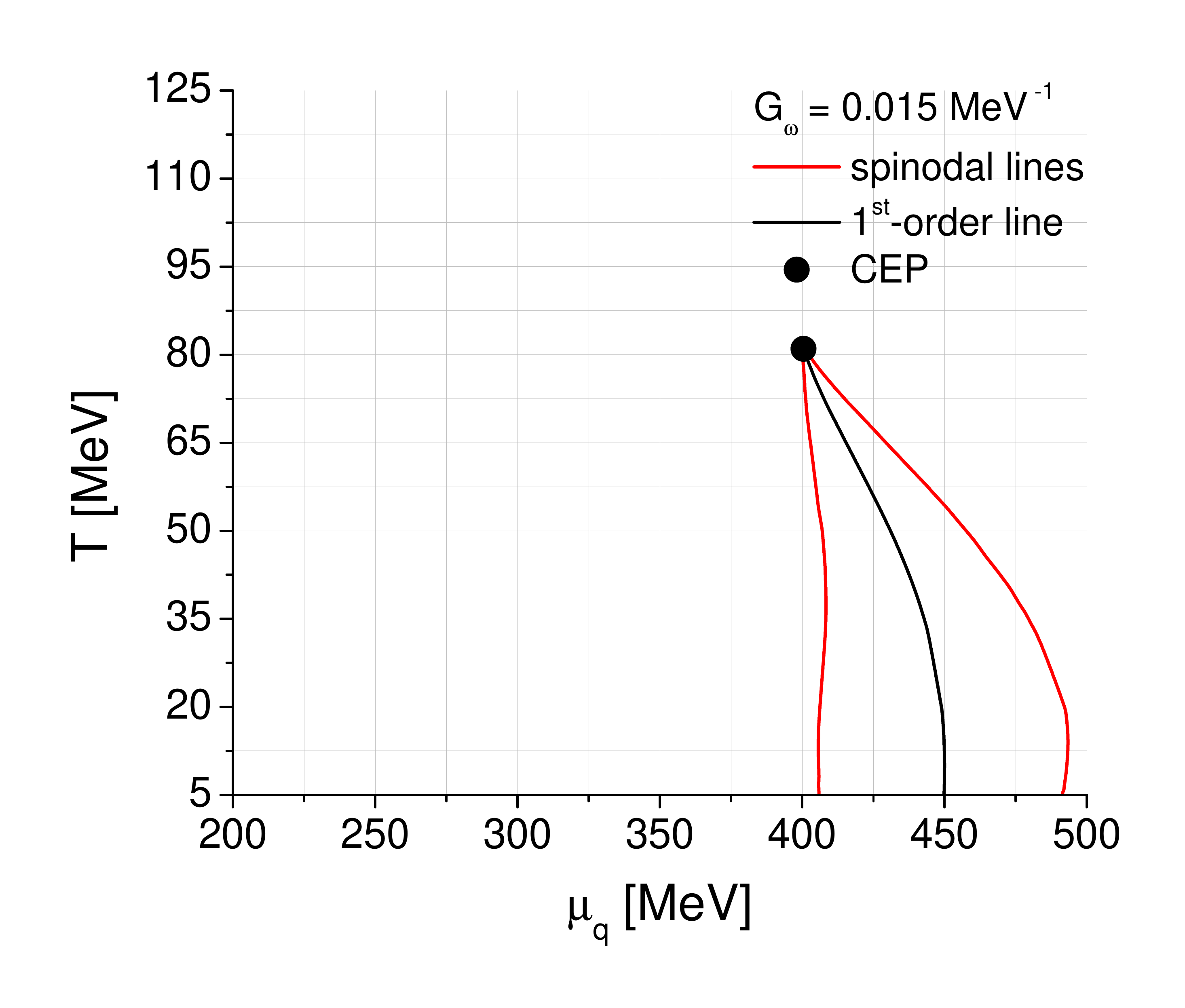}
\caption{}
\label{PD_gw_0015}
\end{subfigure}
\caption{First order phase transition of the QM model, for increasing values of $G_\omega$ and fixed $G_\rho=0$. The red, black and blue lines are the spinodals, first order phase transition and the $s_{k=\text{IR}}=0$ lines, respectively, for each value of $G_\omega$. The CEPs are represented by the black dots. Entropy density is negative below the $s_{k=\text{IR}}=0$ line.}
\label{phase_diagram_QM_model_increasing_Gomega}
\end{figure}

The next step in our study is to consider the effect of finite vector interactions. First we just consider the effect of the $\tilde{\omega}_{0}$ field, by setting $G_\rho=0$, and increasing $G_\omega$. The critical region with increasing $G_\omega$ can be seen in Fig.~\ref{phase_diagram_QM_model_increasing_Gomega}. For increasing vector coupling in the range $G_\omega = \qty[0.001, 0.004 ]$ MeV$^{-1}$ (see Fig. \ref{phase_diagram_QM_model}, panels (a), (b), (c), and (d)), there are two main effects regarding the critical region: the CEP is moved to much higher temperatures and smaller chemical potentials and the extension of the region in-between spinodal lines increases. The low temperature first order phase transition line is slightly shifted to higher chemical potentials while for higher temperatures the first order line is dragged along with the CEP to smaller chemical potentials. Further increasing the vector coupling, $G_\omega = \qty[0.006, 0.015]$ MeV$^{-1}$ (see Fig. \ref{phase_diagram_QM_model}, panels (e), (f), (g), (h), and (i)), a very different behaviour is observed: the CEP is moved towards smaller temperatures and higher chemical potentials while the region in-between spinodal lines gets imperceptibly smaller. The behaviour of the CEP for these values of $G_\omega$ is very similar to the one found in \cite{Zhang:2017icm} even tough in that study, the chiral limit is used.

The behaviour of the negative entropy density region and $s_{k=\text{IR}}=0$ line is very interesting: increasing the vector  coupling $G_\omega$ pushes this region to lower values of temperature. In fact, there is a critical value of $G_\omega$ to which there is no more negative entropy density region on the phase diagram of the model.

As already discussed, we expect that the range of magnitudes that we considered for the vector couplings to be within acceptable and physical ranges. Specially since these couplings may be density dependent. Nonetheless, the vanishing of the negative entropy region for a given vector coupling is not a signal that such a coupling is physical. The critical vector coupling in which we do not observe a negative entropy density region is not unique, since it should be different for another parameter set (different values for $\Lambda$, $m_\Lambda$, $\lambda_\Lambda$, $c$ and $g_S$). Also, we were only able to solve the flow equations down to a minimum temperature of $5~\mathrm{MeV}$. Hence, a given critical value of $G_\omega$ and $G_\rho$, where no negative entropy density is found above $T=5~\mathrm{MeV}$ does not guarantee that, for lower temperatures, the negative entropy density region is not present.

From MF studies one expects that the inclusion of repulsive vector interactions would push the CEP towards lower values of temperature, making it disappear for a high enough vector coupling. However we observe a rather different and complex behaviour when including quantum fluctuations with the FRG. Indeed the CEP and first order phase transition do not disappear for the range of considered vector couplings and the previous unphysical negative entropy density region disappears for increasing $G_\omega$.

As already stated, in order to study the effect of the $\tilde{\rho}_{0}^3$ vector field on the first order phase transition, the two flavour quark system must be on an asymmetric state. As already discussed, we will consider a finite isospin chemical potential of $\delta \mu = - 30$ MeV.

\begin{figure}[ht!]
\includegraphics[width=0.45\linewidth]{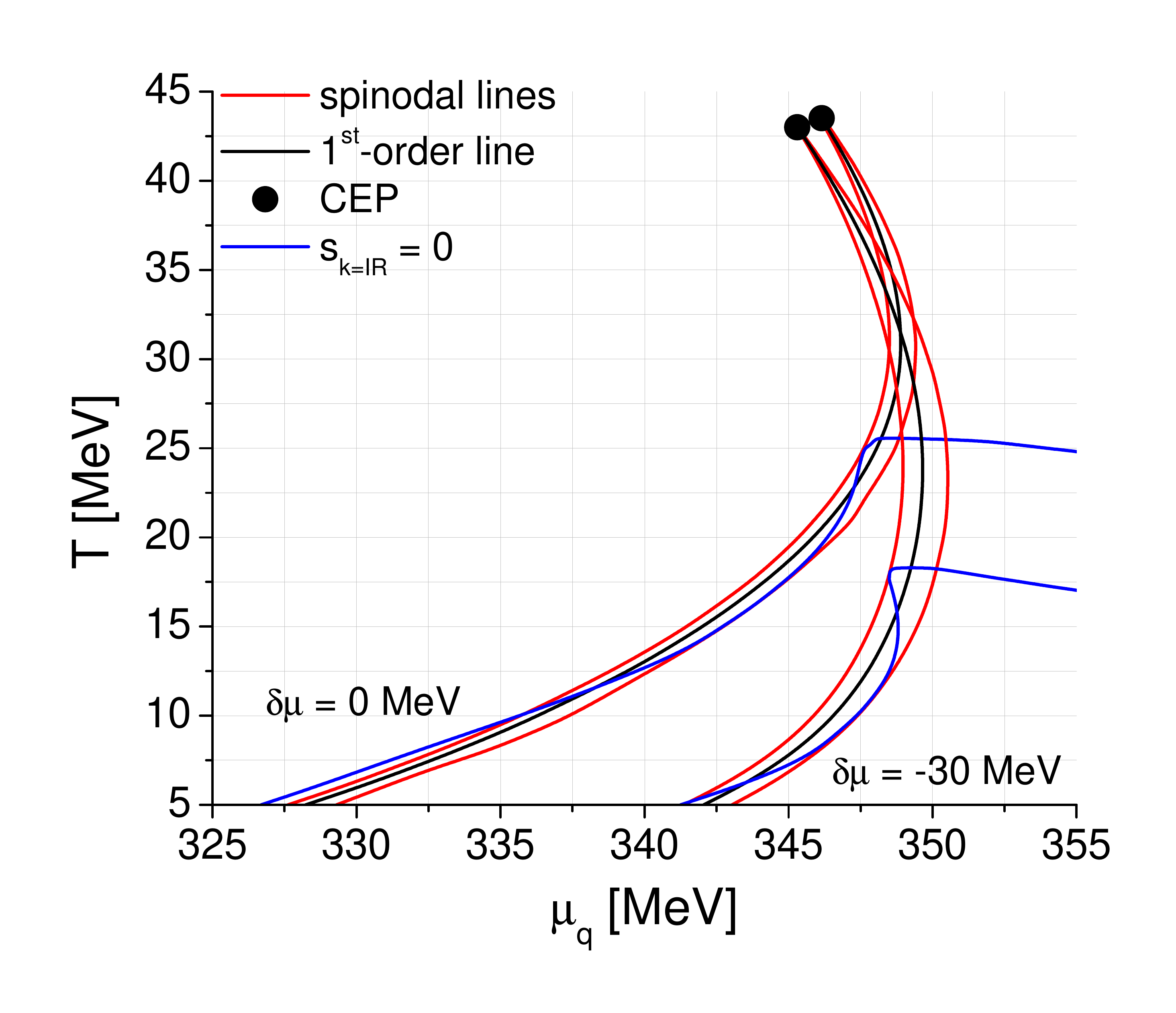}
\caption{First order phase transition of the QM model without vector interactions, for $\delta \mu=0$ and $\delta \mu=-30$ MeV. The red, black and blue lines are the spinodals, first order phase transitions and the $s_{k=\text{IR}}=0$ lines, respectively. The CEPs are represented by the black dots. Entropy density is negative below the $s_{k=\text{IR}}=0$ line.}
\label{phase_diagram_QM_model_isospin_effect}
\end{figure}

In Fig.~\ref{phase_diagram_QM_model_isospin_effect}, we show the results of comparing the critical region of the model with $\delta \mu = 0$ and $\delta \mu = -30$ MeV, without vector interactions i.e., $G_\omega = G_\rho=0$. The effect of considering a finite isospin is the following: the first order line is shifted to higher chemical potentials (at lower temperatures) and the CEP is marginally moved to lower quark chemical potentials but its temperature remains the same (within our level of numerical accuracy). Since the $s_{k=\text{IR}}=0$ isentropic line is connected to the spinodal region, moving the first order line to higher chemical potentials also moves the unphysical negative entropy density region. The inclusion of a finite $\delta \mu$ also enlarges the region in-between the spinodal lines.

\begin{figure}[t!]
\begin{subfigure}[b]{0.35\textwidth}
\includegraphics[width=\textwidth]{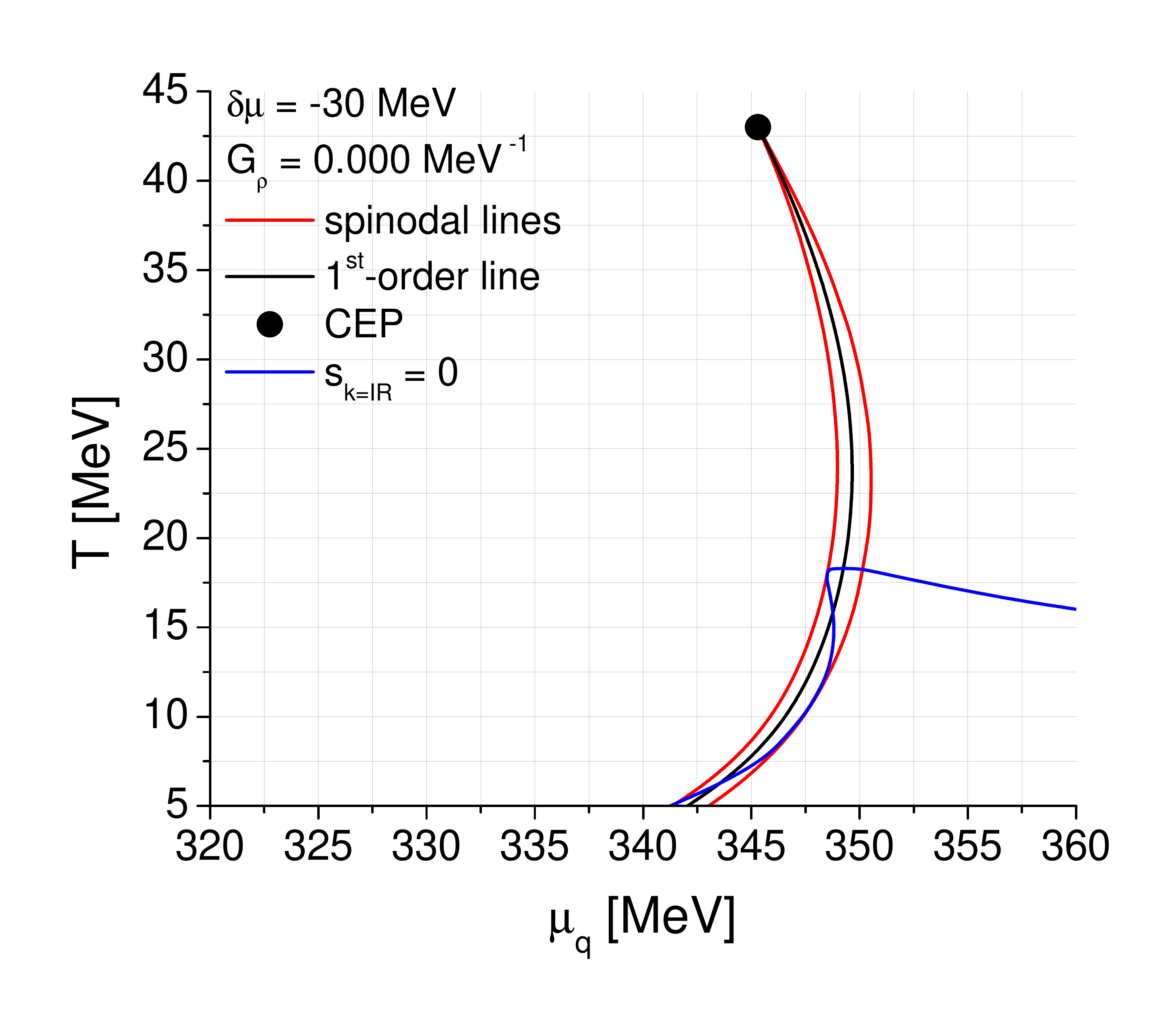}
\caption{}
\label{PD_gr_0000}
\end{subfigure}
\hspace{-0.05\textwidth}
\begin{subfigure}[b]{0.35\textwidth}
\includegraphics[width=\textwidth]{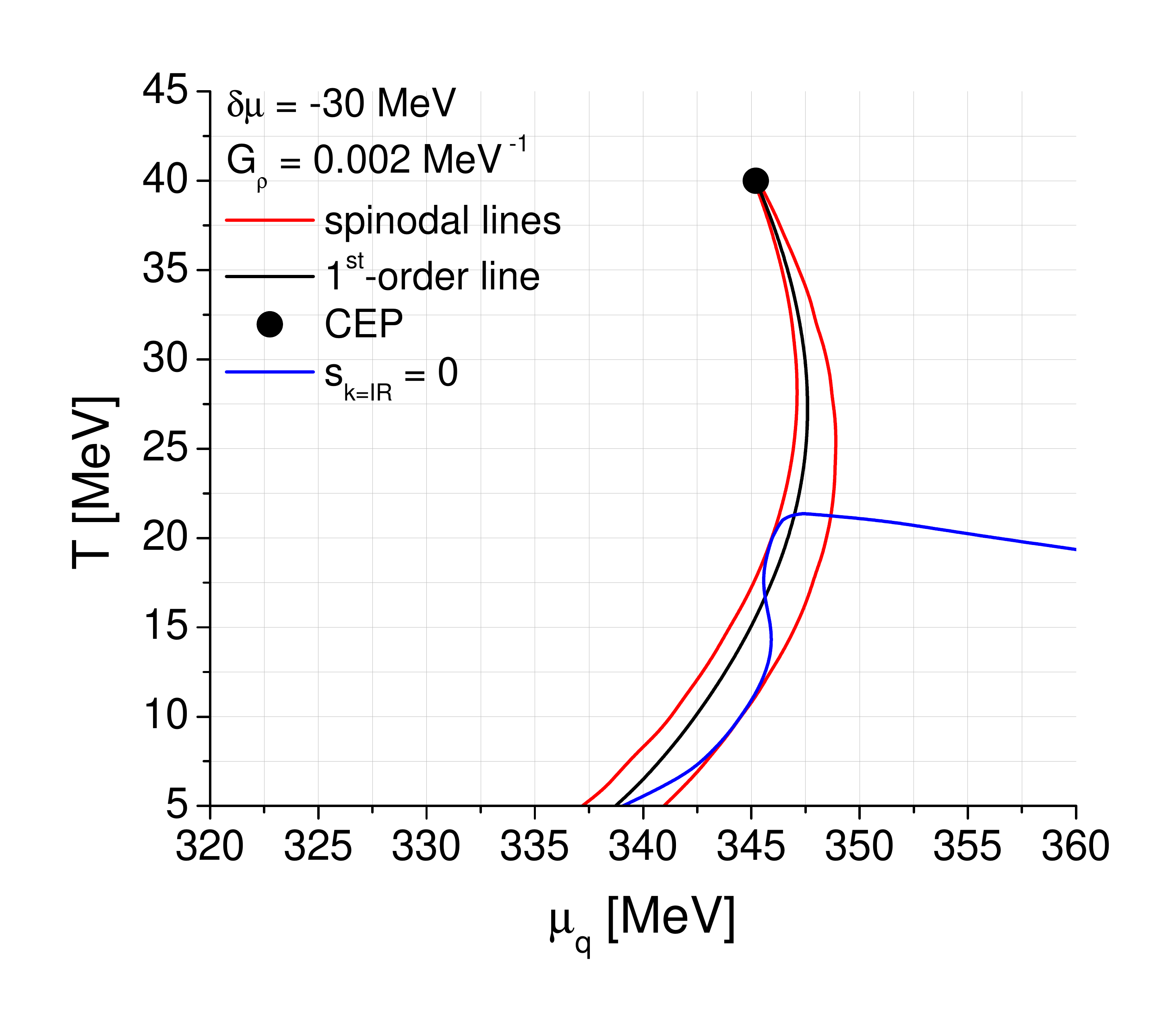}
\caption{}
\label{PD_gr_0002}
\end{subfigure}
\hspace{-0.05\textwidth}
\begin{subfigure}[b]{0.35\textwidth}
\includegraphics[width=\textwidth]{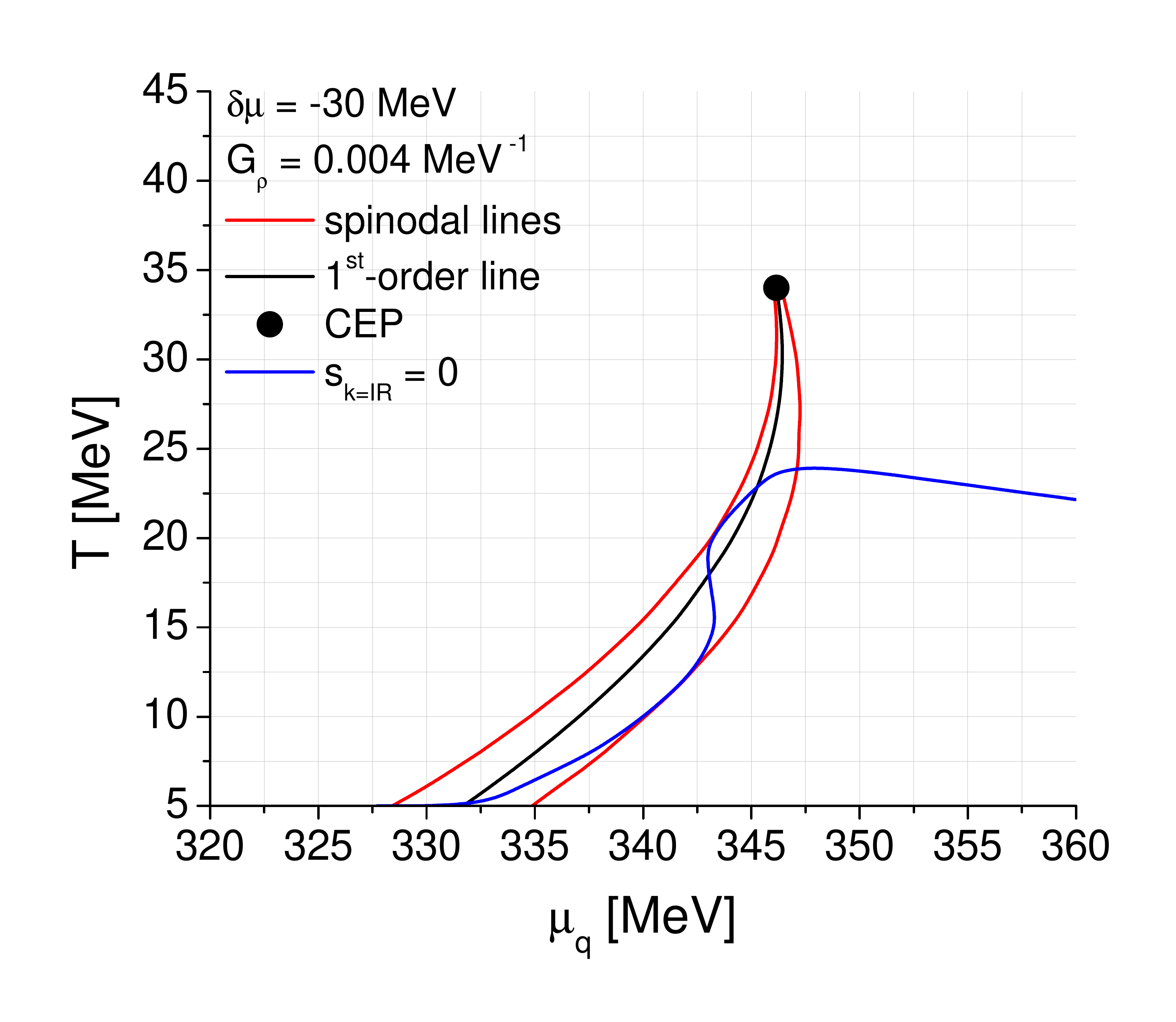}
\caption{}
\label{PD_gr_0004}
\end{subfigure}
\\ 
\begin{subfigure}[b]{0.35\textwidth}
\includegraphics[width=\textwidth]{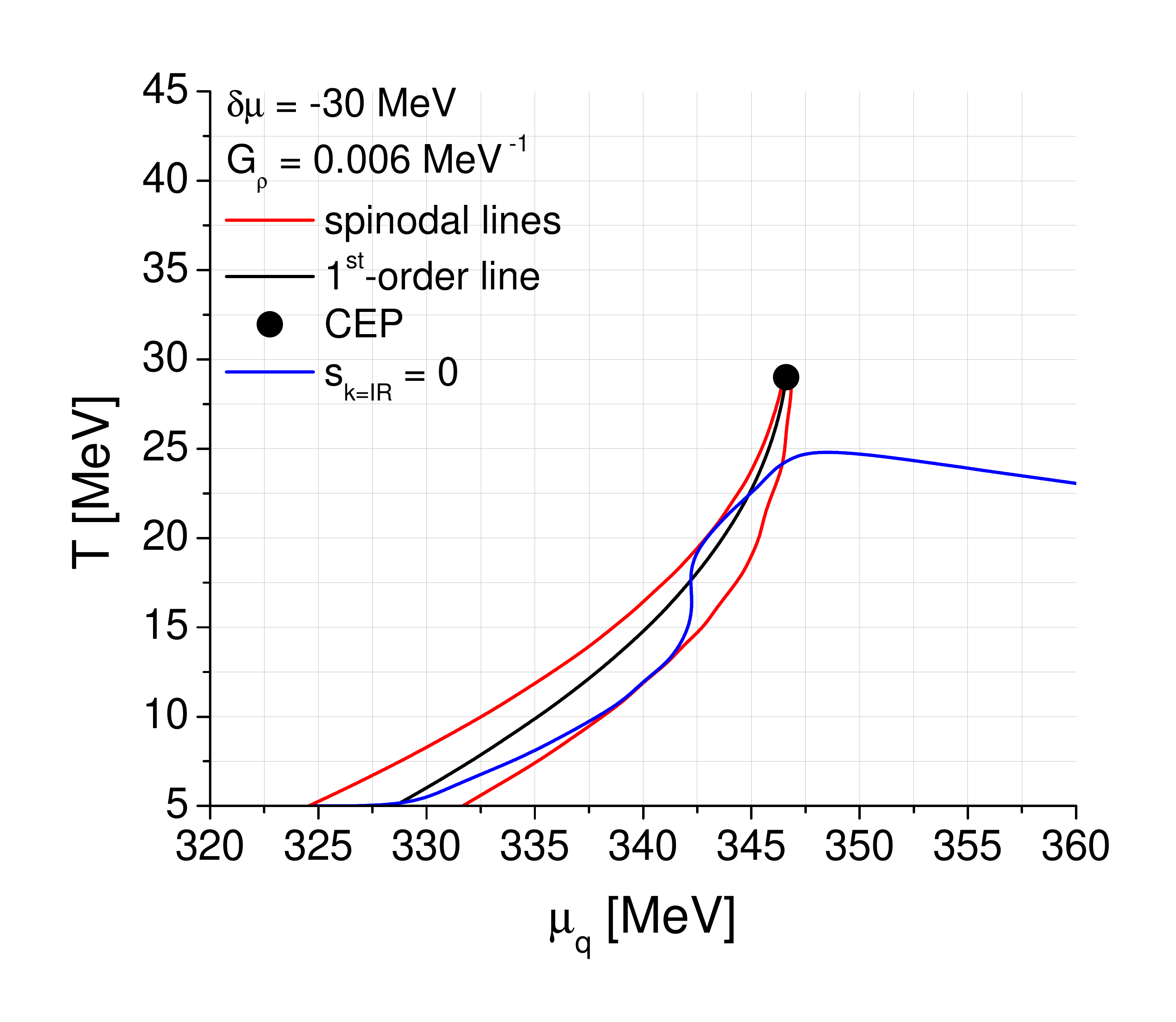}
\caption{}
\label{PD_gr_0006}
\end{subfigure}
\hspace{-0.05\textwidth}
\begin{subfigure}[b]{0.35\textwidth}
\includegraphics[width=\textwidth]{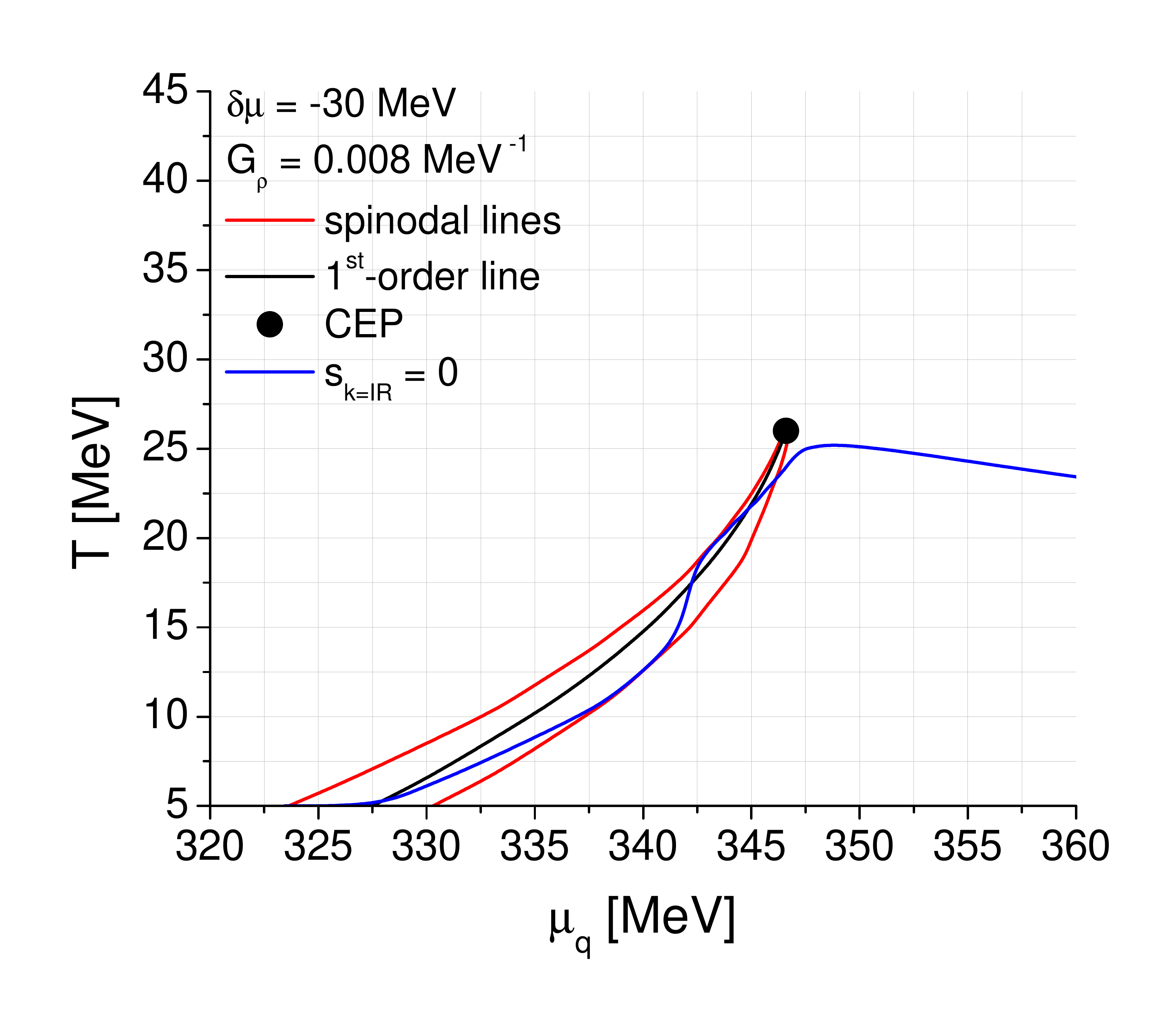}
\caption{}
\label{PD_gr_0008}
\end{subfigure}
\caption{First order phase transition of the QM model with $\delta \mu=-30$ MeV, for increasing $G_\rho$ with fixed $G_\omega=0$. The red, black and blue lines are the spinodals, first order phase transitions and the $s_{k=\text{IR}}=0$ lines, respectively. The CEPs are represented by the black dots. Entropy density is negative below the $s_{k=\text{IR}}=0$ line.}
\label{phase_diagram_QM_model_increasing_Grho}
\end{figure}

In order to study the isolated effect of the $\tilde{\rho}_{0}^3$ vector field with $\delta \mu = - 30$ MeV, we set $G_\omega=0$ and calculated the phase diagram for increasing values of $G_\rho$. The results can be seen in Fig.~\ref{phase_diagram_QM_model_increasing_Grho}. Increasing the coupling $G_\rho$, has the opposite behaviour of considering a finite $\delta \mu$: it shifts the first order line to smaller chemical potentials while the CEP is slightly moved to higher chemical potentials and low temperatures. The region in between spinodal lines is also larger with finite $G_\rho$ when compared to the case without vector interactions, even tough the effect is much less noticeable than when considering finite $G_\omega$. The first order phase transition line at low temperatures is very close to its original location with $\delta \mu=0$ for $G_\rho = 0.008$ MeV$^{-1}$. Thus, increasing this coupling is effectively restoring the isospin symmetry, broken by the finite $\delta \mu$. Indeed, in nuclear relativistic mean field models, the $\tilde{\rho}_{0}^3$ vector field can be added to the theory as an isospin restoring interaction, mirroring the Bethe–Weizsäcker mass formula and the valley of beta stability in nuclear physics \cite{norman1997compact}.

\begin{figure}[ht!]
\includegraphics[width=0.45\linewidth]{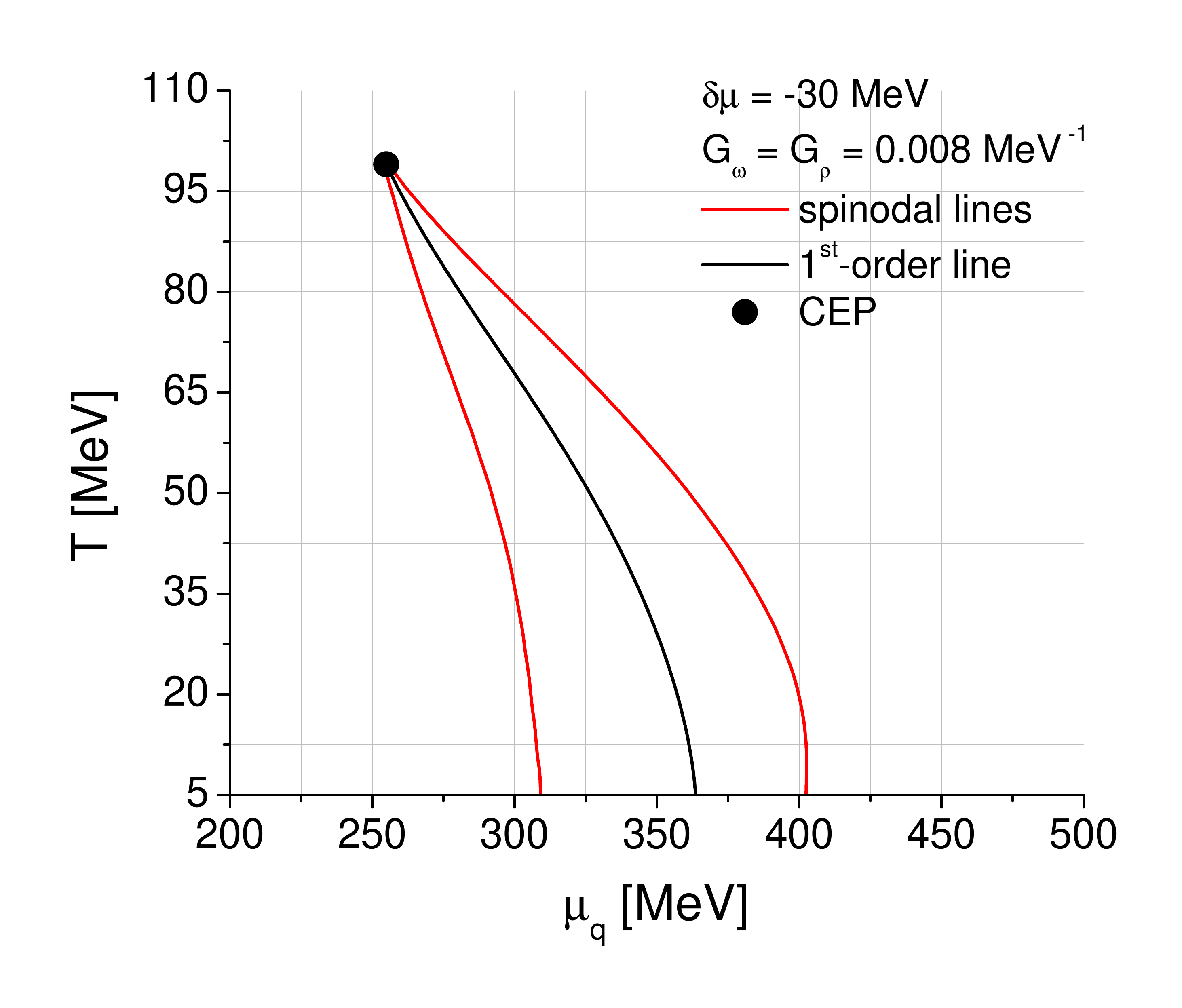}
\caption{First order phase transition of the QM model with $\delta \mu= -30$ MeV, for $G_\omega=G_\rho=0.008$ MeV$^{-1}$. The red lines are the spinodals, the black line is the first order phase transition line and the black dot is the CEP.}
\label{phase_diagram_QM_model_increasing_Gomega_Grho}
\end{figure}

Finally in Fig.~\ref{phase_diagram_QM_model_increasing_Gomega_Grho} we consider $G_\omega=G_\rho=0.008$ MeV$^{-1}$,  with $\delta \mu= -30$ MeV. In this scenario we are taking into account the combined effect of the $\tilde{\omega}_{0}$ and $\tilde{\rho}_{0}^3$ vector fields.  The obtained phase diagram is extremely similar to the one obtained in the Fig. \ref{phase_diagram_QM_model_increasing_Gomega} panel (f), with $G_\omega=0.008$ MeV$^{-1}$ and $\delta \mu= 0$. The only difference is on the location of the first order line which is negligibly dislocated to smaller chemical potentials. Taking the previous results into account, this behaviour is expected: the $\tilde{\rho}_{0}^3$ field is restoring the isospin symmetry while the influence of the $\tilde{\omega}_{0}$ field is identical to the one observed in the isospin symmetric case.

\section{Conclusions}
\label{conclusions}

We have calculated the critical region near the first order phase transition of the two flavour QM model with vector interactions, within the FRG approach to include quantum fluctuations. Besides the first order chiral transition and the CEP, the spinodal lines were presented. The unphysical region of negative entropy density reported by \cite{Tripolt:2017zgc} was also found and its behaviour due to the presence of vector interactions was studied.

The behaviour of the critical region under finite vector interactions is different from mean field calculations: increasing the repulsive vector interaction pushes the CEP towards higher values of temperature and lower values of chemical potential. Further increasing the vector interaction, drives the CEP to smaller temperatures and higher chemical potentials. Another important conclusion is that the region in-between spinodal lines increases in chemical potential with increasing vector couplings. Matter inside the spinodal region corresponds to unstable matter which can only be reached in a non-equilibrium evolution of the system in the form of clusterized matter. Very different from the case without vector interactions, the increase of this region in chemical potential, due to finite vector interactions, indicates that it is possible to have clusterized chiral symmetric matter in a wider region of densities.

We also found that the region of negative entropy density is present on both sides of the first order phase transition line. The positive entropy density region and the negative entropy density region is, trivially separated by the $s_{k_{\text{IR}}}=0$ line. However, this line behaves like an isentropic line: it passes through the first order line, touches one spinodal line, changes direction crossing the first order line again, touches the other spinodal and changes direction again. This leads us to conclude that the appearance of the negative entropy density region is a consequence of the displacement of the the $s=0$ isentropic line from its $T=0$ location. For a high enough vector interaction the negative entropy density regions disappears leaving a physical phase diagram with a first order phase transition and CEP and without negative entropy.

Considering a difference of up and down quark chemical potentials $\delta \mu$, so a finite isospin chemical potential, has a big effect on the chemical potential of the first order line but the location of the CEP is unchanged in temperature and marginally changed to smaller chemical potentials. Increasing at a finite $\delta \mu$ the coupling of the $\tilde{\rho}_{0}^3$ vector field, $G_\rho$, is equivalent to restore isospin symmetry while pushing the CEP to lower values of temperature, leading to a phase structure similar to the one with $\delta \mu = 0$ with a CEP at smaller temperatures.

To better understand the origin of the unphysical negative entropy density region, as previously found by \cite{Tripolt:2017zgc}, a flow equation beyond the LPA could be derived and the phase diagram and entropy density calculated. A different regulator function could also influence the results. Due to the mathematical nature of the QM flow equation, we were only able to calculate the phase diagram down to low temperatures ($T=5$ MeV) but not at zero temperature. Solving the $T=0$ flow equation exactly could also provide some new analytical and numerical insights. Some efforts in this direction have been done in \cite{Barnafoldi:2016tkd}, where the authors try to solve the flow equation at $T=0$ by executing a mathematical transformation to the differential equations in order to transform the rectangular initial condition on a circular one, due to the Fermi sphere.

Another possible source for the appearance of the negative entropy density region is the fact that the UV potential is temperature and chemical potential independent. As future work we plan to explore how different, temperature and chemical potential dependent UV potentials affect the phase diagram and the negative density entropy region.

\section{Acknowledgements}

This work was supported by national funds from FCT (Fundação para a Ciência e a Tecnologia, I.P, Portugal) under the IDPASC Ph.D. program (International Doctorate Network in Particle Physics, Astrophysics and Cosmology), with the Grant No. PD/BD/128234/2016 (R.C.P.), and under the Projects No. UID/FIS/04564/2019, No. UID/04564/2020, and No. POCI-01-0145-FEDER-029912 with financial support from POCI “Programa Operacional Competitividade e Internacionalização (COMPETE 2020)”, in its FEDER component, as well as by an INFN Post Doctoral Fellowship (competition INFN notice n.~18372/2016, RS).  We wish to thank Constança Providência for useful comments. One of the authors (R.C.P.) would like to thank Konstantin Otto for useful email correspondence regarding the numerical approach and Marcos Gouveia for help with the OpenMP interface. 
We acknowledge the Laboratory for Advanced Computing at the University of Coimbra for providing CPU time with the Navigator cluster. The authors also acknowledge the INFN (Istituto Nazionale di Fisica Nucleare) - Sezione di Torino and the COST Action CA16214 {``PHAROS: The multi-messenger physics and astrophysics of neutron stars''} for funding a Short Term Scientific Mission (STSM) at the INFN - Sezione di Torino.

\section{Appendix}
\label{appendix}

\subsection{Numerical details}
\label{appendix_numerical_details}

In \cite{Yokota:2016tip} it was demonstrated how to ensure numerical stability during the integration of a generalized version of Eq.~(\ref{eq:QM_pot_vec_flow_EQ}), through an optimal step-size. To derive such optimal step, for simplicity, it was considered that the function derivatives are calculated with low order finite-difference methods: forward difference for the RG time variable, $t$, and three-point rule for the $\sigma-$direction. It is supposed that the numerical stability condition derived within this simpler scheme, is also valid for the fourth-order Runge-Kutta method, used in the $t$ variable and higher order finite-differences used for the $\sigma$ derivatives. Following this approach the following conditions for the step-size $\Delta t$ was derived:
\begin{align}
\abs{ \Delta t } & \leq \frac{ 2 \abs{G} }{ \abs{F^2} } ,
\label{delta_t_1}
\\
\abs{ \Delta t } & \leq \frac{ \Delta \sigma^2 }{ 2\abs{G} } .
\label{delta_t_2}
\end{align}
Here, $G$ and $F$ are given by:
\begin{align}
G & = 
-
\frac{k^5}{ 24 \pi^2 E_\sigma^3 }
\qty[
\coth \qty( \frac{ E_\sigma }{ 2T } ) +
\frac{ E_\sigma }{ 2T }
\csch^2 \qty( \frac{ E_\sigma }{ 2T } ) 
] ,
\label{condition_G}
\\
F & = 
-
\frac{k^5}{ 8 \pi^2 \sigma E_\pi^3 }
\qty[
\coth \qty( \frac{ E_\pi }{ 2T } ) +
\frac{ E_\pi }{ 2T }
\csch^2 \qty( \frac{ E_\pi }{ 2T } ) 
] .
\label{condition_F}
\end{align}

As in \cite{Yokota:2016tip}, we do not consider these conditions for $\sigma \sim 0$. Since these conditions only depend on the bosonic sector of the flow equation, the fact that one is dealing with effective finite chemical potentials does not change the conditions directly. The effect of finite chemical potential and finite vector mesons only change these conditions indirectly, since the potential and its derivatives will be different during the flow.

To solve the set of coupled differential equations, in such a way to get full access to the full effective potential, we employed the grid method. In this method, the field variable $\sigma$ is discretized in an one-dimensional grid, and the first and second derivatives of the effective potential with respect to $\sigma$ are calculated using finite differences. The five point midpoint rule  was used except in the grid endpoints where the forward and backward rules were used.

One starts the calculation in the UV scale i.e., at $k=\Lambda$. At this momentum scale the effective potential and entropy density are calculated using the initial conditions provided in Eqs.~(\ref{UV_potential}) and (\ref{UV_entropy}), respectively. The needed derivatives with respect to $\sigma$ are calculated for every $\sigma$-grid point, using finite differences. Next, an optimal step size in the renormalization group time, $\Delta t$, is calculated using Eqs.~(\ref{delta_t_1}) and (\ref{delta_t_2}): the smaller $\Delta t$ is used. The flow equations are then solved using the fourth-order Runge-Kutta method, to provide the $\sigma$-dependent effective potential and entropy density in the next step $t-\Delta t$, i.e., $U_{k=\Lambda \exp\qty( t- \Delta t ) }\qty(\sigma)$ and $s_{k=\Lambda \exp\qty( t- \Delta t ) }\qty(\sigma)$. This process is repeated until the infrared scale is reached at $k={\mathrm{IR}}$. After reaching the infrared scale, one is in possession of the $U_{k=\mathrm{IR}}\qty(\sigma)$ and $s_{k=\mathrm{IR}}\qty(\sigma)$ and can then calculate the minimum of the effective potential, in which all observables are defined.

When considering finite vector interactions, the self-consistent Eqs.~(\ref{omega_equation}) and (\ref{rho_equation}) have to be solved at every $\sigma$-grid point, at every momentum scale $k$. Hence, for a given $k$, for every $\sigma$-grid point, a 2-dimensional root finding algorithm is used to find the values of $g_\omega \tilde{\omega}_{0,k}$ and $ g_\rho \tilde{\rho}_{0,k}^3$ that fulfil this system of equations. To speed up the root finding process, the solutions at the momentum scale $k$ are provided as guesses for the next momentum shell. Since we are using the forth-order Runge-Kutta method this process has to be performed four times to be able to calculate the effective potential and entropy density in a given momentum scale.

The computing time is then related to the $\sigma$ grid size, the infrared cutoff, $k_{\mathrm{IR}}$, and the root fiding precision, when considering vector interactions. The complexity of the flow equations also dictates the computing time since the step-size $\Delta t$ dictates how fast one goes from the UV down to the IR and different values of temperature and chemical potential influence the overall magnitude of the adaptive step-size. 

One very important observation is that the optimal step-size calculated using Eqs.~(\ref{condition_G}) and (\ref{condition_F}) does not depend on the chemical potential and can still be used in the calculation with finite vector interactions.

In order to arrive at the phase diagram, the flow equation was solved multiple times for different values of temperature and chemical potential. In order to speed-up calculations, the OpenMP interface was used to run the computer code in parallel.

\subsection{Temperature derivative of the vector fields}
\label{vector_temperature_derivatives}

In order to calculate the entropy flow equation, it is necessary to calculate, at each momentum shell $k$ the following quantity\footnote{In this section we use, $\tilde{\omega}_k=g_\omega \tilde{\omega}_{0,k}$, $\tilde{\rho}_k=g_\rho \tilde{\rho}_{0,k}^3$, $G_\omega=\frac{ g_\omega }{ m_\omega }$ and $G_\rho=\frac{ g_\rho }{ m_\rho }$.}:
\begin{align}
\pdv{ v_{k,i} }{ T } = 
\pdv{ \tilde{\omega}_k }{ T } +
\qty(-1)^{i}
\pdv{ \tilde{\rho}_k }{ T } .
\end{align}
Considering that the vector stationary conditions hold, for a given momentum shell, $k$, we can use Eqs.~(\ref{omega_equation}), (\ref{rho_equation}) and (\ref{I_k,eta}), to write the temperature derivatives of the vector fields as:
\begin{align}
\pdv{\tilde{\omega}_k ( T ; \tilde{\omega}_k , \tilde{\rho}_k ) }{T} 
& = 
a_\omega
\sumeta
\sumi
\pdv{T} I_{k,\eta i}( T ; \tilde{\omega}_k , \tilde{\rho}_k ) ,
\\
\pdv{\tilde{\rho}_k \qty( T ; \tilde{\omega}_k , \tilde{\rho}_k ) }{T} 
& = 
a_\rho
\sumeta
\sumi
\qty(-1)^i
\pdv{T} I_{k,\eta i}\qty( T ; \tilde{\omega}_k , \tilde{\rho}_k )  . 
\end{align}
With $a_\omega = {N_c G_\omega^2}/{ 3 \pi^2 }$ and $a_\rho = {N_c G_\rho^2}/{ 3 \pi^2 }$. We only need to calculate $\pdv{T} I_{k,\eta i}$, which can be written as:
\begin{align}
\pdv{T} I_{k,\eta i}\qty( T ; \tilde{\omega}_k , \tilde{\rho}_k ) 
=
J^{(1)}_{k,\eta i}\qty( T ; \tilde{\omega}_k , \tilde{\rho}_k ) 
-
J^{(2)}_{k,\eta i}\qty( T ; \tilde{\omega}_k , \tilde{\rho}_k ) 
\qty[
\pdv{\tilde{\omega}_k \qty( T ; \tilde{\omega}_k , \tilde{\rho}_k ) }{T} 
+
\qty(-1)^i
\pdv{\tilde{\rho}_k \qty( T ; \tilde{\omega}_k , \tilde{\rho}_k ) }{T} 
] .
\end{align}
Here,
\begin{align*}
J^{(1)}_{k,\eta i}\qty( T ; \tilde{\omega}_k , \tilde{\rho}_k ) 
& =
3\int_k^\Lambda \dd{p} 
\eta p^2 
\frac{ n_\fermi \qty( E_\psi - \eta \tilde{\mu}_{k,i} ) }{ T^2 }
\Big( 1 - n_\fermi \qty( E_\psi - \eta \tilde{\mu}_{k,i} ) \Big)
\Big( E_\psi - \eta \tilde{\mu}_{k,i} \Big)
\\
& 
-
\qty[
\eta p^3 
\frac{ n_\fermi \qty( E_\psi - \eta \tilde{\mu}_{k,i} ) }{ T^2 }
\Big( 1 - n_\fermi \qty( E_\psi - \eta \tilde{\mu}_{k,i} ) \Big)
\Big( E_\psi - \eta \tilde{\mu}_{k,i} \Big)
]_k^\Lambda ,
\numberthis
\\
J^{(2)}_{k,\eta i}\qty( T ; \tilde{\omega}_k , \tilde{\rho}_k ) 
& =
3\int_k^\Lambda \dd{p} p^2 
\frac{ n_\fermi \qty( E_\psi - \eta \tilde{\mu}_{k,i} ) }{ T }
\Big( 1 - n_\fermi \qty( E_\psi - \eta \tilde{\mu}_{k,i} ) \Big)
\\
& -
\qty[
p^3 
\frac{ n_\fermi \qty( E_\psi - \eta \tilde{\mu}_{k,i} ) }{ T }
\Big( 1 - n_\fermi \qty( E_\psi - \eta \tilde{\mu}_{k,i} ) \Big)
]_k^\Lambda .
\numberthis
\end{align*}

The derivatives of the vector fields with respect to temperature are given by:
\begin{align*}
\tilde{\omega}_k^\p \qty( T; \tilde{\omega}_k , \tilde{\rho}_k )  
& = 
a_\omega
\sumeta
\sumi
\qty{
J^{(1)}_{k,\eta i}\qty( T; \tilde{\omega}_k , \tilde{\rho}_k ) 
-
J^{(2)}_{k,\eta i}\qty( T; \tilde{\omega}_k , \tilde{\rho}_k ) 
\qty[
\tilde{\omega}_k^\p \qty( T; \tilde{\omega}_k , \tilde{\rho}_k )
+
\qty(-1)^i
\tilde{\rho}_k^\p \qty( T; \tilde{\omega}_k , \tilde{\rho}_k ) 
] } ,
\\
\tilde{\rho}_k^\p \qty( T; \tilde{\omega}_k , \tilde{\rho}_k ) 
& = 
a_\rho
\sumeta
\sumi
\qty(-1)^i
\qty{
J^{(1)}_{k,\eta i}\qty( T; \tilde{\omega}_k , \tilde{\rho}_k ) 
-
J^{(2)}_{k,\eta i}\qty( T; \tilde{\omega}_k , \tilde{\rho}_ka ) 
\qty[
\tilde{\omega}_k^\p \qty( T; \tilde{\omega}_k , \tilde{\rho}_k )
+
\qty(-1)^i
\tilde{\rho}_k^\p \qty( T; \tilde{\omega}_k , \tilde{\rho}_k ) 
] } . 
\end{align*}

This system of equations can be solved analytically for $\tilde{\omega}_k^\p$ and $\tilde{\rho}_k^\p$. Neglecting the variable dependences, we can write:
\begin{align}
\tilde{\omega}_k^\p 
& = 
A_k - B_k \tilde{\omega}_k^\p - C_k \tilde{\rho}_k^\p   ,
\\
\tilde{\rho}_k^\p 
& = 
D_k - E_k \tilde{\omega}_k^\p - F_k \tilde{\rho}_k^\p   . 
\end{align}
Where we have defined,
\begin{align}
A_k \qty( T; \tilde{\omega}_k , \tilde{\rho}_k ) & = 
a_\omega
\sumeta
\sumi
J^{(1)}_{k,\eta i}\qty( T; \tilde{\omega}_k , \tilde{\rho}_k ) ,
\\
B_k \qty( T; \tilde{\omega}_k , \tilde{\rho}_k ) & = 
a_\omega
\sumeta
\sumi
J^{(2)}_{k,\eta i}\qty( T; \tilde{\omega}_k , \tilde{\rho}_k ) ,
\\
C_k \qty( T; \tilde{\omega}_k , \tilde{\rho}_k ) & = 
a_\omega
\sumeta
\sumi
\qty(-1)^i
J^{(2)}_{k,\eta i}\qty( T; \tilde{\omega}_k , \tilde{\rho}_k )  ,
\\
D_k \qty( T; \tilde{\omega}_k , \tilde{\rho}_k ) & = 
a_\rho
\sumeta
\sumi
\qty(-1)^i
J^{(1)}_{k,\eta i}\qty( T; \tilde{\omega}_k , \tilde{\rho}_k ) ,
\\
E_k \qty( T; \tilde{\omega}_k , \tilde{\rho}_k ) & =
a_\rho
\sumeta
\sumi
\qty(-1)^i
J^{(2)}_{k,\eta i}\qty( T; \tilde{\omega}_k , \tilde{\rho}_k )=
\frac{ a_\rho }{ a_\omega } 
C_k \qty( T; \tilde{\omega}_k , \tilde{\rho}_k ) ,
\\
F_k \qty( T; \tilde{\omega}_k , \tilde{\rho}_k ) & =
a_\rho
\sumeta
\sumi
J^{(2)}_{k,\eta i}\qty( T; \tilde{\omega}_k , \tilde{\rho}_k ) =
\frac{ a_\rho }{ a_\omega } B_k \qty( T; \tilde{\omega}_k , \tilde{\rho}_k ) .
\end{align}

Very easily one can solve the system of linear equations to get:
\begin{align}
\pdv{ \tilde{\omega}_k }{ T } 
& = 
\frac
{ A_k - C_k D_k + A_k F_k }
{ 1 + B_k - C_k E_k + F_k + B_k F_k } ,
\\
\pdv{ \tilde{\rho}_k  }{ T }
& = 
\frac
{ D_k + B_k D_k - A_k E_k }
{ 1 + B_k - C_k E_k + F_k + B_k F_k } .
\end{align}

When considering only one vector field i.e., if $G_\rho=0$ or $G_\omega=0$, the temperature derivatives are much simpler:
\begin{align}
G_\rho = 0 
\to
\pdv{ \tilde{\omega}_k }{ T }  & = 
\frac
{ A_k }
{ 1 + B_k } ,
\\
G_\omega = 0 
\to
\pdv{ \tilde{\rho}_k  }{ T } & = 
\frac
{ D_k }
{ 1 + F_k } .
\end{align}

\bibliography{article}

\begin{thebibliography}{67}%
\makeatletter
\providecommand \@ifxundefined [1]{%
 \@ifx{#1\undefined}
}%
\providecommand \@ifnum [1]{%
 \ifnum #1\expandafter \@firstoftwo
 \else \expandafter \@secondoftwo
 \fi
}%
\providecommand \@ifx [1]{%
 \ifx #1\expandafter \@firstoftwo
 \else \expandafter \@secondoftwo
 \fi
}%
\providecommand \natexlab [1]{#1}%
\providecommand \enquote  [1]{``#1''}%
\providecommand \bibnamefont  [1]{#1}%
\providecommand \bibfnamefont [1]{#1}%
\providecommand \citenamefont [1]{#1}%
\providecommand \href@noop [0]{\@secondoftwo}%
\providecommand \href [0]{\begingroup \@sanitize@url \@href}%
\providecommand \@href[1]{\@@startlink{#1}\@@href}%
\providecommand \@@href[1]{\endgroup#1\@@endlink}%
\providecommand \@sanitize@url [0]{\catcode `\\12\catcode `\$12\catcode
  `\&12\catcode `\#12\catcode `\^12\catcode `\_12\catcode `\%12\relax}%
\providecommand \@@startlink[1]{}%
\providecommand \@@endlink[0]{}%
\providecommand \url  [0]{\begingroup\@sanitize@url \@url }%
\providecommand \@url [1]{\endgroup\@href {#1}{\urlprefix }}%
\providecommand \urlprefix  [0]{URL }%
\providecommand \Eprint [0]{\href }%
\providecommand \doibase [0]{http://dx.doi.org/}%
\providecommand \selectlanguage [0]{\@gobble}%
\providecommand \bibinfo  [0]{\@secondoftwo}%
\providecommand \bibfield  [0]{\@secondoftwo}%
\providecommand \translation [1]{[#1]}%
\providecommand \BibitemOpen [0]{}%
\providecommand \bibitemStop [0]{}%
\providecommand \bibitemNoStop [0]{.\EOS\space}%
\providecommand \EOS [0]{\spacefactor3000\relax}%
\providecommand \BibitemShut  [1]{\csname bibitem#1\endcsname}%
\let\auto@bib@innerbib\@empty
\bibitem [{\citenamefont {Cabibbo}\ and\ \citenamefont
  {Parisi}(1975)}]{CABIBBO197567}%
  \BibitemOpen
  \bibfield  {author} {\bibinfo {author} {\bibfnamefont {N.}~\bibnamefont
  {Cabibbo}}\ and\ \bibinfo {author} {\bibfnamefont {G.}~\bibnamefont
  {Parisi}},\ }\href {\doibase https://doi.org/10.1016/0370-2693(75)90158-6}
  {\bibfield  {journal} {\bibinfo  {journal} {Physics Letters B}\ }\textbf
  {\bibinfo {volume} {59}},\ \bibinfo {pages} {67 } (\bibinfo {year}
  {1975})}\BibitemShut {NoStop}%
\bibitem [{\citenamefont {Adamczyk}\ \emph {et~al.}(2014)\citenamefont
  {Adamczyk} \emph {et~al.}}]{Adamczyk:2014fia}%
  \BibitemOpen
  \bibfield  {author} {\bibinfo {author} {\bibfnamefont {L.}~\bibnamefont
  {Adamczyk}} \emph {et~al.} (\bibinfo {collaboration} {STAR}),\ }\href
  {\doibase 10.1103/PhysRevLett.113.092301} {\bibfield  {journal} {\bibinfo
  {journal} {Phys.Rev.Lett.}\ }\textbf {\bibinfo {volume} {113}},\ \bibinfo
  {pages} {092301} (\bibinfo {year} {2014})},\ \Eprint
  {http://arxiv.org/abs/1402.1558} {arXiv:1402.1558 [nucl-ex]} \BibitemShut
  {NoStop}%
\bibitem [{\citenamefont {Adamczyk}\ \emph {et~al.}(2018)\citenamefont
  {Adamczyk} \emph {et~al.}}]{Adamczyk:2017wsl}%
  \BibitemOpen
  \bibfield  {author} {\bibinfo {author} {\bibfnamefont {L.}~\bibnamefont
  {Adamczyk}} \emph {et~al.} (\bibinfo {collaboration} {STAR}),\ }\href
  {\doibase 10.1016/j.physletb.2018.07.066} {\bibfield  {journal} {\bibinfo
  {journal} {Phys. Lett.}\ }\textbf {\bibinfo {volume} {B785}},\ \bibinfo
  {pages} {551} (\bibinfo {year} {2018})},\ \Eprint
  {http://arxiv.org/abs/1709.00773} {arXiv:1709.00773 [nucl-ex]} \BibitemShut
  {NoStop}%
\bibitem [{\citenamefont {Adamczyk}\ \emph {et~al.}(2017)\citenamefont
  {Adamczyk} \emph {et~al.}}]{Adamczyk:2017iwn}%
  \BibitemOpen
  \bibfield  {author} {\bibinfo {author} {\bibfnamefont {L.}~\bibnamefont
  {Adamczyk}} \emph {et~al.} (\bibinfo {collaboration} {STAR}),\ }\href
  {\doibase 10.1103/PhysRevC.96.044904} {\bibfield  {journal} {\bibinfo
  {journal} {Phys. Rev.}\ }\textbf {\bibinfo {volume} {C96}},\ \bibinfo {pages}
  {044904} (\bibinfo {year} {2017})},\ \Eprint
  {http://arxiv.org/abs/1701.07065} {arXiv:1701.07065 [nucl-ex]} \BibitemShut
  {NoStop}%
\bibitem [{\citenamefont {Aduszkiewicz}\ \emph {et~al.}(2016)\citenamefont
  {Aduszkiewicz} \emph {et~al.}}]{Aduszkiewicz:2015jna}%
  \BibitemOpen
  \bibfield  {author} {\bibinfo {author} {\bibfnamefont {A.}~\bibnamefont
  {Aduszkiewicz}} \emph {et~al.} (\bibinfo {collaboration} {NA61/SHINE}),\
  }\href {\doibase 10.1140/epjc/s10052-016-4450-9} {\bibfield  {journal}
  {\bibinfo  {journal} {Eur. Phys. J.}\ }\textbf {\bibinfo {volume} {C76}},\
  \bibinfo {pages} {635} (\bibinfo {year} {2016})},\ \Eprint
  {http://arxiv.org/abs/1510.00163} {arXiv:1510.00163 [hep-ex]} \BibitemShut
  {NoStop}%
\bibitem [{\citenamefont {Grebieszkow}(2017)}]{Grebieszkow:2017gqx}%
  \BibitemOpen
  \bibfield  {author} {\bibinfo {author} {\bibfnamefont {K.}~\bibnamefont
  {Grebieszkow}} (\bibinfo {collaboration} {NA61/SHINE}),\ }\bibfield
  {booktitle} {\emph {\bibinfo {booktitle} {{Proceedings, 2017 European
  Physical Society Conference on High Energy Physics (EPS-HEP 2017): Venice,
  Italy, July 5-12, 2017}}},\ }\href {\doibase 10.22323/1.314.0167} {\bibfield
  {journal} {\bibinfo  {journal} {PoS}\ }\textbf {\bibinfo {volume}
  {EPS-HEP2017}},\ \bibinfo {pages} {167} (\bibinfo {year} {2017})},\ \Eprint
  {http://arxiv.org/abs/1709.10397} {arXiv:1709.10397 [nucl-ex]} \BibitemShut
  {NoStop}%
\bibitem [{\citenamefont {{Blaschke, David}}\ \emph {et~al.}(2016)\citenamefont
  {{Blaschke, David}}, \citenamefont {{Aichelin, J\"org}}, \citenamefont
  {{Bratkovskaya, Elena}}, \citenamefont {{Friese, Volker}}, \citenamefont
  {{Gazdzicki, Marek}}, \citenamefont {{Randrup, J\o{}rgen}}, \citenamefont
  {{Rogachevsky, Oleg}}, \citenamefont {{Teryaev, Oleg}},\ and\ \citenamefont
  {{Toneev, Viacheslav}}}]{NICAWP}%
  \BibitemOpen
  \bibfield  {author} {\bibinfo {author} {\bibnamefont {{Blaschke, David}}},
  \bibinfo {author} {\bibnamefont {{Aichelin, J\"org}}}, \bibinfo {author}
  {\bibnamefont {{Bratkovskaya, Elena}}}, \bibinfo {author} {\bibnamefont
  {{Friese, Volker}}}, \bibinfo {author} {\bibnamefont {{Gazdzicki, Marek}}},
  \bibinfo {author} {\bibnamefont {{Randrup, J\o{}rgen}}}, \bibinfo {author}
  {\bibnamefont {{Rogachevsky, Oleg}}}, \bibinfo {author} {\bibnamefont
  {{Teryaev, Oleg}}}, \ and\ \bibinfo {author} {\bibnamefont {{Toneev,
  Viacheslav}}},\ }\href {\doibase 10.1140/epja/i2016-16267-x} {\bibfield
  {journal} {\bibinfo  {journal} {Eur. Phys. J. A}\ }\textbf {\bibinfo {volume}
  {52}},\ \bibinfo {pages} {267} (\bibinfo {year} {2016})}\BibitemShut
  {NoStop}%
\bibitem [{\citenamefont {Ablyazimov}\ \emph {et~al.}(2017)\citenamefont
  {Ablyazimov} \emph {et~al.}}]{Ablyazimov:2017guv}%
  \BibitemOpen
  \bibfield  {author} {\bibinfo {author} {\bibfnamefont {T.}~\bibnamefont
  {Ablyazimov}} \emph {et~al.} (\bibinfo {collaboration} {CBM}),\ }\href
  {\doibase 10.1140/epja/i2017-12248-y} {\bibfield  {journal} {\bibinfo
  {journal} {Eur. Phys. J.}\ }\textbf {\bibinfo {volume} {A53}},\ \bibinfo
  {pages} {60} (\bibinfo {year} {2017})},\ \Eprint
  {http://arxiv.org/abs/1607.01487} {arXiv:1607.01487 [nucl-ex]} \BibitemShut
  {NoStop}%
\bibitem [{\citenamefont {Sako}\ \emph {et~al.}(2014)\citenamefont {Sako},
  \citenamefont {Chujo}, \citenamefont {Gunji}, \citenamefont {Harada},
  \citenamefont {Imai}, \citenamefont {Kaneta}, \citenamefont {Kinsho},
  \citenamefont {Liu}, \citenamefont {Nagamiya}, \citenamefont {Nishio},
  \citenamefont {Ozawa}, \citenamefont {Saha}, \citenamefont {Sakaguchi},
  \citenamefont {Sato},\ and\ \citenamefont {Tamura}}]{Sako:J-PARC}%
  \BibitemOpen
  \bibfield  {author} {\bibinfo {author} {\bibfnamefont {H.}~\bibnamefont
  {Sako}}, \bibinfo {author} {\bibfnamefont {T.}~\bibnamefont {Chujo}},
  \bibinfo {author} {\bibfnamefont {T.}~\bibnamefont {Gunji}}, \bibinfo
  {author} {\bibfnamefont {H.}~\bibnamefont {Harada}}, \bibinfo {author}
  {\bibfnamefont {K.}~\bibnamefont {Imai}}, \bibinfo {author} {\bibfnamefont
  {M.}~\bibnamefont {Kaneta}}, \bibinfo {author} {\bibfnamefont
  {M.}~\bibnamefont {Kinsho}}, \bibinfo {author} {\bibfnamefont
  {Y.}~\bibnamefont {Liu}}, \bibinfo {author} {\bibfnamefont {S.}~\bibnamefont
  {Nagamiya}}, \bibinfo {author} {\bibfnamefont {K.}~\bibnamefont {Nishio}},
  \bibinfo {author} {\bibfnamefont {K.}~\bibnamefont {Ozawa}}, \bibinfo
  {author} {\bibfnamefont {P.}~\bibnamefont {Saha}}, \bibinfo {author}
  {\bibfnamefont {T.}~\bibnamefont {Sakaguchi}}, \bibinfo {author}
  {\bibfnamefont {S.}~\bibnamefont {Sato}}, \ and\ \bibinfo {author}
  {\bibfnamefont {J.}~\bibnamefont {Tamura}},\ }\href {\doibase
  https://doi.org/10.1016/j.nuclphysa.2014.08.065} {\bibfield  {journal}
  {\bibinfo  {journal} {Nuclear Physics A}\ }\textbf {\bibinfo {volume}
  {931}},\ \bibinfo {pages} {1158 } (\bibinfo {year} {2014})}\BibitemShut
  {NoStop}%
\bibitem [{\citenamefont {Weber}\ \emph {et~al.}(2007)\citenamefont {Weber},
  \citenamefont {Negreiros},\ and\ \citenamefont {Rosenfield}}]{Weber:2007ch}%
  \BibitemOpen
  \bibfield  {author} {\bibinfo {author} {\bibfnamefont {F.}~\bibnamefont
  {Weber}}, \bibinfo {author} {\bibfnamefont {R.}~\bibnamefont {Negreiros}}, \
  and\ \bibinfo {author} {\bibfnamefont {P.}~\bibnamefont {Rosenfield}}\
  }(\bibinfo {year} {2007})\ \Eprint {http://arxiv.org/abs/0705.2708}
  {arXiv:0705.2708 [astro-ph]} \BibitemShut {NoStop}%
\bibitem [{\citenamefont {Vida{\~n}a}(2018)}]{Vidana:2018lqp}%
  \BibitemOpen
  \bibfield  {author} {\bibinfo {author} {\bibfnamefont {I.}~\bibnamefont
  {Vida{\~n}a}},\ }\href {\doibase 10.1140/epjp/i2018-12329-x} {\bibfield
  {journal} {\bibinfo  {journal} {Eur. Phys. J. Plus}\ }\textbf {\bibinfo
  {volume} {133}},\ \bibinfo {pages} {445} (\bibinfo {year} {2018})},\ \Eprint
  {http://arxiv.org/abs/1805.00837} {arXiv:1805.00837 [nucl-th]} \BibitemShut
  {NoStop}%
\bibitem [{\citenamefont {Schmidt}\ and\ \citenamefont
  {Sharma}(2017)}]{Schmidt:2017bjt}%
  \BibitemOpen
  \bibfield  {author} {\bibinfo {author} {\bibfnamefont {C.}~\bibnamefont
  {Schmidt}}\ and\ \bibinfo {author} {\bibfnamefont {S.}~\bibnamefont
  {Sharma}},\ }\href {\doibase 10.1088/1361-6471/aa824a} {\bibfield  {journal}
  {\bibinfo  {journal} {J. Phys.}\ }\textbf {\bibinfo {volume} {G44}},\
  \bibinfo {pages} {104002} (\bibinfo {year} {2017})},\ \Eprint
  {http://arxiv.org/abs/1701.04707} {arXiv:1701.04707 [hep-lat]} \BibitemShut
  {NoStop}%
\bibitem [{\citenamefont {Seiler}(2018)}]{Seiler:2017wvd}%
  \BibitemOpen
  \bibfield  {author} {\bibinfo {author} {\bibfnamefont {E.}~\bibnamefont
  {Seiler}},\ }\bibfield  {booktitle} {\emph {\bibinfo {booktitle}
  {{Proceedings, 35th International Symposium on Lattice Field Theory (Lattice
  2017): Granada, Spain, June 18-24, 2017}}},\ }\href {\doibase
  10.1051/epjconf/201817501019} {\bibfield  {journal} {\bibinfo  {journal} {EPJ
  Web Conf.}\ }\textbf {\bibinfo {volume} {175}},\ \bibinfo {pages} {01019}
  (\bibinfo {year} {2018})},\ \Eprint {http://arxiv.org/abs/1708.08254}
  {arXiv:1708.08254 [hep-lat]} \BibitemShut {NoStop}%
\bibitem [{\citenamefont {Iwami}\ \emph {et~al.}(2015)\citenamefont {Iwami},
  \citenamefont {Ejiri}, \citenamefont {Kanaya}, \citenamefont {Nakagawa},
  \citenamefont {Yamamoto},\ and\ \citenamefont {Umeda}}]{Iwami:2015eba}%
  \BibitemOpen
  \bibfield  {author} {\bibinfo {author} {\bibfnamefont {R.}~\bibnamefont
  {Iwami}}, \bibinfo {author} {\bibfnamefont {S.}~\bibnamefont {Ejiri}},
  \bibinfo {author} {\bibfnamefont {K.}~\bibnamefont {Kanaya}}, \bibinfo
  {author} {\bibfnamefont {Y.}~\bibnamefont {Nakagawa}}, \bibinfo {author}
  {\bibfnamefont {D.}~\bibnamefont {Yamamoto}}, \ and\ \bibinfo {author}
  {\bibfnamefont {T.}~\bibnamefont {Umeda}},\ }\href {\doibase
  10.1103/PhysRevD.92.094507} {\bibfield  {journal} {\bibinfo  {journal} {Phys.
  Rev.}\ }\textbf {\bibinfo {volume} {D92}},\ \bibinfo {pages} {094507}
  (\bibinfo {year} {2015})},\ \Eprint {http://arxiv.org/abs/1508.01747}
  {arXiv:1508.01747 [hep-lat]} \BibitemShut {NoStop}%
\bibitem [{\citenamefont {Fischer}\ \emph
  {et~al.}(2014{\natexlab{a}})\citenamefont {Fischer}, \citenamefont {Fister},
  \citenamefont {Luecker},\ and\ \citenamefont {Pawlowski}}]{Fischer:2013eca}%
  \BibitemOpen
  \bibfield  {author} {\bibinfo {author} {\bibfnamefont {C.~S.}\ \bibnamefont
  {Fischer}}, \bibinfo {author} {\bibfnamefont {L.}~\bibnamefont {Fister}},
  \bibinfo {author} {\bibfnamefont {J.}~\bibnamefont {Luecker}}, \ and\
  \bibinfo {author} {\bibfnamefont {J.~M.}\ \bibnamefont {Pawlowski}},\ }\href
  {\doibase 10.1016/j.physletb.2014.03.057} {\bibfield  {journal} {\bibinfo
  {journal} {Phys. Lett.}\ }\textbf {\bibinfo {volume} {B732}},\ \bibinfo
  {pages} {273} (\bibinfo {year} {2014}{\natexlab{a}})},\ \Eprint
  {http://arxiv.org/abs/1306.6022} {arXiv:1306.6022 [hep-ph]} \BibitemShut
  {NoStop}%
\bibitem [{\citenamefont {Fischer}\ \emph
  {et~al.}(2014{\natexlab{b}})\citenamefont {Fischer}, \citenamefont
  {Luecker},\ and\ \citenamefont {Welzbacher}}]{Fischer:2014ata}%
  \BibitemOpen
  \bibfield  {author} {\bibinfo {author} {\bibfnamefont {C.~S.}\ \bibnamefont
  {Fischer}}, \bibinfo {author} {\bibfnamefont {J.}~\bibnamefont {Luecker}}, \
  and\ \bibinfo {author} {\bibfnamefont {C.~A.}\ \bibnamefont {Welzbacher}},\
  }\href {\doibase 10.1103/PhysRevD.90.034022} {\bibfield  {journal} {\bibinfo
  {journal} {Phys. Rev.}\ }\textbf {\bibinfo {volume} {D90}},\ \bibinfo {pages}
  {034022} (\bibinfo {year} {2014}{\natexlab{b}})},\ \Eprint
  {http://arxiv.org/abs/1405.4762} {arXiv:1405.4762 [hep-ph]} \BibitemShut
  {NoStop}%
\bibitem [{\citenamefont {Herbst}\ \emph {et~al.}(2011)\citenamefont {Herbst},
  \citenamefont {Pawlowski},\ and\ \citenamefont {Schaefer}}]{Herbst:2010rf}%
  \BibitemOpen
  \bibfield  {author} {\bibinfo {author} {\bibfnamefont {T.~K.}\ \bibnamefont
  {Herbst}}, \bibinfo {author} {\bibfnamefont {J.~M.}\ \bibnamefont
  {Pawlowski}}, \ and\ \bibinfo {author} {\bibfnamefont {B.-J.}\ \bibnamefont
  {Schaefer}},\ }\href {\doibase 10.1016/j.physletb.2010.12.003} {\bibfield
  {journal} {\bibinfo  {journal} {Phys. Lett.}\ }\textbf {\bibinfo {volume}
  {B696}},\ \bibinfo {pages} {58} (\bibinfo {year} {2011})},\ \Eprint
  {http://arxiv.org/abs/1008.0081} {arXiv:1008.0081 [hep-ph]} \BibitemShut
  {NoStop}%
\bibitem [{\citenamefont {Gupta}\ and\ \citenamefont
  {Tiwari}(2012)}]{Gupta:2011ez}%
  \BibitemOpen
  \bibfield  {author} {\bibinfo {author} {\bibfnamefont {U.~S.}\ \bibnamefont
  {Gupta}}\ and\ \bibinfo {author} {\bibfnamefont {V.~K.}\ \bibnamefont
  {Tiwari}},\ }\href {\doibase 10.1103/PhysRevD.85.014010} {\bibfield
  {journal} {\bibinfo  {journal} {Phys. Rev.}\ }\textbf {\bibinfo {volume}
  {D85}},\ \bibinfo {pages} {014010} (\bibinfo {year} {2012})},\ \Eprint
  {http://arxiv.org/abs/1107.1312} {arXiv:1107.1312 [hep-ph]} \BibitemShut
  {NoStop}%
\bibitem [{\citenamefont {Costa}\ \emph {et~al.}(2010)\citenamefont {Costa},
  \citenamefont {Ruivo}, \citenamefont {de~Sousa},\ and\ \citenamefont
  {Hansen}}]{Costa:2010zw}%
  \BibitemOpen
  \bibfield  {author} {\bibinfo {author} {\bibfnamefont {P.}~\bibnamefont
  {Costa}}, \bibinfo {author} {\bibfnamefont {M.~C.}\ \bibnamefont {Ruivo}},
  \bibinfo {author} {\bibfnamefont {C.~A.}\ \bibnamefont {de~Sousa}}, \ and\
  \bibinfo {author} {\bibfnamefont {H.}~\bibnamefont {Hansen}},\ }\href
  {\doibase 10.3390/sym2031338} {\bibfield  {journal} {\bibinfo  {journal}
  {Symmetry}\ }\textbf {\bibinfo {volume} {2}},\ \bibinfo {pages} {1338}
  (\bibinfo {year} {2010})},\ \Eprint {http://arxiv.org/abs/1007.1380}
  {arXiv:1007.1380 [hep-ph]} \BibitemShut {NoStop}%
\bibitem [{\citenamefont {Stiele}\ and\ \citenamefont
  {Schaffner-Bielich}(2016)}]{Stiele:2016cfs}%
  \BibitemOpen
  \bibfield  {author} {\bibinfo {author} {\bibfnamefont {R.}~\bibnamefont
  {Stiele}}\ and\ \bibinfo {author} {\bibfnamefont {J.}~\bibnamefont
  {Schaffner-Bielich}},\ }\href {\doibase 10.1103/PhysRevD.93.094014}
  {\bibfield  {journal} {\bibinfo  {journal} {Phys.~Rev.}\ }\textbf {\bibinfo
  {volume} {D 93}},\ \bibinfo {pages} {094014} (\bibinfo {year} {2016})},\
  \Eprint {http://arxiv.org/abs/1601.05731} {arXiv:1601.05731 [hep-ph]}
  \BibitemShut {NoStop}%
\bibitem [{\citenamefont {Costa}\ and\ \citenamefont
  {Pereira}(2019)}]{Costa:2019bua}%
  \BibitemOpen
  \bibfield  {author} {\bibinfo {author} {\bibfnamefont {P.}~\bibnamefont
  {Costa}}\ and\ \bibinfo {author} {\bibfnamefont {R.~C.}\ \bibnamefont
  {Pereira}},\ }\href {\doibase 10.3390/sym11040507} {\bibfield  {journal}
  {\bibinfo  {journal} {Symmetry}\ }\textbf {\bibinfo {volume} {11}},\ \bibinfo
  {pages} {507} (\bibinfo {year} {2019})},\ \Eprint
  {http://arxiv.org/abs/1904.05805} {arXiv:1904.05805 [hep-ph]} \BibitemShut
  {NoStop}%
\bibitem [{\citenamefont {Nikolov}\ \emph {et~al.}(1996)\citenamefont
  {Nikolov}, \citenamefont {Broniowski}, \citenamefont {Christov},
  \citenamefont {Ripka},\ and\ \citenamefont {Goeke}}]{Nikolov:1996jj}%
  \BibitemOpen
  \bibfield  {author} {\bibinfo {author} {\bibfnamefont {E.~N.}\ \bibnamefont
  {Nikolov}}, \bibinfo {author} {\bibfnamefont {W.}~\bibnamefont {Broniowski}},
  \bibinfo {author} {\bibfnamefont {C.~V.}\ \bibnamefont {Christov}}, \bibinfo
  {author} {\bibfnamefont {G.}~\bibnamefont {Ripka}}, \ and\ \bibinfo {author}
  {\bibfnamefont {K.}~\bibnamefont {Goeke}},\ }\href {\doibase
  10.1016/0375-9474(96)00231-X} {\bibfield  {journal} {\bibinfo  {journal}
  {Nucl. Phys.}\ }\textbf {\bibinfo {volume} {A608}},\ \bibinfo {pages} {411}
  (\bibinfo {year} {1996})},\ \Eprint {http://arxiv.org/abs/hep-ph/9602274}
  {arXiv:hep-ph/9602274 [hep-ph]} \BibitemShut {NoStop}%
\bibitem [{\citenamefont {Nemoto}\ \emph {et~al.}(2000)\citenamefont {Nemoto},
  \citenamefont {Naito},\ and\ \citenamefont {Oka}}]{Nemoto:1999qf}%
  \BibitemOpen
  \bibfield  {author} {\bibinfo {author} {\bibfnamefont {Y.}~\bibnamefont
  {Nemoto}}, \bibinfo {author} {\bibfnamefont {K.}~\bibnamefont {Naito}}, \
  and\ \bibinfo {author} {\bibfnamefont {M.}~\bibnamefont {Oka}},\ }\href
  {\doibase 10.1007/s100500070042} {\bibfield  {journal} {\bibinfo  {journal}
  {Eur. Phys. J.}\ }\textbf {\bibinfo {volume} {A9}},\ \bibinfo {pages} {245}
  (\bibinfo {year} {2000})},\ \Eprint {http://arxiv.org/abs/hep-ph/9911431}
  {arXiv:hep-ph/9911431 [hep-ph]} \BibitemShut {NoStop}%
\bibitem [{\citenamefont {Oertel}\ \emph {et~al.}(2001)\citenamefont {Oertel},
  \citenamefont {Buballa},\ and\ \citenamefont {Wambach}}]{Oertel:2000jp}%
  \BibitemOpen
  \bibfield  {author} {\bibinfo {author} {\bibfnamefont {M.}~\bibnamefont
  {Oertel}}, \bibinfo {author} {\bibfnamefont {M.}~\bibnamefont {Buballa}}, \
  and\ \bibinfo {author} {\bibfnamefont {J.}~\bibnamefont {Wambach}},\ }\href
  {\doibase 10.1134/1.1368226} {\bibfield  {journal} {\bibinfo  {journal}
  {Phys. Atom. Nucl.}\ }\textbf {\bibinfo {volume} {64}},\ \bibinfo {pages}
  {698} (\bibinfo {year} {2001})},\ \bibinfo {note} {[Yad. Fiz.64,757(2001)]},\
  \Eprint {http://arxiv.org/abs/hep-ph/0008131} {arXiv:hep-ph/0008131 [hep-ph]}
  \BibitemShut {NoStop}%
\bibitem [{\citenamefont {Baacke}\ and\ \citenamefont
  {Michalski}(2003)}]{Baacke:2002pi}%
  \BibitemOpen
  \bibfield  {author} {\bibinfo {author} {\bibfnamefont {J.}~\bibnamefont
  {Baacke}}\ and\ \bibinfo {author} {\bibfnamefont {S.}~\bibnamefont
  {Michalski}},\ }\href {\doibase 10.1103/PhysRevD.67.085006} {\bibfield
  {journal} {\bibinfo  {journal} {Phys. Rev.}\ }\textbf {\bibinfo {volume}
  {D67}},\ \bibinfo {pages} {085006} (\bibinfo {year} {2003})},\ \Eprint
  {http://arxiv.org/abs/hep-ph/0210060} {arXiv:hep-ph/0210060 [hep-ph]}
  \BibitemShut {NoStop}%
\bibitem [{\citenamefont {Andersen}\ and\ \citenamefont
  {Brauner}(2008)}]{Andersen:2008qk}%
  \BibitemOpen
  \bibfield  {author} {\bibinfo {author} {\bibfnamefont {J.~O.}\ \bibnamefont
  {Andersen}}\ and\ \bibinfo {author} {\bibfnamefont {T.}~\bibnamefont
  {Brauner}},\ }\href {\doibase 10.1103/PhysRevD.78.014030} {\bibfield
  {journal} {\bibinfo  {journal} {Phys. Rev.}\ }\textbf {\bibinfo {volume}
  {D78}},\ \bibinfo {pages} {014030} (\bibinfo {year} {2008})},\ \Eprint
  {http://arxiv.org/abs/0804.4604} {arXiv:0804.4604 [hep-ph]} \BibitemShut
  {NoStop}%
\bibitem [{\citenamefont {Muller}\ \emph {et~al.}(2010)\citenamefont {Muller},
  \citenamefont {Buballa},\ and\ \citenamefont {Wambach}}]{Muller:2010am}%
  \BibitemOpen
  \bibfield  {author} {\bibinfo {author} {\bibfnamefont {D.}~\bibnamefont
  {Muller}}, \bibinfo {author} {\bibfnamefont {M.}~\bibnamefont {Buballa}}, \
  and\ \bibinfo {author} {\bibfnamefont {J.}~\bibnamefont {Wambach}},\ }\href
  {\doibase 10.1103/PhysRevD.81.094022} {\bibfield  {journal} {\bibinfo
  {journal} {Phys. Rev.}\ }\textbf {\bibinfo {volume} {D81}},\ \bibinfo {pages}
  {094022} (\bibinfo {year} {2010})},\ \Eprint {http://arxiv.org/abs/1002.4252}
  {arXiv:1002.4252 [hep-ph]} \BibitemShut {NoStop}%
\bibitem [{\citenamefont {Yamazaki}\ and\ \citenamefont
  {Matsui}(2013)}]{YAMAZAKI201319}%
  \BibitemOpen
  \bibfield  {author} {\bibinfo {author} {\bibfnamefont {K.}~\bibnamefont
  {Yamazaki}}\ and\ \bibinfo {author} {\bibfnamefont {T.}~\bibnamefont
  {Matsui}},\ }\href {\doibase https://doi.org/10.1016/j.nuclphysa.2013.05.018}
  {\bibfield  {journal} {\bibinfo  {journal} {Nuclear Physics A}\ }\textbf
  {\bibinfo {volume} {913}},\ \bibinfo {pages} {19 } (\bibinfo {year}
  {2013})}\BibitemShut {NoStop}%
\bibitem [{\citenamefont {Zacchi}\ and\ \citenamefont
  {Schaffner-Bielich}(2018)}]{Zacchi:2017ahv}%
  \BibitemOpen
  \bibfield  {author} {\bibinfo {author} {\bibfnamefont {A.}~\bibnamefont
  {Zacchi}}\ and\ \bibinfo {author} {\bibfnamefont {J.}~\bibnamefont
  {Schaffner-Bielich}},\ }\href {\doibase 10.1103/PhysRevD.97.074011}
  {\bibfield  {journal} {\bibinfo  {journal} {Phys. Rev.}\ }\textbf {\bibinfo
  {volume} {D97}},\ \bibinfo {pages} {074011} (\bibinfo {year} {2018})},\
  \Eprint {http://arxiv.org/abs/1712.01629} {arXiv:1712.01629 [hep-ph]}
  \BibitemShut {NoStop}%
\bibitem [{\citenamefont {C{\^a}mara~Pereira}\ and\ \citenamefont
  {Costa}(2020)}]{CamaraPereira:2020ipu}%
  \BibitemOpen
  \bibfield  {author} {\bibinfo {author} {\bibfnamefont {R.}~\bibnamefont
  {C{\^a}mara~Pereira}}\ and\ \bibinfo {author} {\bibfnamefont
  {P.}~\bibnamefont {Costa}},\ }\href {\doibase 10.1103/PhysRevD.101.054025}
  {\bibfield  {journal} {\bibinfo  {journal} {Phys. Rev.}\ }\textbf {\bibinfo
  {volume} {D101}},\ \bibinfo {pages} {054025} (\bibinfo {year} {2020})},\
  \Eprint {http://arxiv.org/abs/2003.08430} {arXiv:2003.08430 [hep-ph]}
  \BibitemShut {NoStop}%
\bibitem [{\citenamefont {{\'A}vila}\ and\ \citenamefont
  {Birse}(2015)}]{Avila:2015iza}%
  \BibitemOpen
  \bibfield  {author} {\bibinfo {author} {\bibfnamefont {B.~J.}\ \bibnamefont
  {{\'A}vila}}\ and\ \bibinfo {author} {\bibfnamefont {M.~C.}\ \bibnamefont
  {Birse}},\ }\href {\doibase 10.1103/PhysRevA.92.023601} {\bibfield  {journal}
  {\bibinfo  {journal} {Phys. Rev.}\ }\textbf {\bibinfo {volume} {A92}},\
  \bibinfo {pages} {023601} (\bibinfo {year} {2015})},\ \Eprint
  {http://arxiv.org/abs/1506.04949} {arXiv:1506.04949 [cond-mat.quant-gas]}
  \BibitemShut {NoStop}%
\bibitem [{\citenamefont {Saueressig}\ \emph {et~al.}(2016)\citenamefont
  {Saueressig}, \citenamefont {Alkofer}, \citenamefont {D'Odorico},\ and\
  \citenamefont {Vidotto}}]{Saueressig:2015xua}%
  \BibitemOpen
  \bibfield  {author} {\bibinfo {author} {\bibfnamefont {F.}~\bibnamefont
  {Saueressig}}, \bibinfo {author} {\bibfnamefont {N.}~\bibnamefont {Alkofer}},
  \bibinfo {author} {\bibfnamefont {G.}~\bibnamefont {D'Odorico}}, \ and\
  \bibinfo {author} {\bibfnamefont {F.}~\bibnamefont {Vidotto}},\ }\bibfield
  {booktitle} {\emph {\bibinfo {booktitle} {{Proceedings, 14th International
  Symposium Frontiers of Fundamental Physics (FFP14): Marseille, France, July
  15-18, 2014}}},\ }\href@noop {} {\bibfield  {journal} {\bibinfo  {journal}
  {PoS}\ }\textbf {\bibinfo {volume} {FFP14}},\ \bibinfo {pages} {174}
  (\bibinfo {year} {2016})},\ \Eprint {http://arxiv.org/abs/1503.06472}
  {arXiv:1503.06472 [hep-th]} \BibitemShut {NoStop}%
\bibitem [{\citenamefont {Safari}\ and\ \citenamefont
  {Vacca}(2017)}]{Safari:2017irw}%
  \BibitemOpen
  \bibfield  {author} {\bibinfo {author} {\bibfnamefont {M.}~\bibnamefont
  {Safari}}\ and\ \bibinfo {author} {\bibfnamefont {G.~P.}\ \bibnamefont
  {Vacca}},\ }\href@noop {} {\  (\bibinfo {year} {2017})},\ \Eprint
  {http://arxiv.org/abs/1708.09795} {arXiv:1708.09795 [hep-th]} \BibitemShut
  {NoStop}%
\bibitem [{\citenamefont {Fukushima}\ and\ \citenamefont
  {Pawlowski}(2012)}]{Fukushima:2012xw}%
  \BibitemOpen
  \bibfield  {author} {\bibinfo {author} {\bibfnamefont {K.}~\bibnamefont
  {Fukushima}}\ and\ \bibinfo {author} {\bibfnamefont {J.~M.}\ \bibnamefont
  {Pawlowski}},\ }\href {\doibase 10.1103/PhysRevD.86.076013} {\bibfield
  {journal} {\bibinfo  {journal} {Phys. Rev.}\ }\textbf {\bibinfo {volume}
  {D86}},\ \bibinfo {pages} {076013} (\bibinfo {year} {2012})},\ \Eprint
  {http://arxiv.org/abs/1203.4330} {arXiv:1203.4330 [hep-ph]} \BibitemShut
  {NoStop}%
\bibitem [{\citenamefont {Braun}(2012)}]{Braun:2011pp}%
  \BibitemOpen
  \bibfield  {author} {\bibinfo {author} {\bibfnamefont {J.}~\bibnamefont
  {Braun}},\ }\href {\doibase 10.1088/0954-3899/39/3/033001} {\bibfield
  {journal} {\bibinfo  {journal} {J. Phys.}\ }\textbf {\bibinfo {volume}
  {G39}},\ \bibinfo {pages} {033001} (\bibinfo {year} {2012})},\ \Eprint
  {http://arxiv.org/abs/1108.4449} {arXiv:1108.4449 [hep-ph]} \BibitemShut
  {NoStop}%
\bibitem [{\citenamefont {Aoki}\ \emph {et~al.}(2014)\citenamefont {Aoki},
  \citenamefont {Kumamoto},\ and\ \citenamefont {Sato}}]{Aoki:2014ola}%
  \BibitemOpen
  \bibfield  {author} {\bibinfo {author} {\bibfnamefont {K.-I.}\ \bibnamefont
  {Aoki}}, \bibinfo {author} {\bibfnamefont {S.-I.}\ \bibnamefont {Kumamoto}},
  \ and\ \bibinfo {author} {\bibfnamefont {D.}~\bibnamefont {Sato}},\ }\href
  {\doibase 10.1093/ptep/ptu039} {\bibfield  {journal} {\bibinfo  {journal}
  {PTEP}\ }\textbf {\bibinfo {volume} {2014}},\ \bibinfo {pages} {043B05}
  (\bibinfo {year} {2014})},\ \Eprint {http://arxiv.org/abs/1403.0174}
  {arXiv:1403.0174 [hep-th]} \BibitemShut {NoStop}%
\bibitem [{\citenamefont {Aoki}\ and\ \citenamefont
  {Yamada}(2015)}]{Aoki:2015hsa}%
  \BibitemOpen
  \bibfield  {author} {\bibinfo {author} {\bibfnamefont {K.-I.}\ \bibnamefont
  {Aoki}}\ and\ \bibinfo {author} {\bibfnamefont {M.}~\bibnamefont {Yamada}},\
  }\href {\doibase 10.1142/S0217751X15501808} {\bibfield  {journal} {\bibinfo
  {journal} {Int. J. Mod. Phys.}\ }\textbf {\bibinfo {volume} {A30}},\ \bibinfo
  {pages} {1550180} (\bibinfo {year} {2015})},\ \Eprint
  {http://arxiv.org/abs/1504.00749} {arXiv:1504.00749 [hep-ph]} \BibitemShut
  {NoStop}%
\bibitem [{\citenamefont {Schaefer}\ and\ \citenamefont
  {Wambach}(2005)}]{Schaefer:2004en}%
  \BibitemOpen
  \bibfield  {author} {\bibinfo {author} {\bibfnamefont {B.-J.}\ \bibnamefont
  {Schaefer}}\ and\ \bibinfo {author} {\bibfnamefont {J.}~\bibnamefont
  {Wambach}},\ }\href {\doibase 10.1016/j.nuclphysa.2005.04.012} {\bibfield
  {journal} {\bibinfo  {journal} {Nucl. Phys.}\ }\textbf {\bibinfo {volume}
  {A757}},\ \bibinfo {pages} {479} (\bibinfo {year} {2005})},\ \Eprint
  {http://arxiv.org/abs/nucl-th/0403039} {arXiv:nucl-th/0403039 [nucl-th]}
  \BibitemShut {NoStop}%
\bibitem [{\citenamefont {Herbst}\ \emph {et~al.}(2013)\citenamefont {Herbst},
  \citenamefont {Pawlowski},\ and\ \citenamefont {Schaefer}}]{Herbst:2013ail}%
  \BibitemOpen
  \bibfield  {author} {\bibinfo {author} {\bibfnamefont {T.~K.}\ \bibnamefont
  {Herbst}}, \bibinfo {author} {\bibfnamefont {J.~M.}\ \bibnamefont
  {Pawlowski}}, \ and\ \bibinfo {author} {\bibfnamefont {B.-J.}\ \bibnamefont
  {Schaefer}},\ }\href {\doibase 10.1103/PhysRevD.88.014007} {\bibfield
  {journal} {\bibinfo  {journal} {Phys. Rev.}\ }\textbf {\bibinfo {volume}
  {D88}},\ \bibinfo {pages} {014007} (\bibinfo {year} {2013})},\ \Eprint
  {http://arxiv.org/abs/1302.1426} {arXiv:1302.1426 [hep-ph]} \BibitemShut
  {NoStop}%
\bibitem [{\citenamefont {Fu}\ and\ \citenamefont
  {Pawlowski}(2015)}]{Fu:2015naa}%
  \BibitemOpen
  \bibfield  {author} {\bibinfo {author} {\bibfnamefont {W.-j.}\ \bibnamefont
  {Fu}}\ and\ \bibinfo {author} {\bibfnamefont {J.~M.}\ \bibnamefont
  {Pawlowski}},\ }\href {\doibase 10.1103/PhysRevD.92.116006} {\bibfield
  {journal} {\bibinfo  {journal} {Phys. Rev.}\ }\textbf {\bibinfo {volume}
  {D92}},\ \bibinfo {pages} {116006} (\bibinfo {year} {2015})},\ \Eprint
  {http://arxiv.org/abs/1508.06504} {arXiv:1508.06504 [hep-ph]} \BibitemShut
  {NoStop}%
\bibitem [{\citenamefont {Herbst}\ \emph {et~al.}(2014)\citenamefont {Herbst},
  \citenamefont {Mitter}, \citenamefont {Pawlowski}, \citenamefont {Schaefer},\
  and\ \citenamefont {Stiele}}]{Herbst:2013ufa}%
  \BibitemOpen
  \bibfield  {author} {\bibinfo {author} {\bibfnamefont {T.~K.}\ \bibnamefont
  {Herbst}}, \bibinfo {author} {\bibfnamefont {M.}~\bibnamefont {Mitter}},
  \bibinfo {author} {\bibfnamefont {J.~M.}\ \bibnamefont {Pawlowski}}, \bibinfo
  {author} {\bibfnamefont {B.-J.}\ \bibnamefont {Schaefer}}, \ and\ \bibinfo
  {author} {\bibfnamefont {R.}~\bibnamefont {Stiele}},\ }\href {\doibase
  10.1016/j.physletb.2014.02.045} {\bibfield  {journal} {\bibinfo  {journal}
  {Phys. Lett.}\ }\textbf {\bibinfo {volume} {B731}},\ \bibinfo {pages} {248}
  (\bibinfo {year} {2014})},\ \Eprint {http://arxiv.org/abs/1308.3621}
  {arXiv:1308.3621 [hep-ph]} \BibitemShut {NoStop}%
\bibitem [{\citenamefont {Tripolt}\ \emph {et~al.}(2014)\citenamefont
  {Tripolt}, \citenamefont {Strodthoff}, \citenamefont {von Smekal},\ and\
  \citenamefont {Wambach}}]{Tripolt:2013jra}%
  \BibitemOpen
  \bibfield  {author} {\bibinfo {author} {\bibfnamefont {R.-A.}\ \bibnamefont
  {Tripolt}}, \bibinfo {author} {\bibfnamefont {N.}~\bibnamefont {Strodthoff}},
  \bibinfo {author} {\bibfnamefont {L.}~\bibnamefont {von Smekal}}, \ and\
  \bibinfo {author} {\bibfnamefont {J.}~\bibnamefont {Wambach}},\ }\href
  {\doibase 10.1103/PhysRevD.89.034010} {\bibfield  {journal} {\bibinfo
  {journal} {Phys. Rev.}\ }\textbf {\bibinfo {volume} {D89}},\ \bibinfo {pages}
  {034010} (\bibinfo {year} {2014})},\ \Eprint {http://arxiv.org/abs/1311.0630}
  {arXiv:1311.0630 [hep-ph]} \BibitemShut {NoStop}%
\bibitem [{\citenamefont {Jung}\ \emph {et~al.}(2017)\citenamefont {Jung},
  \citenamefont {Rennecke}, \citenamefont {Tripolt}, \citenamefont {von
  Smekal},\ and\ \citenamefont {Wambach}}]{Jung:2016yxl}%
  \BibitemOpen
  \bibfield  {author} {\bibinfo {author} {\bibfnamefont {C.}~\bibnamefont
  {Jung}}, \bibinfo {author} {\bibfnamefont {F.}~\bibnamefont {Rennecke}},
  \bibinfo {author} {\bibfnamefont {R.-A.}\ \bibnamefont {Tripolt}}, \bibinfo
  {author} {\bibfnamefont {L.}~\bibnamefont {von Smekal}}, \ and\ \bibinfo
  {author} {\bibfnamefont {J.}~\bibnamefont {Wambach}},\ }\href {\doibase
  10.1103/PhysRevD.95.036020} {\bibfield  {journal} {\bibinfo  {journal} {Phys.
  Rev.}\ }\textbf {\bibinfo {volume} {D95}},\ \bibinfo {pages} {036020}
  (\bibinfo {year} {2017})},\ \Eprint {http://arxiv.org/abs/1610.08754}
  {arXiv:1610.08754 [hep-ph]} \BibitemShut {NoStop}%
\bibitem [{\citenamefont {Andersen}\ \emph {et~al.}(2014)\citenamefont
  {Andersen}, \citenamefont {Naylor},\ and\ \citenamefont
  {Tranberg}}]{Andersen:2013swa}%
  \BibitemOpen
  \bibfield  {author} {\bibinfo {author} {\bibfnamefont {J.~O.}\ \bibnamefont
  {Andersen}}, \bibinfo {author} {\bibfnamefont {W.~R.}\ \bibnamefont
  {Naylor}}, \ and\ \bibinfo {author} {\bibfnamefont {A.}~\bibnamefont
  {Tranberg}},\ }\href {\doibase 10.1007/JHEP04(2014)187} {\bibfield  {journal}
  {\bibinfo  {journal} {JHEP}\ }\textbf {\bibinfo {volume} {04}},\ \bibinfo
  {pages} {187} (\bibinfo {year} {2014})},\ \Eprint
  {http://arxiv.org/abs/1311.2093} {arXiv:1311.2093 [hep-ph]} \BibitemShut
  {NoStop}%
\bibitem [{\citenamefont {Gies}(2012)}]{Gies:2006wv}%
  \BibitemOpen
  \bibfield  {author} {\bibinfo {author} {\bibfnamefont {H.}~\bibnamefont
  {Gies}},\ }\bibfield  {booktitle} {\emph {\bibinfo {booktitle}
  {{Renormalization group and effective field theory approaches to many-body
  systems}}},\ }\href {\doibase 10.1007/978-3-642-27320-9_6} {\bibfield
  {journal} {\bibinfo  {journal} {Lect. Notes Phys.}\ }\textbf {\bibinfo
  {volume} {852}},\ \bibinfo {pages} {287} (\bibinfo {year} {2012})},\ \Eprint
  {http://arxiv.org/abs/hep-ph/0611146} {arXiv:hep-ph/0611146 [hep-ph]}
  \BibitemShut {NoStop}%
\bibitem [{\citenamefont {Pawlowski}(2007)}]{Pawlowski:2005xe}%
  \BibitemOpen
  \bibfield  {author} {\bibinfo {author} {\bibfnamefont {J.~M.}\ \bibnamefont
  {Pawlowski}},\ }\href {\doibase 10.1016/j.aop.2007.01.007} {\bibfield
  {journal} {\bibinfo  {journal} {Annals Phys.}\ }\textbf {\bibinfo {volume}
  {322}},\ \bibinfo {pages} {2831} (\bibinfo {year} {2007})},\ \Eprint
  {http://arxiv.org/abs/hep-th/0512261} {arXiv:hep-th/0512261 [hep-th]}
  \BibitemShut {NoStop}%
\bibitem [{\citenamefont {Delamotte}(2012)}]{Delamotte:2007pf}%
  \BibitemOpen
  \bibfield  {author} {\bibinfo {author} {\bibfnamefont {B.}~\bibnamefont
  {Delamotte}},\ }\href {\doibase 10.1007/978-3-642-27320-9_2} {\bibfield
  {journal} {\bibinfo  {journal} {Lect. Notes Phys.}\ }\textbf {\bibinfo
  {volume} {852}},\ \bibinfo {pages} {49} (\bibinfo {year} {2012})},\ \Eprint
  {http://arxiv.org/abs/cond-mat/0702365} {arXiv:cond-mat/0702365
  [cond-mat.stat-mech]} \BibitemShut {NoStop}%
\bibitem [{\citenamefont {Tripolt}\ \emph {et~al.}(2018)\citenamefont
  {Tripolt}, \citenamefont {Schaefer}, \citenamefont {von Smekal},\ and\
  \citenamefont {Wambach}}]{Tripolt:2017zgc}%
  \BibitemOpen
  \bibfield  {author} {\bibinfo {author} {\bibfnamefont {R.-A.}\ \bibnamefont
  {Tripolt}}, \bibinfo {author} {\bibfnamefont {B.-J.}\ \bibnamefont
  {Schaefer}}, \bibinfo {author} {\bibfnamefont {L.}~\bibnamefont {von
  Smekal}}, \ and\ \bibinfo {author} {\bibfnamefont {J.}~\bibnamefont
  {Wambach}},\ }\href {\doibase 10.1103/PhysRevD.97.034022} {\bibfield
  {journal} {\bibinfo  {journal} {Phys. Rev.}\ }\textbf {\bibinfo {volume}
  {D97}},\ \bibinfo {pages} {034022} (\bibinfo {year} {2018})},\ \Eprint
  {http://arxiv.org/abs/1709.05991} {arXiv:1709.05991 [hep-ph]} \BibitemShut
  {NoStop}%
\bibitem [{\citenamefont {Zacchi}\ \emph {et~al.}(2015)\citenamefont {Zacchi},
  \citenamefont {Stiele},\ and\ \citenamefont
  {Schaffner-Bielich}}]{Zacchi:2015lwa}%
  \BibitemOpen
  \bibfield  {author} {\bibinfo {author} {\bibfnamefont {A.}~\bibnamefont
  {Zacchi}}, \bibinfo {author} {\bibfnamefont {R.}~\bibnamefont {Stiele}}, \
  and\ \bibinfo {author} {\bibfnamefont {J.}~\bibnamefont
  {Schaffner-Bielich}},\ }\href {\doibase 10.1103/PhysRevD.92.045022}
  {\bibfield  {journal} {\bibinfo  {journal} {Phys. Rev.}\ }\textbf {\bibinfo
  {volume} {D92}},\ \bibinfo {pages} {045022} (\bibinfo {year} {2015})},\
  \Eprint {http://arxiv.org/abs/1506.01868} {arXiv:1506.01868 [astro-ph.HE]}
  \BibitemShut {NoStop}%
\bibitem [{\citenamefont {Pereira}\ \emph {et~al.}(2016)\citenamefont
  {Pereira}, \citenamefont {Costa},\ and\ \citenamefont
  {Provid{\^e}ncia}}]{Pereira:2016dfg}%
  \BibitemOpen
  \bibfield  {author} {\bibinfo {author} {\bibfnamefont {R.~C.}\ \bibnamefont
  {Pereira}}, \bibinfo {author} {\bibfnamefont {P.}~\bibnamefont {Costa}}, \
  and\ \bibinfo {author} {\bibfnamefont {C.}~\bibnamefont {Provid{\^e}ncia}},\
  }\href {\doibase 10.1103/PhysRevD.94.094001} {\bibfield  {journal} {\bibinfo
  {journal} {Phys. Rev.}\ }\textbf {\bibinfo {volume} {D94}},\ \bibinfo {pages}
  {094001} (\bibinfo {year} {2016})},\ \Eprint
  {http://arxiv.org/abs/1610.06435} {arXiv:1610.06435 [nucl-th]} \BibitemShut
  {NoStop}%
\bibitem [{\citenamefont {Otto}\ \emph {et~al.}(2019)\citenamefont {Otto},
  \citenamefont {Oertel},\ and\ \citenamefont {Schaefer}}]{Otto:2019zjy}%
  \BibitemOpen
  \bibfield  {author} {\bibinfo {author} {\bibfnamefont {K.}~\bibnamefont
  {Otto}}, \bibinfo {author} {\bibfnamefont {M.}~\bibnamefont {Oertel}}, \ and\
  \bibinfo {author} {\bibfnamefont {B.-J.}\ \bibnamefont {Schaefer}},\
  }\href@noop {} {\  (\bibinfo {year} {2019})},\ \Eprint
  {http://arxiv.org/abs/1910.11929} {arXiv:1910.11929 [hep-ph]} \BibitemShut
  {NoStop}%
\bibitem [{\citenamefont {Bratovic}\ \emph {et~al.}(2013)\citenamefont
  {Bratovic}, \citenamefont {Hatsuda},\ and\ \citenamefont
  {Weise}}]{Bratovic:2012qs}%
  \BibitemOpen
  \bibfield  {author} {\bibinfo {author} {\bibfnamefont {N.~M.}\ \bibnamefont
  {Bratovic}}, \bibinfo {author} {\bibfnamefont {T.}~\bibnamefont {Hatsuda}}, \
  and\ \bibinfo {author} {\bibfnamefont {W.}~\bibnamefont {Weise}},\ }\href
  {\doibase 10.1016/j.physletb.2013.01.003} {\bibfield  {journal} {\bibinfo
  {journal} {Phys. Lett.}\ }\textbf {\bibinfo {volume} {B719}},\ \bibinfo
  {pages} {131} (\bibinfo {year} {2013})},\ \Eprint
  {http://arxiv.org/abs/1204.3788} {arXiv:1204.3788 [hep-ph]} \BibitemShut
  {NoStop}%
\bibitem [{\citenamefont {Das}(1993)}]{das1993field}%
  \BibitemOpen
  \bibfield  {author} {\bibinfo {author} {\bibfnamefont {A.}~\bibnamefont
  {Das}},\ }\href {https://books.google.pt/books?id=UbzsCgAAQBAJ} {\emph
  {\bibinfo {title} {Field Theory: A Path Integral Approach}}},\ World
  Scientific Lecture Notes in Physics\ (\bibinfo {year} {1993})\BibitemShut
  {NoStop}%
\bibitem [{\citenamefont {Drews}\ and\ \citenamefont
  {Weise}(2015)}]{Drews:2014spa}%
  \BibitemOpen
  \bibfield  {author} {\bibinfo {author} {\bibfnamefont {M.}~\bibnamefont
  {Drews}}\ and\ \bibinfo {author} {\bibfnamefont {W.}~\bibnamefont {Weise}},\
  }\href {\doibase 10.1103/PhysRevC.91.035802} {\bibfield  {journal} {\bibinfo
  {journal} {Phys. Rev.}\ }\textbf {\bibinfo {volume} {C91}},\ \bibinfo {pages}
  {035802} (\bibinfo {year} {2015})},\ \Eprint {http://arxiv.org/abs/1412.7655}
  {arXiv:1412.7655 [nucl-th]} \BibitemShut {NoStop}%
\bibitem [{\citenamefont {Wetterich}(1993)}]{Wetterich:1992yh}%
  \BibitemOpen
  \bibfield  {author} {\bibinfo {author} {\bibfnamefont {C.}~\bibnamefont
  {Wetterich}},\ }\href {\doibase 10.1016/0370-2693(93)90726-X} {\bibfield
  {journal} {\bibinfo  {journal} {Phys. Lett.}\ }\textbf {\bibinfo {volume}
  {B301}},\ \bibinfo {pages} {90} (\bibinfo {year} {1993})},\ \Eprint
  {http://arxiv.org/abs/1710.05815} {arXiv:1710.05815 [hep-th]} \BibitemShut
  {NoStop}%
\bibitem [{\citenamefont {Litim}(2001)}]{Litim:2001up}%
  \BibitemOpen
  \bibfield  {author} {\bibinfo {author} {\bibfnamefont {D.~F.}\ \bibnamefont
  {Litim}},\ }\href {\doibase 10.1103/PhysRevD.64.105007} {\bibfield  {journal}
  {\bibinfo  {journal} {Phys. Rev.}\ }\textbf {\bibinfo {volume} {D64}},\
  \bibinfo {pages} {105007} (\bibinfo {year} {2001})},\ \Eprint
  {http://arxiv.org/abs/hep-th/0103195} {arXiv:hep-th/0103195 [hep-th]}
  \BibitemShut {NoStop}%
\bibitem [{\citenamefont {Kapusta}\ and\ \citenamefont
  {Gale}(2011)}]{Kapusta:2006pm}%
  \BibitemOpen
  \bibfield  {author} {\bibinfo {author} {\bibfnamefont {J.~I.}\ \bibnamefont
  {Kapusta}}\ and\ \bibinfo {author} {\bibfnamefont {C.}~\bibnamefont {Gale}},\
  }\href {\doibase 10.1017/CBO9780511535130} {\emph {\bibinfo {title}
  {{Finite-temperature field theory: Principles and applications}}}},\
  Cambridge Monographs on Mathematical Physics\ (\bibinfo  {publisher}
  {Cambridge University Press},\ \bibinfo {year} {2011})\BibitemShut {NoStop}%
\bibitem [{\citenamefont {Zhang}\ \emph {et~al.}(2017)\citenamefont {Zhang},
  \citenamefont {Hou}, \citenamefont {Kojo},\ and\ \citenamefont
  {Qin}}]{Zhang:2017icm}%
  \BibitemOpen
  \bibfield  {author} {\bibinfo {author} {\bibfnamefont {H.}~\bibnamefont
  {Zhang}}, \bibinfo {author} {\bibfnamefont {D.}~\bibnamefont {Hou}}, \bibinfo
  {author} {\bibfnamefont {T.}~\bibnamefont {Kojo}}, \ and\ \bibinfo {author}
  {\bibfnamefont {B.}~\bibnamefont {Qin}},\ }\href {\doibase
  10.1103/PhysRevD.96.114029} {\bibfield  {journal} {\bibinfo  {journal} {Phys.
  Rev.}\ }\textbf {\bibinfo {volume} {D96}},\ \bibinfo {pages} {114029}
  (\bibinfo {year} {2017})},\ \Eprint {http://arxiv.org/abs/1709.05654}
  {arXiv:1709.05654 [hep-ph]} \BibitemShut {NoStop}%
\bibitem [{\citenamefont {Drews}(2014)}]{drewsthesis}%
  \BibitemOpen
  \bibfield  {author} {\bibinfo {author} {\bibfnamefont {M.}~\bibnamefont
  {Drews}},\ }\emph {\bibinfo {title} {Renormalization group approach to dense
  baryonic matter}},\ \href@noop {} {Ph.D. thesis},\ \bibinfo  {school}
  {Technical University Munich} (\bibinfo {year} {2014})\BibitemShut {NoStop}%
\bibitem [{\citenamefont {Buballa}(2005)}]{Buballa:2003qv}%
  \BibitemOpen
  \bibfield  {author} {\bibinfo {author} {\bibfnamefont {M.}~\bibnamefont
  {Buballa}},\ }\href {\doibase 10.1016/j.physrep.2004.11.004} {\bibfield
  {journal} {\bibinfo  {journal} {Phys. Rept.}\ }\textbf {\bibinfo {volume}
  {407}},\ \bibinfo {pages} {205} (\bibinfo {year} {2005})},\ \Eprint
  {http://arxiv.org/abs/hep-ph/0402234} {arXiv:hep-ph/0402234 [hep-ph]}
  \BibitemShut {NoStop}%
\bibitem [{\citenamefont {Kamikado}\ \emph {et~al.}(2013)\citenamefont
  {Kamikado}, \citenamefont {Strodthoff}, \citenamefont {von Smekal},\ and\
  \citenamefont {Wambach}}]{Kamikado:2012bt}%
  \BibitemOpen
  \bibfield  {author} {\bibinfo {author} {\bibfnamefont {K.}~\bibnamefont
  {Kamikado}}, \bibinfo {author} {\bibfnamefont {N.}~\bibnamefont
  {Strodthoff}}, \bibinfo {author} {\bibfnamefont {L.}~\bibnamefont {von
  Smekal}}, \ and\ \bibinfo {author} {\bibfnamefont {J.}~\bibnamefont
  {Wambach}},\ }\href {\doibase 10.1016/j.physletb.2012.11.055} {\bibfield
  {journal} {\bibinfo  {journal} {Phys. Lett.}\ }\textbf {\bibinfo {volume}
  {B718}},\ \bibinfo {pages} {1044} (\bibinfo {year} {2013})},\ \Eprint
  {http://arxiv.org/abs/1207.0400} {arXiv:1207.0400 [hep-ph]} \BibitemShut
  {NoStop}%
\bibitem [{\citenamefont {Strodthoff}\ and\ \citenamefont {von
  Smekal}(2014)}]{Strodthoff:2013cua}%
  \BibitemOpen
  \bibfield  {author} {\bibinfo {author} {\bibfnamefont {N.}~\bibnamefont
  {Strodthoff}}\ and\ \bibinfo {author} {\bibfnamefont {L.}~\bibnamefont {von
  Smekal}},\ }\href {\doibase 10.1016/j.physletb.2014.03.008} {\bibfield
  {journal} {\bibinfo  {journal} {Phys. Lett.}\ }\textbf {\bibinfo {volume}
  {B731}},\ \bibinfo {pages} {350} (\bibinfo {year} {2014})},\ \Eprint
  {http://arxiv.org/abs/1306.2897} {arXiv:1306.2897 [hep-ph]} \BibitemShut
  {NoStop}%
\bibitem [{\citenamefont {Costa}(2016)}]{Costa:2016vbb}%
  \BibitemOpen
  \bibfield  {author} {\bibinfo {author} {\bibfnamefont {P.}~\bibnamefont
  {Costa}},\ }\href {\doibase 10.1103/PhysRevD.93.114035} {\bibfield  {journal}
  {\bibinfo  {journal} {Phys. Rev.}\ }\textbf {\bibinfo {volume} {D93}},\
  \bibinfo {pages} {114035} (\bibinfo {year} {2016})},\ \Eprint
  {http://arxiv.org/abs/1610.06433} {arXiv:1610.06433 [nucl-th]} \BibitemShut
  {NoStop}%
\bibitem [{\citenamefont {Ferreira}\ \emph {et~al.}(2018)\citenamefont
  {Ferreira}, \citenamefont {Costa},\ and\ \citenamefont
  {Provid{\^e}ncia}}]{Ferreira:2017wtx}%
  \BibitemOpen
  \bibfield  {author} {\bibinfo {author} {\bibfnamefont {M.}~\bibnamefont
  {Ferreira}}, \bibinfo {author} {\bibfnamefont {P.}~\bibnamefont {Costa}}, \
  and\ \bibinfo {author} {\bibfnamefont {C.}~\bibnamefont {Provid{\^e}ncia}},\
  }\href {\doibase 10.1103/PhysRevD.97.014014} {\bibfield  {journal} {\bibinfo
  {journal} {Phys. Rev.}\ }\textbf {\bibinfo {volume} {D97}},\ \bibinfo {pages}
  {014014} (\bibinfo {year} {2018})},\ \Eprint
  {http://arxiv.org/abs/1712.08378} {arXiv:1712.08378 [hep-ph]} \BibitemShut
  {NoStop}%
\bibitem [{\citenamefont {Glendenning}(1997)}]{norman1997compact}%
  \BibitemOpen
  \bibfield  {author} {\bibinfo {author} {\bibfnamefont {N.~K.}\ \bibnamefont
  {Glendenning}},\ }\href {https://books.google.pt/books?id=57XvAAAAMAAJ}
  {\emph {\bibinfo {title} {Compact Stars: Nuclear Physics, Particle Physics,
  and General Relativity}}},\ Astronomy and astrophysics library\ (\bibinfo
  {publisher} {Springer},\ \bibinfo {year} {1997})\BibitemShut {NoStop}%
\bibitem [{\citenamefont {Barnafoldi}\ \emph {et~al.}(2017)\citenamefont
  {Barnafoldi}, \citenamefont {Jakovac},\ and\ \citenamefont
  {Posfay}}]{Barnafoldi:2016tkd}%
  \BibitemOpen
  \bibfield  {author} {\bibinfo {author} {\bibfnamefont {G.~G.}\ \bibnamefont
  {Barnafoldi}}, \bibinfo {author} {\bibfnamefont {A.}~\bibnamefont {Jakovac}},
  \ and\ \bibinfo {author} {\bibfnamefont {P.}~\bibnamefont {Posfay}},\ }\href
  {\doibase 10.1103/PhysRevD.95.025004} {\bibfield  {journal} {\bibinfo
  {journal} {Phys. Rev.}\ }\textbf {\bibinfo {volume} {D95}},\ \bibinfo {pages}
  {025004} (\bibinfo {year} {2017})},\ \Eprint
  {http://arxiv.org/abs/1604.01717} {arXiv:1604.01717 [hep-th]} \BibitemShut
  {NoStop}%
\bibitem [{\citenamefont {Yokota}\ \emph {et~al.}(2016)\citenamefont {Yokota},
  \citenamefont {Kunihiro},\ and\ \citenamefont {Morita}}]{Yokota:2016tip}%
  \BibitemOpen
  \bibfield  {author} {\bibinfo {author} {\bibfnamefont {T.}~\bibnamefont
  {Yokota}}, \bibinfo {author} {\bibfnamefont {T.}~\bibnamefont {Kunihiro}}, \
  and\ \bibinfo {author} {\bibfnamefont {K.}~\bibnamefont {Morita}},\ }\href
  {\doibase 10.1093/ptep/ptw062} {\bibfield  {journal} {\bibinfo  {journal}
  {PTEP}\ }\textbf {\bibinfo {volume} {2016}},\ \bibinfo {pages} {073D01}
  (\bibinfo {year} {2016})},\ \Eprint {http://arxiv.org/abs/1603.02147}
  {arXiv:1603.02147 [hep-ph]} \BibitemShut {NoStop}%
\end{thebibliography}%

\end{document}